\documentclass[useAMS,usenatbib,a4]{mn2e}
\usepackage[]{graphicx}
\usepackage{epstopdf}

\title[Runaway stars as progenitors of  SNe and GRBs]{Runaway stars as progenitors of supernovae and gamma-ray bursts}
\author[John J. Eldridge]{John J. Eldridge$^{1}$ \thanks{E-mail: jje@ast.cam.ac.uk}, Norbert Langer$^{2,3}$ \& Christopher A. Tout$^{1}$ \\
$^{1}$Institute of Astronomy, The Observatories, University of Cambridge, Madingley Road, Cambridge, CB3 0HA\\
$^{2}$Argelander-Institut f\"ur Astronomie, Bonn University, Auf dem H\"ugel 71, 53121 Bonn, Germany\\
$^{3}$Astronomical Institute, University of Utrecht, Postbus 80000, 3508 TA Utrecht, The Netherlands\\}

\pagerange{\pageref{firstpage}--\pageref{lastpage}} \pubyear{2010}
\begin{document}
\maketitle
\label{firstpage}

\begin{abstract}
When a core collapse supernova occurs in a binary system, the
surviving star as well as the compact remnant emerging from the
supernova, may reach a substantial space velocity. With binary
population synthesis modelling at solar and one fifth of solar
metallicity, we predict the velocities of such runaway stars or
binaries.  We compile predictions for runaway OB stars, red
supergiants and Wolf-Rayet stars, either isolated or with a compact
companion.  For those stars or binaries which undergo a second stellar
explosion we compute their further evolution and the distance
travelled until a Type~II or Type~Ibc supernova or a long or short
gamma-ray burst occurs.  We find our predicted population of OB
runaway stars broadly matches the observed population of stars but, to
match the fastest observed Wolf-Rayet runaway stars, we require that
black holes receive an asymmetric kick upon formation.  We find that
at solar metallicity Type~Ic supernova progenitors travel shorter
distances than the progenitors of other supernova types because they
are typically more massive and thus have shorter lifetimes. Those of
Type~IIP supernovae can fly farthest about 48\,pc {\em on average} at
solar metallicity, with about 8 per cent of them reaching 100\,pc.  In
considering the consequences of assuming that the progenitors of long
gamma-ray bursts are spun-up secondary stars that experience
quasi-homogeneous evolution, we find that such evolution has a
dramatic effect on the population of runaway Wolf-Rayet stars and that
some 30 per cent of GRBs could occur a hundred parsecs or more from
their initial positions. We also consider mergers of double compact
object binaries consisting of neutron stars and/or black holes. We
find the most common type of visible mergers are neutron star--black
hole mergers that are roughly ten times more common than neutron
star--neutron star mergers. All compact mergers have a wide range of
merger times from years to Gyrs and are predicted to occur three
hundred times less often than supernovae in the Milky Way. We also
find that there may be a population of low-velocity neutron stars that
are ejected from a binary rather than by their own natal kick. These
neutrons stars need to be included when the distribution of neutron
star kicks is deduced from observations.
\end{abstract}
     

%
 
\section{Introduction}

Runaway stars are isolated stars or binaries which have escaped from
their parent clusters. A satisfactory set of observational
characteristics that defines them is difficult to find. Often, it is
assumed that a runaway star must have a space velocity of $30\,{\rm km
  \, s^{-1}}$ or greater. The most commonly observed massive runaway
stars are OB runaways
\citep{blaauw,gies,stone,hoogerwerf,dewit}. Currently about 56 such
stars are known in the Galaxy, \citep{hoogerwerf} with velocities up
to $200\,{\rm km \, s^{-1}}$. There are some cases known with
velocities greater than $500\,{\rm km \, s^{-1}}$
\citep{highvel}. These high velocities and the stellar lifetimes of a
few million years imply that such stars travel many parsecs from their
initial positions before themselves exploding in core-collapse
supernovae (SNe).

There are two scenarios that create massive runaway stars. The first
is dynamical ejection scenario (DES) where stars are ejected by close
encounters in a dense cluster. The second is the binary supernova
scenario (BSS) when one of the stars in a binary under goes a SN
explosion and the system becomes unbound. The companion star then
travels at roughly its pre-SN orbital velocity. It is thought that
both scenarios contribute similar numbers of runaway stars.  However
evidence suggests that the BSS could be responsible for up to two
thirds of observed runaways \citep{hoogerwerf}.  Methods to
discriminate between the two rely on the observational characteristics
of the runaway star. Typically BSS runaways are expected to have a
surface compositions that indicates they have experienced a binary
interaction, while typical DES runaways might have main-sequence
composition. A further complication is that some runaway stars,
especially those with velocities in excess of about $200\,{\rm km \,
  s^{-1}}$, may come from a binary that was ejected by the DES with
one star being boosted to high velocities after a BSS ejection as
discussed by \citet{twostepejection}. Such binaries would be less
common but could naturally explain the highest velocities observed for
runaway stars.

Interest in the final fate of runaway stars has increased owing to the
observations of \citet{hammer}. They observed that some long gamma-ray
bursts \citep{review} occur a few hundred parsecs away from the
nearest star forming region. \citet{cantiello} suggest that this can
be explained if the progenitor of the GRB was the secondary in a
binary system that was ejected via the BSS. 

Concurrently the observations made by \citet{fruchter}, \citet{kelly} and
\citet{snhalpha} also reveal details of how SNe are distributed in
their host galaxies. They found that type Ic SNe and GRBs tend to be
associated with the most luminous parts, and thus the sites of most
intense star-formation, of galaxies while type~II and type~Ib SNe
tend to be more evenly distributed throughout their host
galaxies. \citet{larsson} and \citet{raskin} put this down to the
difference in progenitor mass for the different SN types.  The
most massive stars explode earliest and thus closest to the sites of
active star formation. While less massive stars have longer lifetimes
and are not as closely associated with the most recent star formation. They did
not consider the effect of runaway stars. It is important to
understand how runaways change the distribution of SNe within a galaxy
to remove any systematic error in determining the nature of different
SN progenitors.

In this paper we concern ourselves with describing the population of
runaway stars from the BSS and their effect on the distribution of SNe
within a galaxy. First we describe our method of simulating the
runaways. Secondly we give our predictions for the space velocities of
runaway stars, binaries and compact remnants and the distances our
model runaways travel before exploding as SNe. Thirdly we outline how
this affects how SNe are distributed with respect to their initial
locations. Fourthly we perform a similar analysis for the merger of
double compact object binaries owing to orbital decay by gravitational
radiation. Finally we discuss our results and outline our conclusions.

\section{Computational Method}

Our method is built upon the models and the population synthesis code
described by \citet{EIT08}. We use the large number of detailed
stellar evolution models that were calculated with the Cambridge STARS
evolution code, created by \citet{E71} and updated by various authors
since \citep{P95,E03,EIT08}. The population synthesis code then uses
our detailed binary models to estimate various details of a binary
population that are of interest and comparable with observations. For
example, the relative numbers of different stellar types, the relative
rates of different SN types and, here, the expected velocities of
runaway stars. Similar studies have used binary population synthesis
to predict the runaway population before
\citep{vanrun1,vanrun2,vanbook,Dray2005}. However this is the first
time the effect of runaway stars on the distribution of SNe within
galaxies has been considered.

While a full description of our detailed models can be found in
\citet{EIT08} we provide a brief overview.  All the models employ our
standard mass-loss prescription because it agrees best with various
observations \citep{ETsne,EIT08}. For pre-WR mass loss, we use the
rates of \citet{dejager} except for OB stars for which we use the
theoretical rates of \citet{VKL2001}. When the star becomes a WR star
[$X_{\rm surface} < 0.4$, $\log (T_{\rm eff}/{\rm K}) > 4.0$], we use
the rates of \citet{NL00}. We scale all rates with the standard factor
$(Z/Z_{\odot})^{0.5}$ \citep{K87,H03}, except for the rates of
\citet{VKL2001} which include their own metallicity scaling.

We have modified our stellar evolution code to model binary evolution.
The details of our binary interaction algorithm are relatively simple
compared to the scheme outlined by \citet{HPT02}. We used their scheme
as a basis but we changed some details which cannot be directly
applied to our detailed stellar evolution calculation. We also make a
number of assumptions to keep our code relatively simple. Our aim was
to investigate the effect of enhanced mass loss due to binary
interactions on stellar lifetimes and populations. Therefore, we
concentrated on this rather than every possible physical process which
would add more uncertainty to our model. We also make assumptions in
calculating our synthetic population to avoid calculating a large
number of models.  For example, we do not model the accretion on to
the secondary in the detailed code. We take the final mass of the
secondary at the end of the primary evolution as the initial mass of
the secondary when we create our detailed secondary model. This avoids
calculating 10 times more secondary models than primary models.

We always define the primary as the initially more massive star and we
only evolve one star at a time with our detailed code. When we evolve
the primary in detail, it has a shorter evolutionary time-scale than
the secondary which remains on the main sequence until after the
primary completes its evolution and so we can determine the state of
the secondary with the single stellar evolution equations of
\citet{HPT00}. When we evolve the secondary in detail, we assume that
its companion is the compact remnant of the primary (a white dwarf,
neutron star or black hole) and treat this as a point mass.

If Roche lobe overflow occurs mass lost from the primary is
transferred to the secondary but not all is necessarily
accreted. Accretion causes the star to expand owing to increased total
mass and therefore an increased energy production rate if $\dot{M_2}\ge
M_2/\tau_{\rm KH}$, where $\tau_{\rm KH}$ is the thermal, or
Kelvin-Helmholtz, time-scale. We assume that the star's maximum
accretion rate is determined by its current mass and its thermal
timescale.  We define a maximum accretion rate for a star such that
$\dot{M}_{2,{\rm max}}= M_2/\tau_{\rm KH}$. If the accretion rate is
greater than this, then any additional mass and its orbital angular
momentum are lost from the system.  In general, stars with lower
masses have longer thermal time-scales than more massive
stars. Efficient transfer is only possible if the two stars are of
nearly equal mass so the thermal time-scales are similar. This is an
approximate treatment but provides a similar result to the more
complex model of \citet{petrovic} who included rotation and found that
it led to inefficient mass transfer. For compact companions, we derive
the maximum accretion rate from the Eddington limit \citep{edding}.

In summary our stellar models are a set of around 15,000 detailed
stellar evolution models of single stars, primary stars, secondary
stars and merged systems \citep{EIT08}. We also include a new series
of models to account for quasi-homogeneous evolution. These are
discussed below. For this work we made several refinements to our
population synthesis code to predict runaway velocities and account
for the possibility of SN reversal. That is when the initially
lower-mass secondary star undergoes its SN first after mass accretion.

\subsection{Population synthesis}

Our population synthesis calculations are built upon those described
by \citet{EIT08} with some improvements. Here we focus on the
evolution after the first SN in the binary. First we estimate the
lifetimes of the primary and the secondary star. Our secondary models
only have a limited companion mass range but this provides a
reasonable estimate because the lifetime varies little with different
amounts of mass loss. We compare the lifetimes of the two stars to
check whether the SN order is reversed. We find that it is more likely
at higher metallicity but only occasional, occurring in 7 per
cent of our binary systems. If the secondary lifetime is shorter than
the primary lifetime we perform a similar calculation as described
below but now the secondary explodes first and may eject the primary
star. This gives rise to more stripped SNe occurring away from their
initial position.

If a star has a carbon/oxygen core mass greater than $1.38M_{\odot}$
and the final mass of the star is greater than $2M_{\odot}$ we assume
it explodes in a SN. We select the SN type as described below.We
estimate what type of compact remnant a stellar model will produce by
using the method outline in \citet{ETsne}. We assume that first a
neutron star is formed at the centre of the star after core collapse
of mass $M_{\rm Ch} = 1.4 M_{\odot}$. This produces about $10^{46}$J
of energy from the release of gravitational binding energy in neutron
star formation. We then assume a hundredth of this energy is
transferred into the envelope by some unknown mechanism. The current
suggestion is the transfer occurs via neutrinos released from forming
the protoneutron star that are thermalized within the envelope or
dense outer parts of the core.  We integrate the binding energy of the
star from the surface towards the centre until we reach
$10^{44}J$. The envelope outside this region is ejected with the
remaining amount forming the remnant.  If we have $M_{\rm rem} > 3
M_{\odot}$, it is a black hole and we set $M_{\rm BH}=M_{\rm
  rem}$. Otherwise we have a neutron star with mass $1.4 M_{\odot}$.
We determine the fate of the binary if a neutron star is formed by the
work of \citet{boundunbound} and \citet{boundunbound2} with the latest
determination for the kick velocity distribution from observations of
\citet{newkick}. If the system is unbound then the velocities of both
stars are calculated by the method of \citet{boundunbound} that
considers every relevant factor. However we neglect the supernova
impact on the companion star.  If the system remains bound then the
velocity of both components is the resultant system velocity
\citep{blackholekicks2}. If the remnant is a black hole, we assume
that it receives a similar kick. Because the masses of black holes are
greater than those of neutron stars we use the kick distribution of
\citet{newkick} but as a momentum distribution. We calculated the
black hole kick velocity, $v_{\rm BH}$ from $v_{\rm BH} =v_{\rm
  NS}(1.4M_{\odot}/M_{\rm BH})$. Where $v_{\rm NS}$ is a kick velocity
selected at random from the neutron star kick velocity distribution
and $M_{\rm BH}$ is the mass of the black hole. Black holes are not
normally considered to have significant kicks although their
importance has been investigated by others
\citep{blackholekicks2,blackholekicks1,mirabel2,mirabel,voss}. We find
that we must include them or we obtain no runaway WR stars with
initial masses greater than $30M_{\odot}$.

The resulting velocities of runaway stars and runaway binaries are
recorded. We also record the SN type and its location at the initial
position of the binary. We weight each event by a Saltpeter IMF and
assume the distributions of initial mass ratios and the logarithm of
separations are flat. Our models have initial separations that take
values between $1 \le \log_{10}(a/R_{\odot}) \le 4$ in steps of 0.25
dex. The mass ratio takes values of $q=0.3$, 0.5, 0.7 and 0.9. We do
not use the $q=0.1$ models employed by \citet{EIT08} because there is
growing evidence that the mass ratio is skewed to larger values in
massive binaries \citep{binaryminq,twins1,twins2,kiminki}.

We next consider the fate of the secondary star. If it accreted
material from the primary we assume the star is rejuvenated and use
its post mass transfer mass as its new initial mass. If a secondary
accreted more than 5 per cent of its initial mass when it was a
main-sequence star we assume it has been spun up and rotationally
induced mixing mixes fresh hydrogen into the core and it is
rejuvenated and take it to be a zero-age main-sequence star. We do
this at all metallicities. This extends the lifetime of some of our
runaway stars. This is a similar assumption as used by
\citet{vanbook}. If the binary was unbound by the first SN we use a
single star model to determine the result of its evolution. If the
system remains bound then we use our secondary models. If the
secondary experiences a SN it does so after it has travelled away from
the location of the primary SN. We determine the distance travelled by
considering in detail the geometry of situation using the runaway
velocity relative to the original plane, the orientation of the binary
to the line of site, the phase of the stars in the binary and the time
that has passed since the first SN. We then record the location of the
SN. Finally we determine the final outcome of the binary and whether
the system is unbound or a double compact object. We record the
velocities of the single and binary compact object.

We perform the above analysis over the full range of our primary
models. Because we select our neutron star kicks at random, we repeat
our calculations a large number of times to cover the full range of
possible outcomes after the first SN.

\subsection{Quasi-homogeneous evolution}

We include one new evolutionary path, not in the standard picture of
binary evolution for our secondary stars. If a secondary accreted more
than 5 per cent of its initial mass as discussed above we assume it
has been spun up and rotationally induced mixing mixes fresh hydrogen
into the core and it is rejuvenated and take it to be a zero-age
main-sequence star. If the secondary's initial metal mass fraction is
less than or equal to $0.004$ and it has a mass after accretion of
more than $10M_{\odot}$, we assume it continues to evolve fully mixed
during its entire main-sequence lifetime. This is referred to as
quasi-homogeneous evolution (QHE). It is the result of rapid rotation
due to the accretion of material from the primary star and is
described by \citet{maederqhe}, \cite{mmgrb}, \citet{yoon1},
\citet{yoon2} and \citet{cantiello}. The star does not spin down as at
lower metallicity stellar winds are weaker so less angular momentum
can be lost.

To include we use simple models in which we assume the stars are fully
mixed during their hydrogen burning evolution. Once hydrogen burning
ends this extra mixing is turned off. We use the models whether the
binary was bound or unbound after the primary SN. We find, at
$Z=0.004$, QHE increases the percentage of SNe that are type Ib/c from
20 to 26 per cent. We find that three per cent of all stars that experience
QHE give rise to a long-GRB as we also require a final CO core mass of
greater than $7M_{\odot}$ for a long-GRB to occur. More metal rich
stars do not experience QHE as we assume the stars rapidly spin-down
as the stronger stellar winds take away angular momentum more rapidly.

Our requirement for QHE to occur may seem quite
relaxed. \citet{cantiello} restricted their study to a Case B,
post-main-sequence mass-transfer system rather than Case A
system. This was to avoid orbital synchronization slowing the rotation
of the secondary. Therefore we may be overestimating the number of
long-GRBs via QHE. However most of the QHE systems in our models are
Case B and also systems with high mass ratios which have similar
thermal timescales so mass-transfer is most efficient. Therefore the
opportunity of tidal synchronization is low. Further more such
synchronization is unlikely as the stars have radiative envelopes
during much of their evolution. Thus any tidal forces are likely to be
weak.

In addition our 5 per cent increase in mass required for QHE may also
seem arbitrary. Only about 10 percent of the mass of a main sequence
star needs to be accreted to bring its spin from zero to
critical. However QHE does not need critical rotation and the initial
rotation rate of these stars is not zero. Therefore 5 per cent is a
reasonable limit. A full treatment would require a more detailed model
of rotation within the star and the binary. However the inclusion of
this would severely limit the stability of the code and thus make it
currently difficult to create the large number of models required for
population synthesis.

\subsection{Non-degenerate mergers}

In Table \ref{mergers} we list the relative fraction of binaries that
merge before the first SN occurs. We find that 7  to 10 per cent of
binaries with primaries massive enough to explode in a SN experience a
merger. Therefore the majority of binaries provide two possible
supernova progenitors and contribute to the runaway stars. However
even stars not massive enough to provide a SN can merge and in some
cases the resulting star is massive enough to produce a SN. We find
more mergers in the higher mass binaries.

\begin{table}
\caption{The relative population of mergers from our binary population
  at different metallicities and with different minimum masses for the
  binary systems.}
\label{mergers}
\begin{tabular}{ccccc}
\hline 
\hline 
Mass Range      & & During  & Post-MS   & No Merger\\
     /$M_{\odot}$ &Z&   MS     &           &           \\
\hline
$5\rightarrow120$ &0.004   & 0.029  &   0.001&     0.970\\
                  &0.020   & 0.036  &   0.007&     0.957\\
\hline
$10\rightarrow120$&0.004   & 0.070  &  0.002  &    0.928\\
                  &0.020   & 0.088  &  0.016  &    0.896\\
\hline
$20\rightarrow120$&0.004   &  0.106 &   0.005  &    0.889\\
                   &0.020  &  0.129 &   0.004 &     0.867\\
\hline
\hline
\end{tabular}
\end{table}

\subsection{Determining SN types}

Determining what SN type a star produces is a difficult task. Here we
use the relative SN rates of \citet{Smartt08} to estimate the
parameters required for the different types. The rates are based on a
volume limited sample of 92 core-collapse SNe within $28\,$Mpc over
$10.5\,$yrs. The range of metallicities for the SNe tend to be an even
mix of LMC and solar like ($Z=0.008$ and 0.020 respectively).  To
match the relative rates, we first select out stars that we expect to
explode in supernovae. We first require that core carbon burning has
occurred and that the total mass of the star is greater than
$2M_{\odot}$ and that the carbon-oxygen core mass is greater than
$1.38M_{\odot}$. Next we vary the total masses of hydrogen to helium
and their relative amounts required for the different SN types. To
reproduce the observed rates of \citet{Smartt08} we find that we must
include one single star for every binary system included in the
code. Otherwise we cannot reproduce enough type IIP SNe while keeping
the maximum type IIP progenitor mass below $20M_{\odot}$. This is the
99 per cent confidence limit on the maximum mass of type IIP
progenitors determined by \citet{Smartt08} from observations. This in
effect increases the number of very wide non-interacting binaries in
our sample. We summarize the parameters required for each SN type in
Table \ref{snpram}. If a type II SN does not produce a IIP event we
include it in a broad class of non-IIP SNe. These events are rare and
the statistics of \citet{Smartt08} are not good enough to determine
the parameters required for the sub-types of IIb, IIn and IIL.

For the hydrogen-poor SN-types we differentiate between type Ib and
Ic, helium rich and poor SN respectively, by considering the amount of
helium in the ejecta. Rather than using the total mass of helium in
the ejecta as other authors \citep{langerbin,dewi,georgy,yoon} we consider
the fraction of helium in the ejecta to differentiate between Ib and
Ic SNe. We prefer this method because otherwise some progenitors with
pure helium ejecta of only a fraction of a Solar mass would be
identified as Ic rather than Ib. This means our results differ from
previous authors. The main difference is that while other authors find
two distinct classes of low and high mass type Ic progenitors with
type Ib progenitors of mainly low masses, we find more of a
overlapping continuum of type Ic and Ib progenitor
masses. \citet{drout} suggested this might be the case from a
systematic study of lightcurves of type Ib/c SNe. At low metallicity
we note more progenitors explode as type Ib SNe. The latter is due to
weaker WR winds removing less helium before core-collapse.

\begin{figure*}
\includegraphics[angle=0,width=84mm,height=90mm]{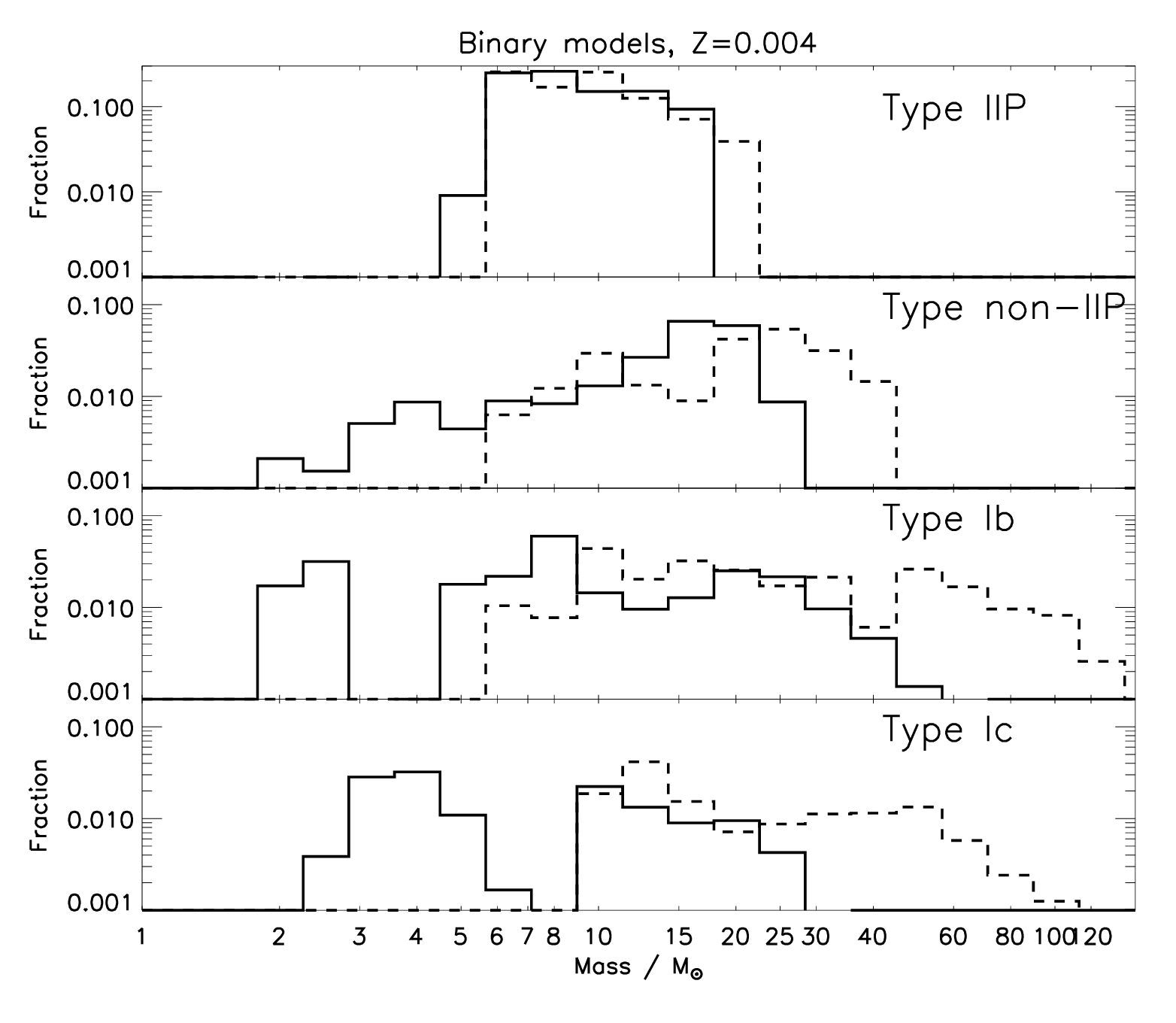}
\includegraphics[angle=0,width=84mm,height=90mm]{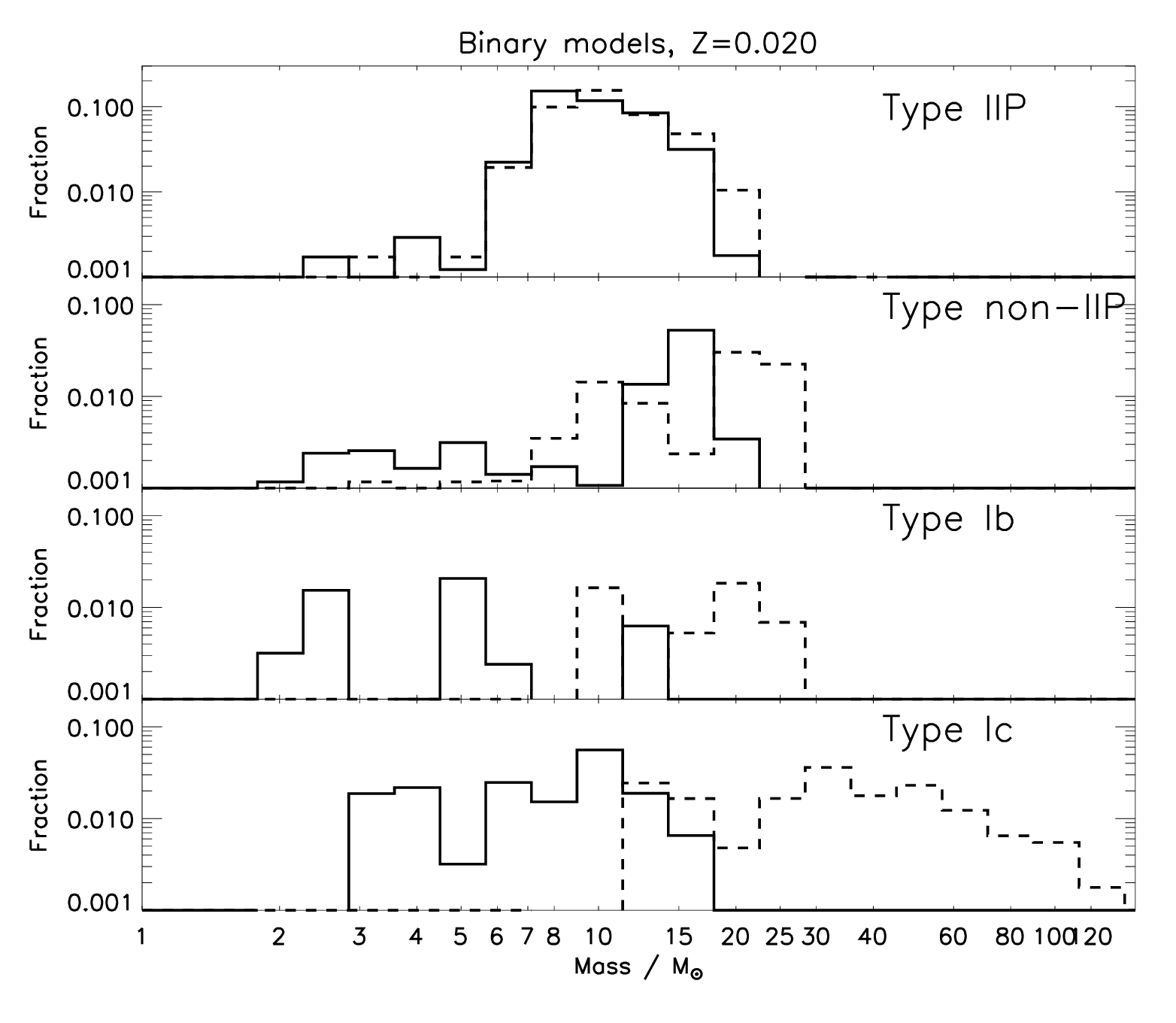}
\caption{The probability distribution of final (pre-explosion) and
  initial masses for different type SN progenitors from our population
  including single stars and binaries. The distribution of initial
  masses are shown by the dashed lines and the final masses are shown
  by the solid lines. The models on the left have a metallicity mass
  fraction of $Z=0.004$ and the models on the right have a metallicity
  mass fraction of $Z=0.020$.}
\label{massranges}
\end{figure*}

The observed relative rates of supernovae from \citet{Smartt08} are
shown along with our synthetic population rates from an equal mix of
solar and LMC metallicity in Table \ref{predictedrates}. The predicted
ratios and mean initial masses for the different supernova types are
given in Section 3.3. We find that the mean initial mass of the
progenitor stars increases through the SN types from IIP being the
least massive to Ic progenitors the most massive. The mass ranges of
the different SN types are shown in Fig. \ref{massranges}. Type IIP SN
progenitors are not only here the lowest mass but also the narrowest
range of masses. The remaining SN types have larger ranges of both
initial and final mass. 

As mentioned above we also record SNe that may have an associated
long-Gamma-Ray Burst (GRB) \citep{review}. While we do not consider
rotation in our models, to produce a GRB we require it must have
experienced QHE as described above, that the CO core mass must be
greater than $7M_{\odot}$ and that the star's rotation axis is within
10 degrees of the line of sight. The direction of the rotation axis is
taken to be random.  We assume that the rotation axis is perpendicular
to the orbital plane. This is important to consider when we calculate
the apparent distance travelled by progenitors before they explode as
discussed below. Due to our assumptions about QHE, long-GRBs only
occur at a metallicity of $Z=0.004$. One in ten QHE stars leads to a
long-GRB and there is one long GRB for every 175 type Ib/c SNe or 685
of all core-collapse SNe. Also we find that the mean initial mass is
lower than the mean mass for type Ib and Ic SNe. This is because the
stars must accrete mass to be spun up and this biases the progenitors
towards lower masses.

\begin{table}
\caption[]{The required parameters for a star to give rise to a specific SN type.}
\label{snpram}
\begin{tabular}{lccccc}
\hline
\hline
     & Final   & CO core &   &   & \\
   SN  & Mass   & Mass &  $M({\rm H})$  &   $M({\rm H})$ & $M({\rm He})$ \\
  Type       & $/M_{\odot}$    & $/M_{\odot}$    &$/M({\rm He})$ &  $/M_{\odot}$&  $/M({\rm ejecta})$    \\
\hline
IIP          &$> 2$&$> 1.38$&$\ge 1.1$  &$> 0.05$  &  -\\
II           &$> 2$&$> 1.38$&$< 1.1$  &$> 0.05$  &  -\\
Ib           &$> 2$&$> 1.38$& - &$\le 0.05$&$\ge 0.58$\\
Ic           &$> 2$&$> 1.38$& - &$\le 0.05$&$< 0.58$\\
\hline
\hline
\end{tabular}
\end{table}

\begin{table}
  \caption{Relative fractions of different SN types from observations
    of \citet{Smartt08} and output from our population synthesis with
    a mix of single stars and binaries at a mix of metallicities.}
  \label{predictedrates}
\begin{tabular}{@{}lcccc@{}}
\hline
\hline
Z  & IIP  & non-IIP  & Ib  &Ic  \\  
\hline
0.008 \& 0.020 & 0.588 &   0.122 &   0.095  &   0.195\\
Smartt et al.  & 0.587 &   0.120 &   0.098  &   0.195 \\
\hline
\hline
\end{tabular}
\end{table}

\section{Results}

While our code can run at any metallicity we restrict ourselves to
just two, solar metallicity with a metal mass fraction $Z=0.02$ and
SMC-like metallicity with $Z=0.004$. This enables us to decouple the
effects of metallicity on our results. 

\begin{table}
\caption[]{The fractions of OB stars, red supergiants and Wolf-Rayet
  stars that are runaways in our synthetic population. We list the
  fractions from two selection criteria for runaways. First any star
  that has a peculiar velocity above $ 5 {\rm km \, s^{-1}}$and a
  second with a higher velocity requirement of $ 30 {\rm km \,
    s^{-1}}$. The former, lower limit, would provide results records
  runaways in a method similar to the observational method of
  \citet{stone}. We list two populations, one from our standard
  population including single stars and binaries, and a second
  including some of the single stars as DES runaways, we assume half
  the single stars are runaways for half their O star lifetime.}
\label{runawaynumbers}
\begin{tabular}{lcccc}
\hline
\hline
    & $Z=0.004$& & $Z=0.020$ &  \\
 $v_{\rm runaway}\ge$ & $ 5 {\rm km \, s^{-1}}$ & $ 30 {\rm km \, s^{-1}}$ & $ 5 {\rm km \, s^{-1}}$  & $ 30 {\rm km \, s^{-1}}$ \\
\hline
\hline
S. \& B.\\
O      &  0.161  &  0.059  &  0.022  & 0.005 \\
B      &  0.069  &  0.025  &  0.071  & 0.022 \\ 
RSG    &  0.105  &  0.020  &  0.186  & 0.043 \\
WR     &  0.496  &  0.207  &  0.179  & 0.029 \\
\hline
S., B, \& DES\\
O      &  0.190  &  0.052  &  0.065  &  0.015\\
B      &  0.093  &  0.034  &  0.090  &  0.028\\ 
RSG    &  0.227  &  0.046  &  0.298  &  0.072\\
WR     &  0.506  &  0.170  &  0.227  &  0.020\\
\hline
\hline
\end{tabular}
\end{table}

\subsection{Single-star Runaways}

\begin{figure}
\includegraphics[angle=0,width=80mm]{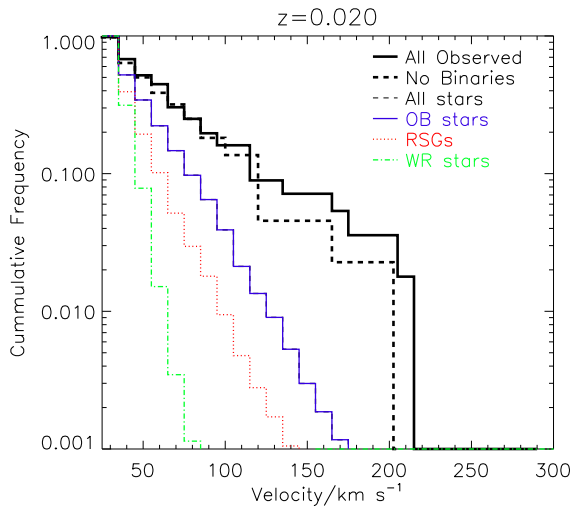}
\caption{The cumulative frequency of stellar runaway velocities. The
  thick black line is the observations from \citet{hoogerwerf}, the
  thick dashed line is for the same data but with the known binary
  stars removed, the thin dashed line is the line predicted by our
  models for all runaway stars, the blue line for all OB stars, the
  red line for all red supergiants and the green line for Wolf-Rayet
  stars. The predictions are for solar metallicity.}
\label{mainrunaways}
\end{figure}

\begin{figure*}
\includegraphics[angle=0,width=80mm]{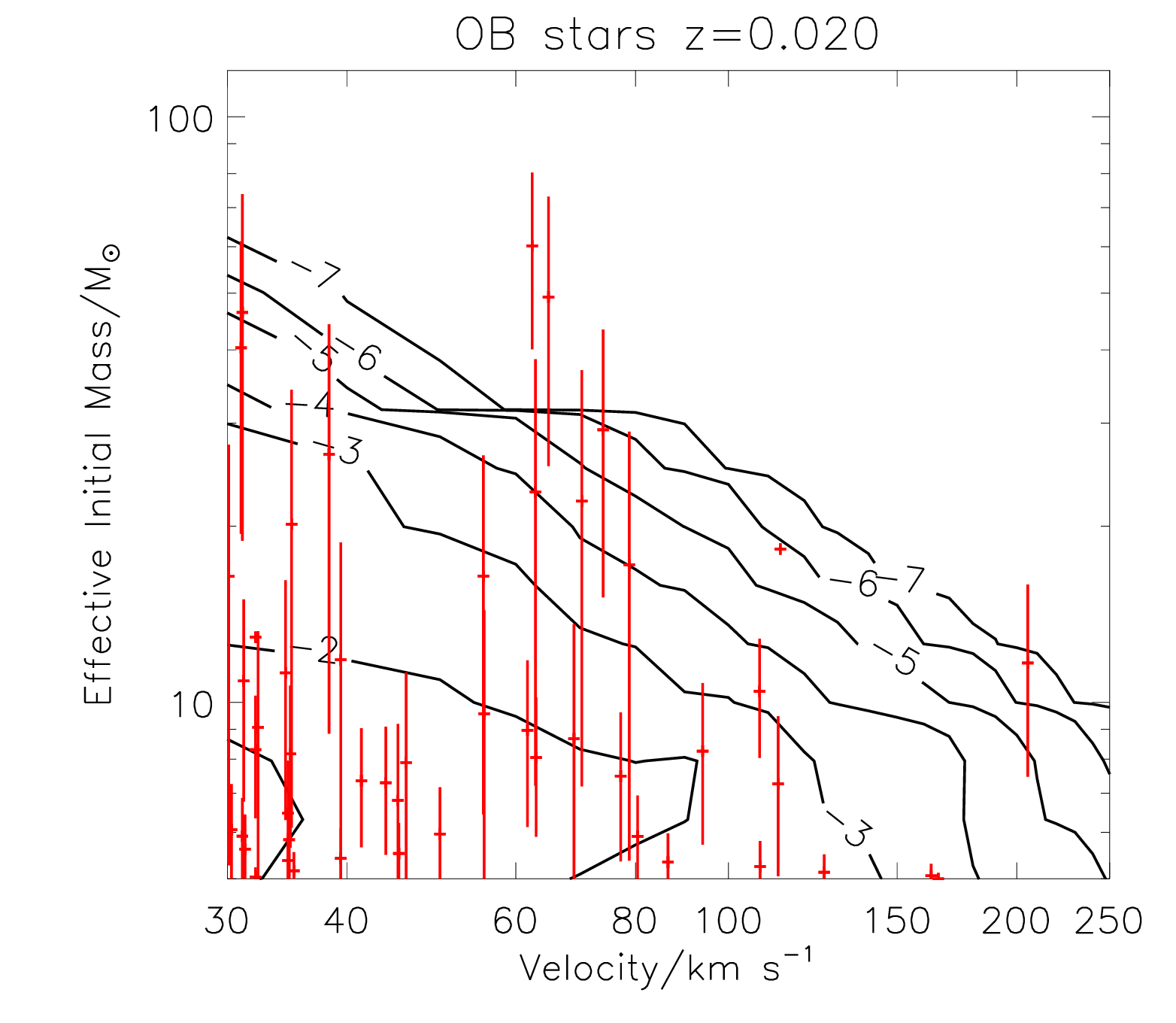}
\includegraphics[angle=0,width=80mm]{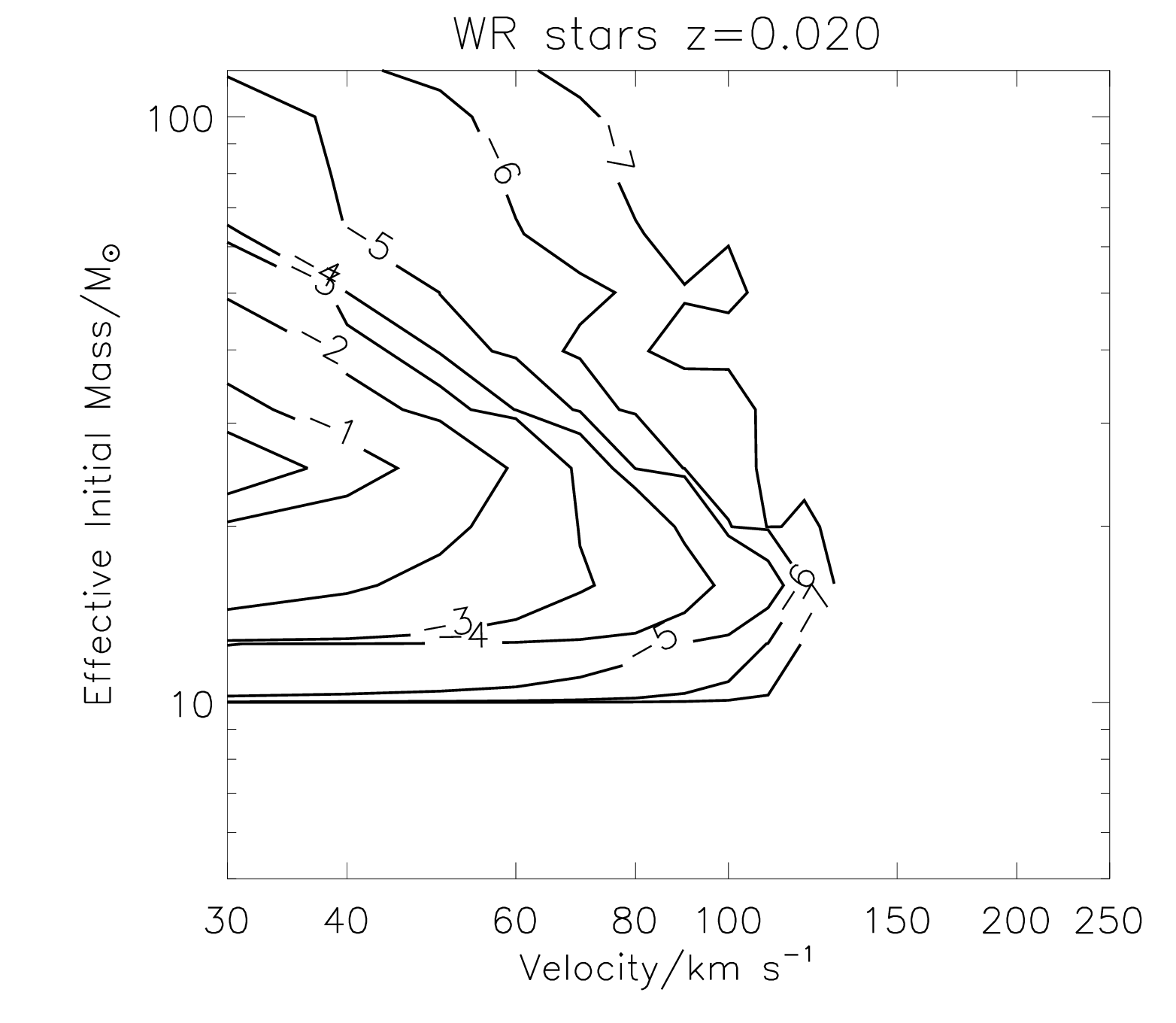}

\includegraphics[angle=0,width=80mm]{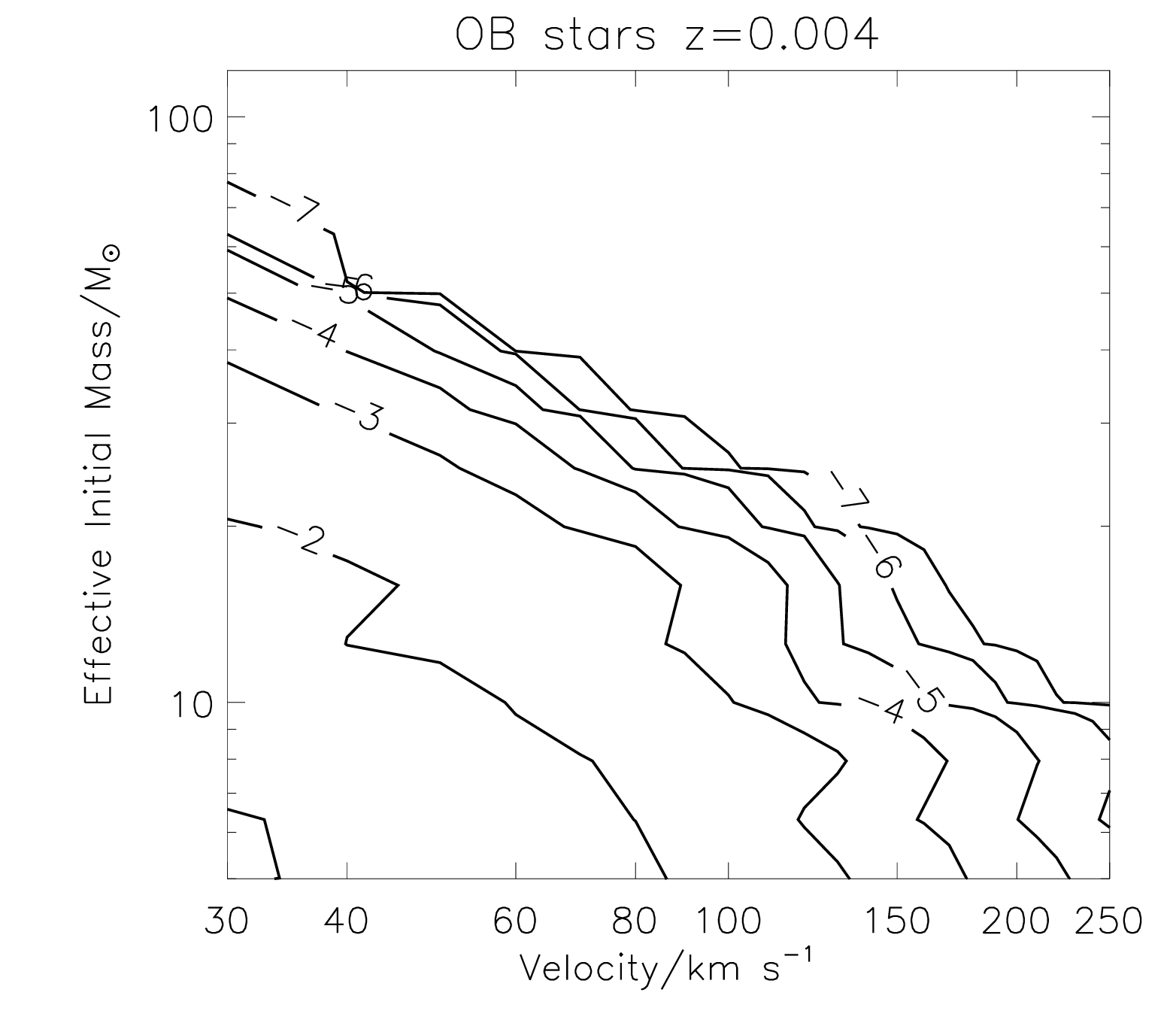}
\includegraphics[angle=0,width=80mm]{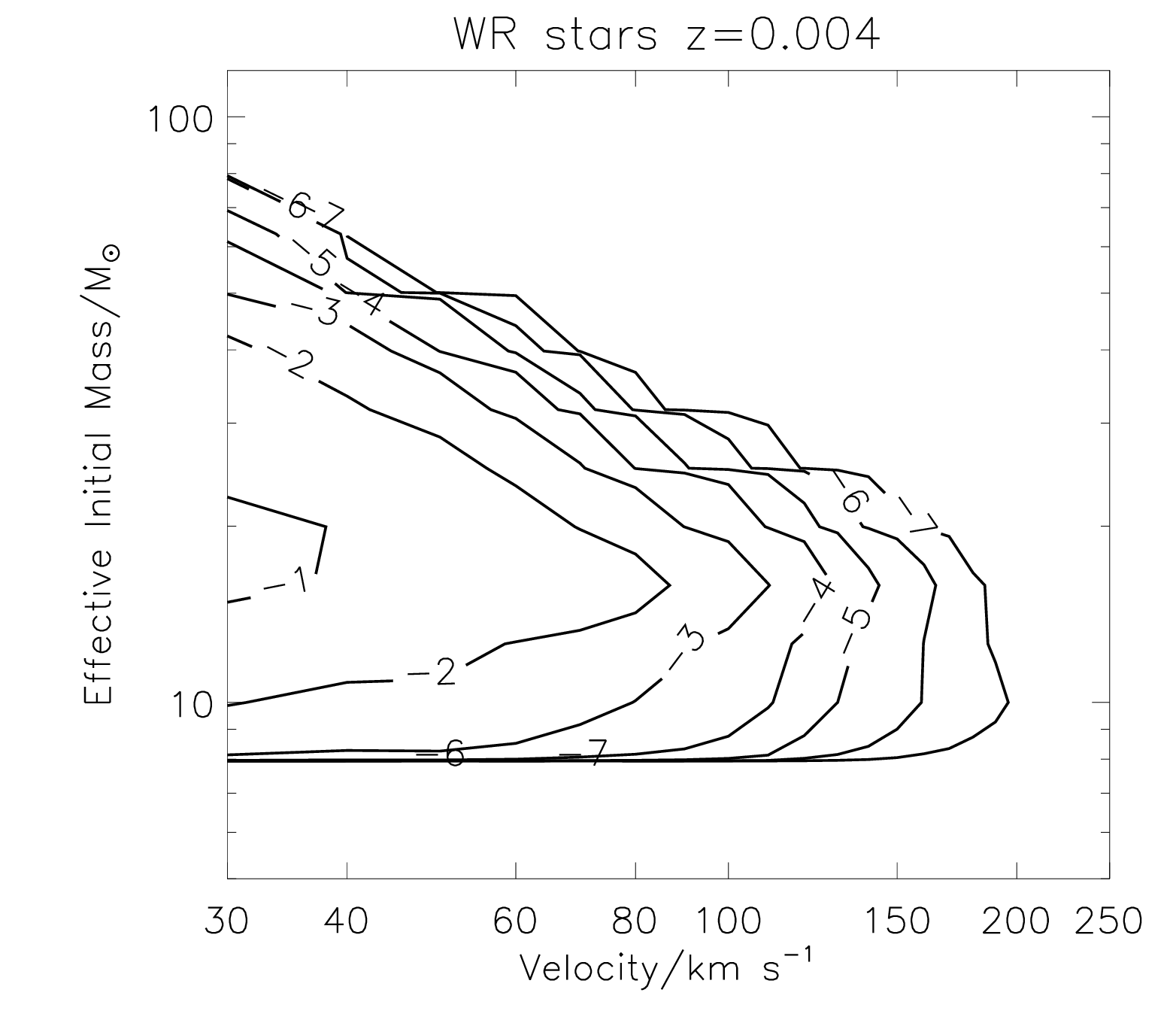}
\caption{Contour plots of the probability for runaway star to have a
  certain mass and velocity. The units of the contours are probability
  per $10\, {\rm km\,s^{-1}}$ and per 0.1 in
  $\log_{10}(M/M_{\odot})$. The mass referred to is the actual mass so
  includes any mass gained by a runaway during mass transfer. The red
  points are the observations of OB runaways from \citet{hoogerwerf}
  with stellar masses derived from fitting the spectral energy
  distribution. The top two panels are for solar metallicity and the
  bottom two panels are for SMC-like metallicity. The left hand panels
  are for OB stars while the right hand panels are for WR stars. The
  mass axis is either the initial mass of the runaway star or its mass
  after accretion due to any binary interaction. A similar figure
  showing the number of stars below the velocity limit for a star to
  be observed as a runaway, $30 {\rm km \, s^{-1}}$, is included in
  the appendix (Fig. \ref{mainrunaways2appendix}). }
\label{mainrunaways2}
\end{figure*}

The main output of our code is the distribution of runaway
velocities. Observations of runaways have been used to infer that
around 10--30 per cent of O stars and 2--10 per cent of B stars are
observed to be runaways \citep{gies,stone,dewit,zinnecker}. These
observed fractions are very uncertain as the number of well observed
runaways is relatively small and the definition of what a runaway is
varies between different studies. For example the above observed
runaway fractions vary because the runaway definition varies between
authors.\citet{gies}, \citet{stone} and \citet{dewit} state that a
runaway must have a velocity greater than $30 {\rm km \, s^{-1}}$ or
be a large distance above the Galactic plane. Also in some cases is it
not clear if the percentage given is the number runaways in the field
O star population or the number of runaways relative to the entire O
star population.

The recent study of \citet{dewit} suggest the 50 per cent of field O
stars are runaways. They also state that since 70 per cent of O stars
are in clusters. These facts suggest that possibly 15 per cent of O
stars are runaways. They identify runaways by peculiar velocities above
$40 {\rm km \, s^{-1}}$, a high distance from the galactic plane or
proximity to young clusters. These mixed definitions make it difficult
to identify a velocity cut-off to use with our model population to
predict the number of runaways. For example even a runaway with a
velocity of $5 {\rm km \, s^{-1}}$ would travel approximately 10 pc in
1 Myr and would appear as a runaway according to
\citet{dewit}. Therefore we estimate a range of the fraction of
runaways from our models. We use the high velocity of $30 {\rm km \,
  s^{-1}}$ to give a strict minimum runaway fraction estimate and a
lower velocity limit of $5 {\rm km \, s^{-1}}$ to provide a more
relaxed maximum runaway fraction estimate.

In Table \ref{runawaynumbers}, we list the fraction of the predicted
populations that are runaways. The numbers are calculated weighting
each model by its lifetime in the phase of evolution and with a
Salpeter IMF. For comparison we list the results for O stars, B stars,
red supergiants (RSGs) and WR stars. B stars are taken to be stars
with $\log_{10}(T_{\rm eff}/K) \ge 4.18$ (B5 and earlier only includes
stars with initial masses greater than $5M_{\odot}$, the minimum mass
of our binary models) and a surface hydrogen mass fraction greater
than 0.4, O stars have hotter effective temperatures of
$\log_{10}(T_{\rm eff}/K) \ge 4.48$ and $4.52$ at $Z=0.020$ and 0.004
respectively. RSGs are defined by $\log_{10}(T_{\rm eff}/K) \le 3.66$
and $\log_{10}(L/L_{\odot}) \ge 4.9$ and WR stars are taken to have a
surface hydrogen mass fraction less than 0.4, $\log_{10}(T_{\rm
  eff}/K) \ge 4$ and the same luminosity limit as the RSGs.

In calculating these numbers we consider count each O, B, RSG or WR
star including both the primary or secondary in the binary. It is
unclear if in the observed samples an O star binary is counted as one
or two O stars. We find 0.5 to 2.2 per cent of O stars are runaways
and 2.2 to 7.1 per cent of B stars are runaways at Solar metallicity.

We note that uncertainties in our model may mean the O runaway
fraction is an underestimate. For example, the absence of rotational
mixing in our stellar evolution models. Rotational mixing mixes fresh
hydrogen into the core of a main-sequence star and this would increase
our number of O star runaways by increasing their lifetimes. The
evidence that rotation has an importance in the number of O star
runaways if we include QHE, the most extreme form of rotational
mixing, at solar metallicity the number of O star runaways increases
to 2.2 to 16 per cent. We note the dramatic increase in number of O
runaways is due to stars, of around $15M_{\odot}$ that have 20 Myrs O
star lifetimes with QHE but are not O stars without it. Rotation will
have a similar but less dramatic effect at Solar metallicity. The
effect of rotation on a stars lifetime is only to increase
it by approximately 10 per cent. Including a detailed model of
accretion on to the secondary, especially following rotation would
allow more accurate inclusion of the effects of rejuvenation on the
secondary stars that accrete material from their primaries.

The number of B star runaways is similar to that suggested from
observations. We seem to under-predict the number of O star runaways
if we compare to the typical numbers quoted for the number of runaway
O stars. The work of \citet{dewit} used various O star catalogues
including that of \citet{ocat}. We use the same catalogue to estimate
from the catalogue the fraction of O star runaways. It is important to
note that we count O star binaries in the catalogue as 2 O stars. We
find that $76\pm5$ per cent of O stars are found in clusters with
$6\pm1$ per cent of O stars being runaways.  The list of runaways
includes some binaries, we assume these are DES in origin and find
that there are $4\pm1$ per cent of single O star runaways. Considering
the uncertainties in classification of runaways and the approximations
within our population model our predicted runaway population is
comparable to the runaway population in the catalogue of \citet{ocat}.

Further complicating matters is that we have not considered in our
model the contribution of DES runaways. We have made a toy-population
assuming that our single star population provide a similar number of
DES runaways to the total number of BSS, approximately 30 percent of
the binary produce BSS. We then assume that they are runaways for half
their O star lifetime and have a similar velocity distribution to the
BSS runaways. We see in Table \ref{runawaynumbers} these modest
assumptions increase the number of O runaways to between 1.5 to 6.6 and
the number of B runaways to between 2.8 and 9 percent. These numbers
are also comparable to the observed numbers.

After considering the population of runaways we now consider the
velocity distribution and parameters of individual runaways. The best
observed sample of OB runaways to date is that from \citet{hoogerwerf}
who list the velocities of 56 runaway stars. \citet{hoogerwerf} give
masses for some of these estimated by \citet{runmass} and
\citet{vanrun0}. By using available UBVJHK photometry we have employed
the methods outlined in \citet{ecol3} to estimate stellar masses with
Cambridge STARS models. Our masses broadly agree with those used by
\citet{hoogerwerf} and are used in Fig. \ref{mainrunaways2}. The mass
range we find is between 5 to $70M_{\odot}$ with half the stars having
masses less than $10M_{\odot}$.

In Fig. \ref{mainrunaways} we plot the cumulative frequency of our
runaway population velocities. In general there is agreement between
the observed runaway velocity distribution and that predicted by our
models. We find that the assumed initial binary separation
distribution largely determines the shape of the runaway
velocities. Here we use a distribution that is flat in $\log_{10}
a$. If we limit ourselves to only close binaries we over predict the
number of fast runaways, while considering the widest binaries leads
to only the slowest runaways. It appears in the figure that we are
under-predicting the number of fast runaways.

Using our estimated initial masses from the runaways we can extend
Fig. \ref{mainrunaways} along another axis as in Figure
\ref{mainrunaways2} of \textit{effective} initial mass of the runaway
star. This is to take account of the increase in its initial mass
due to accretion of mass from its primary companion. The general shape
of our predictions is similar to that of \citet{pz00} who presents a
similar figure. In general it is clear that less massive stars are
more likely to achieve higher runaway velocities. The observed runaway
stars agree with this prediction however there are several outliers
with high masses and moderate velocities. Observations of these stars
indicate they also have large rotational velocities, $v \sin i \approx
200 \,{\rm km \, s^{-1}}$ \citep{hoogerwerf}, and therefore their
masses from SED fitting should be considered upper limits because they
were derived with non-rotating single star models. Rotational mixing
can increase the luminosity of a star and the main-sequence
lifetime. This means their mass would be overestimated and make them
more likely to be observed. Alternatively, they could be DES runaways,
runaways that experience both DES and BSS or our binary models are
producing mergers in very close binaries rather than producing fast
BSS runaways from close binaries.

We must be wary of the selection effects of the observed sample. It is
easier to observed higher velocity and high mass runaway stars. These
selection effects should be more carefully understood before any
detailed comparison between observations and theory. For example most
runaways are likely to have masses less than $10M_{\odot}$ with
velocities below $100\,{\rm km \, s^{-1}}$. However such stars and
velocities are difficult to detect within the \textit{Hipparcos}
catalogue used by \citet{hoogerwerf}. Meanwhile more massive, more
luminous stars are easier to detect, especially if they have large
proper motions. Quantifying these selection effects is difficult. We
may need to wait for the Gaia satellite before obtaining a less biased
catalogue of runaway stars. Because the observed population is
uncertain we do not attempt further tuning of our initial binary
distribution.

We note that at both metallicities the fraction of secondary stars
that have accreted some material during mass transfer is 50 per
cent. In some cases the amount accreted is very little, only 17 per
cent of secondary stars accrete more than 5 per cent their original
mass due to assumptions on the maximum accretion rate for the
secondary. Because of this we estimate that no more than 50 per cent of BSS
runaways should be observed to have high rotation velocities and/or
enriched compositions. Therefore, absence of these observable features
should not be inferred to mean the runaways did not come from BSS.

Finally we again note that our predictions are for the BSS only.
However, the main detail that determines the velocity distribution in
both cases is the initial binary separation distribution assumed. We
therefore suggest that the expected velocity distribution from the DES
should be similar to that of the BSS.  Stars that are ejected by DES
normally follow the interaction of a binary with another binary or a
single star. \citet{desrun} have recently investigated the runaway
velocities possible from three body interactions. They found that the
mean velocities are similar to those we show here in Figure
\ref{mainrunaways}. However they also find the interactions can lead
to velocities in excess of $100\, {\rm km \, s^{-1}}$ for
$80M_{\odot}$ stars in 10 per cent of triple body encounters. This
adds weight to the suggestion that runaways more massive than
$40M_{\odot}$ with runaway velocities faster than $40\, {\rm km \,
  s^{-1}}$ are in fact DES or a DES+BSS combination.

\subsubsection{Wolf-Rayet runaways}

Fig. \ref{mainrunaways2} shows our results for WR stars. The sharp cut
off at low masses in these panels is due to our luminosity limit that
WR stars must have $\log_{10}(L/L_{\odot})\ge4.9$ to be included. The
difference between the two metallicity plots at higher masses is due
to the differing lifetimes of WR stars at the two
metallicities. \citet{evink} found that, at SMC-like metallicity, the
lifetimes of WR stars vary between 3 to 4$\times10^5$ yrs while at
solar metallicity the lifetimes are 4 to 8$\times10^5$
yrs. Furthermore QHE also skews this plot because such stars become WR
stars while still on the main sequence and so the plot begins to
resemble the OB star panel as the time-scales are millions of years
rather than a few hundred thousand years. If we do not include QHE
then the distribution becomes the same as for the solar metallicity
plot. Thus a large population of runaway WR stars in low-metallicity
galaxies with an average velocity above $50{\rm \,km\,s^{-1}}$ would
be circumstantial evidence for the occurrence of QHE in
nature. Surveys for runaways of the SMC have be undertaken but have
not yet provided detailed statistics \citep{smcrunpaper}.

The fraction of WR and RSG stars that are runaways is larger than the
fractions of OB runaways (Table \ref{runawaynumbers}).  This is
because BSS runaway stars spend a large fraction of their
main-sequence life stationary, while they spend most or all of the
lives as RSGs or WR stars as runaways. This makes their relative
number appear greater.  At SMC metallicity, the WR runaway fraction is
particularly high due to QHE scenario, which produces more WR stars
from secondary stars.

The velocities of Wolf-Rayet stars differ strongly between the two
metallicities we study. At solar metallicity we find WR runaways have
velocities less than about $150\,{\rm km \, s^{-1}}$ but runaways with
velocities above $80\,{\rm km \, s^{-1}}$ are rare. If we do not
included black hole kicks we only find WR runaways with initial masses
less than $30M_{\odot}$ and 0.4 per cent of WR stars have velocities
above $30\,{\rm km \, s^{-1}}$. This is because secondaries that
become WR stars owing to stellar wind mass loss have primaries that
form black holes at core-collapse and remain bound because very little
mass is ejected in such supernovae. We find that WR stars eject
relatively little of their final mass, between 2 to $8M_{\odot}$. With
such low ejecta masses without a strong kick it is difficult to unbind
the binary. It is possible that, due to the sparseness of our binary
grid, we may be missing WR runaways from binaries in which both
components have initial masses in the range 20 to $25M_{\odot}$.  We
might also assume too low an initial mass limit for a black hole to be
formed in core collapse. Binaries with component masses between 20 and
$25M_{\odot}$ may give rise to high velocity runaways as shown by
\citet{Dray2005}, fig. 2(c), which refers to a similar population
synthesis with no black hole kicks. In these systems the binary mass
ratio at core collapse is more even so the secondary may achieve a
higher velocity. Such systems are still rare and dominated by binary
WR stars remaining in their binary systems.

Our results broadly agree with similar theoretical predictions of
\citet{Dray2005}. It is difficult to perform a quantitative comparison
but in their fig. 2(b), which describes their model most similar to
ours as it includes black hole kicks, we find broadly similar maximum
velocities for WR stars of up to around $120 \, {\rm km\,
  s^{-1}}$. However most WR stars have velocities below $80\, {\rm
  km\, s^{-1}}$. \citet{Dray2005} also find a restricted number of WR
runaways when they assume that black holes do not have kicks.

We note that \citet{Dray2005}, in fig. 2(a) predict a large number of
runaway WR stars. In this model they ignored stellar-wind mass
loss. So more of the secondary stars accrete enough material to become
WR stars as single stars, when the stellar wind mass-loss rates are
reapplied after the first SN. Such an arrangement is unlikely to occur
in nature because, if the metallicity were low enough to reduce the
main-sequence mass-loss rates, the minimum mass for a WR star would
also increase. The only way for WR stars to occur in such a situation
would be via QHE.

Another reason for mostly slow WR runaways in our synthetic population
is demonstrated by the example that $50M_{\odot}$ primary stars at
solar metallicity form neutron stars in core collapse. These binaries,
even with large neutron star kicks, only obtain slow runaway
velocities because the secondaries are much more massive than their
companions at core-collapse and so their orbital velocities are
smaller.

Finally we may also predict more WR runaways if we were to relax our
luminosity constraint for a WR star. This number is highly uncertain
and if it is set too low we would overpredict the number of WN
stars. This would suggest that the higher velocity a runaway WR star
has, the lower its initial mass.

Observations of runaway O and WR stars by \citet{wrrunaways1} imply
that there are a similar fraction of runaways for both types of
star. They also suggest that mean kinetic ages based on displacement
and motion away from the Galactic plane tend to slightly favour the
DES over BSS.  There are a few WR stars with higher velocities such as
WR 124 with a space velocity of $180{\rm \,km\,s^{-1}}$
\citep{wrrunaways3}. Given our results (Fig.~\ref{mainrunaways2}), it
seems unlikely that WR~124 was ejected through the BSS. Conceivably,
it may be the result of both DES and BSS combined. Or, as suggested by
\citet{wrrunaways1}, the more massive a runaway the more its
probability of coming from DES increases. Recent radio observation by
\citet{wrrunaways2} give velocities of about $30\,{\rm \,km\,s^{-1}}$
for a few WR stars, in agreement with our predictions. However they
also give a runaway velocity of more than $300\,{\rm \,km\,s^{-1}}$
for WR112, a possible WR+O binary. The high velocity for a binary
argues for the highest velocity systems arising from a combination of
DES and BSS.

A qualitative method to determine the WR runaway velocity distribution
is to consider the orbital velocities of WR binaries with massive O
star companions. Using the catalogue of \citet{vanderhucht} which
lists the masses and periods of many WR binaries we find that, without
black hole kicks, many of these binaries are difficult to unbind
because the O star is the more massive of the two stars. From the
binaries we find typically low velocities, similar to those predicted
in Fig. 3, \ref{mainrunaways2} of 10 to 50\,km\,s$^{-1}$ for a
$30M_{\odot}$ companion and up to 70\,km\,s$^{-1}$ for a $10M_{\odot}$
companion. Again this leads us to assume that fast and massive WR
runaways are most likely to be the result of DES.

\subsubsection{Red supergiant runaways}

Runaway red supergiants have a similar distribution to the OB runaway
stars. It is an interesting question whether such stars could be
observed and what would be the structural effects on their
envelopes. The most obvious difference may be the distribution of the
mass loss in a trail as the star moves in space. A trail has recently
been discovered for the red giant Mira which is a low-to
intermediate-mass star with a space velocity of $130\,{\rm km \,
  s^{-1}}$ \citep{mira}.  Also the closest red supergiant to the Sun,
$\alpha$\,Ori or Betelgeuse, is a runaway star, which
moves with a velocity of about $30{\rm \,km\,s^{-1}}$ through its
local ISM \citep{ueta}.

\subsubsection{White dwarfs, neutron stars and black holes}
\label{wdnsbh}

\begin{figure}
\includegraphics[angle=0,width=80mm]{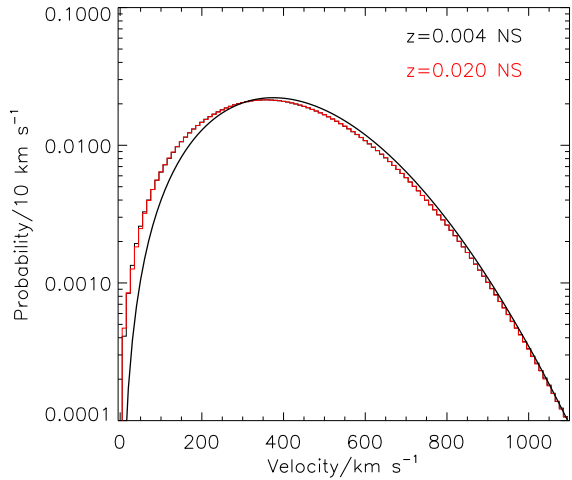}
\caption{The distribution of runaway velocities for neutron stars. The
  smooth black line is the input distribution of neutron star
  kicks. The lines have been normalised so the total probability is
  1.}
\label{NSvelocities}
\end{figure}

\begin{figure}
\includegraphics[angle=0,width=80mm]{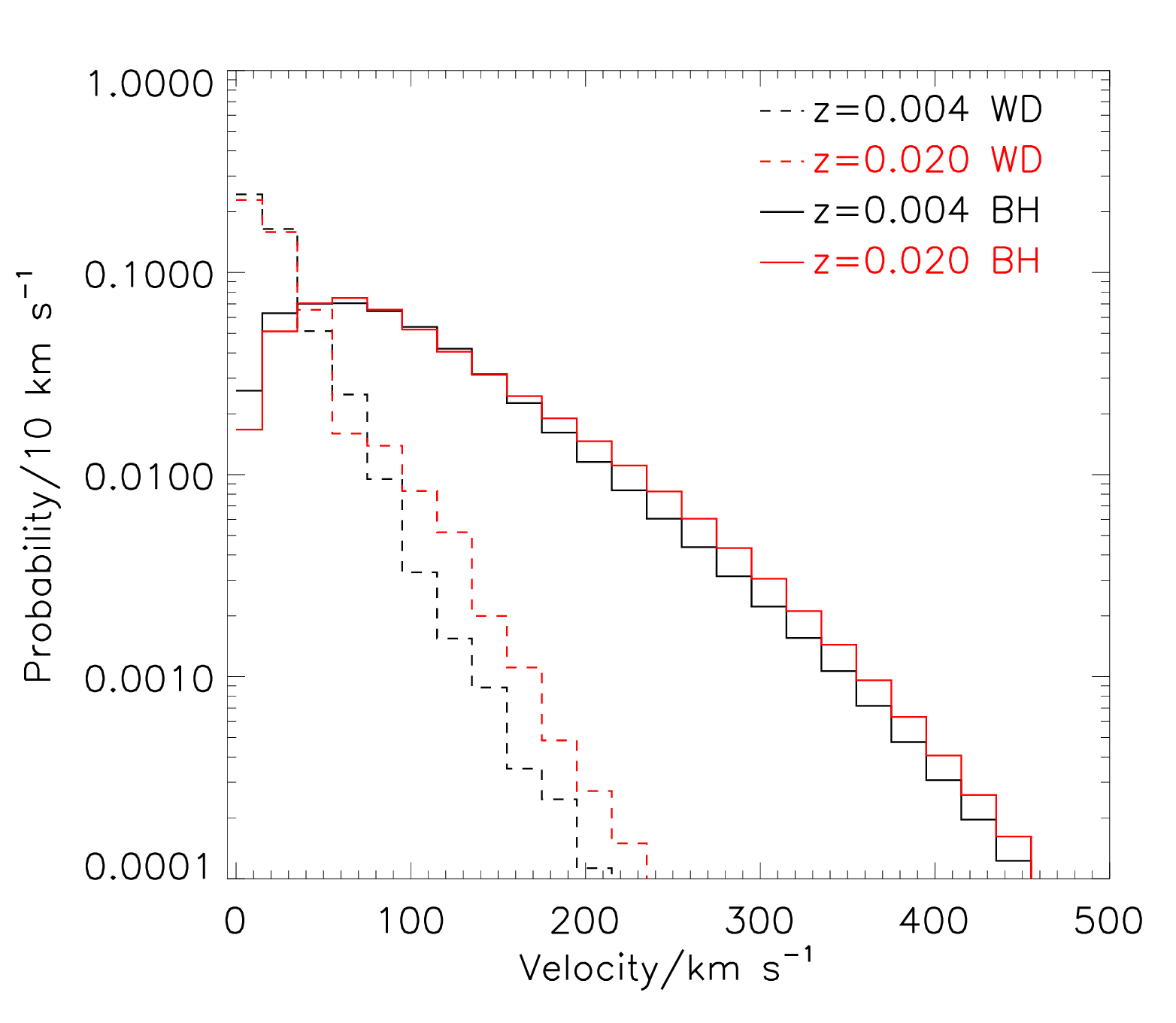}
\caption{Similar to Fig. \ref{NSvelocities} but for black holes and
  white dwarfs.}
\label{WDBHvelocities}
\end{figure}

Finally we consider the velocity distributions for single compact
remnants (Fig. \ref{NSvelocities} and \ref{WDBHvelocities}). Their
relative formation rates and their average space velocities are shown
in Tables \ref{remz020}, \ref{remvelz020}, \ref{remz004} and
\ref{remvelz004}. We see that it is most likely that stellar remnants
are single objects.

The neutron star velocity distribution is dominated by the kick
velocity distribution we assume for these objects. At low velocities
there is a small excess of neutron stars compared to the input kick
velocity distribution. These stars come from binary systems that are
not disrupted in the first SN. Their velocities reflect the orbital
velocity of the neutron star in a binary. Thus, when the neutron star
kick distribution is estimated from observations, account must be
taken of the fact that some of the observed neutron star space
velocities are not determined by the neutron star kick. If this is not
considered the mean kick velocity may be underestimated and the
variance overestimated. Also the upper velocity side of the population
is lower than expected. This is because when a binary is unbound as
the components separate their mutual gravitational attraction must be
overcome. Therefore the neutron stars appear to be travelling slower
than they were at birth. We find that we would have to increase the
root mean squared velocity used in the Maxwell-Distribution from $265
{\rm km \, s^{-1}}$ to $275 {\rm km \, s^{-1}}$. We note that here, we
do not consider the possibility of a population of low-kick neutron
stars emerging from electron-capture supernovae, as advocated by
\citet{podsi03}.

Neutron stars are our most common runaway compact remnant. Black holes
are less populous. We see from Fig. \ref{WDBHvelocities} that black
holes have a distribution with a peak at around $60\,{\rm km \,
  s^{-1}}$ and most black holes below $250\,{\rm km \, s^{-1}}$. Our
rarest runaway remnants are white dwarfs. The distribution is similar
to that for all stars in Fig.  \ref{mainrunaways}. This is because
white dwarfs do not receive any kick and their velocity distribution
is representative of their initial binary separations.

We note that with a typical galactic escape velocity of a few
times $100\,{\rm km \, s^{-1}}$ we can conclude that most white dwarfs
and black holes are retained within their host galaxies, while neutron
stars are most likely to escape to the intergalactic medium.

\subsection{Runaway binaries}

The next objects to consider are the runaway binary systems that must
contain one or more compact objects. Only 20 per cent of binaries
survive the first supernova. Typically these objects have lower
velocities than single stars because any kick that did not disrupt the
binary must be small and had to transfer momentum to the system rather
than just the neutron star.

\begin{figure}
\includegraphics[angle=0,width=80mm]{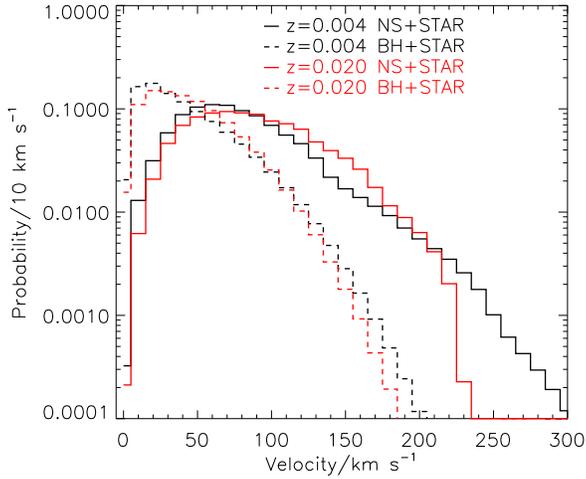}
\caption{Similar to Fig. \ref{NSvelocities} but for binaries that
  contain either a neutron star or black hole. The lifetime of these
  systems is taken into consideration when calculating the
  distribution.}
\label{compactvelocities}
\end{figure}

\begin{figure}
\includegraphics[angle=0,width=80mm]{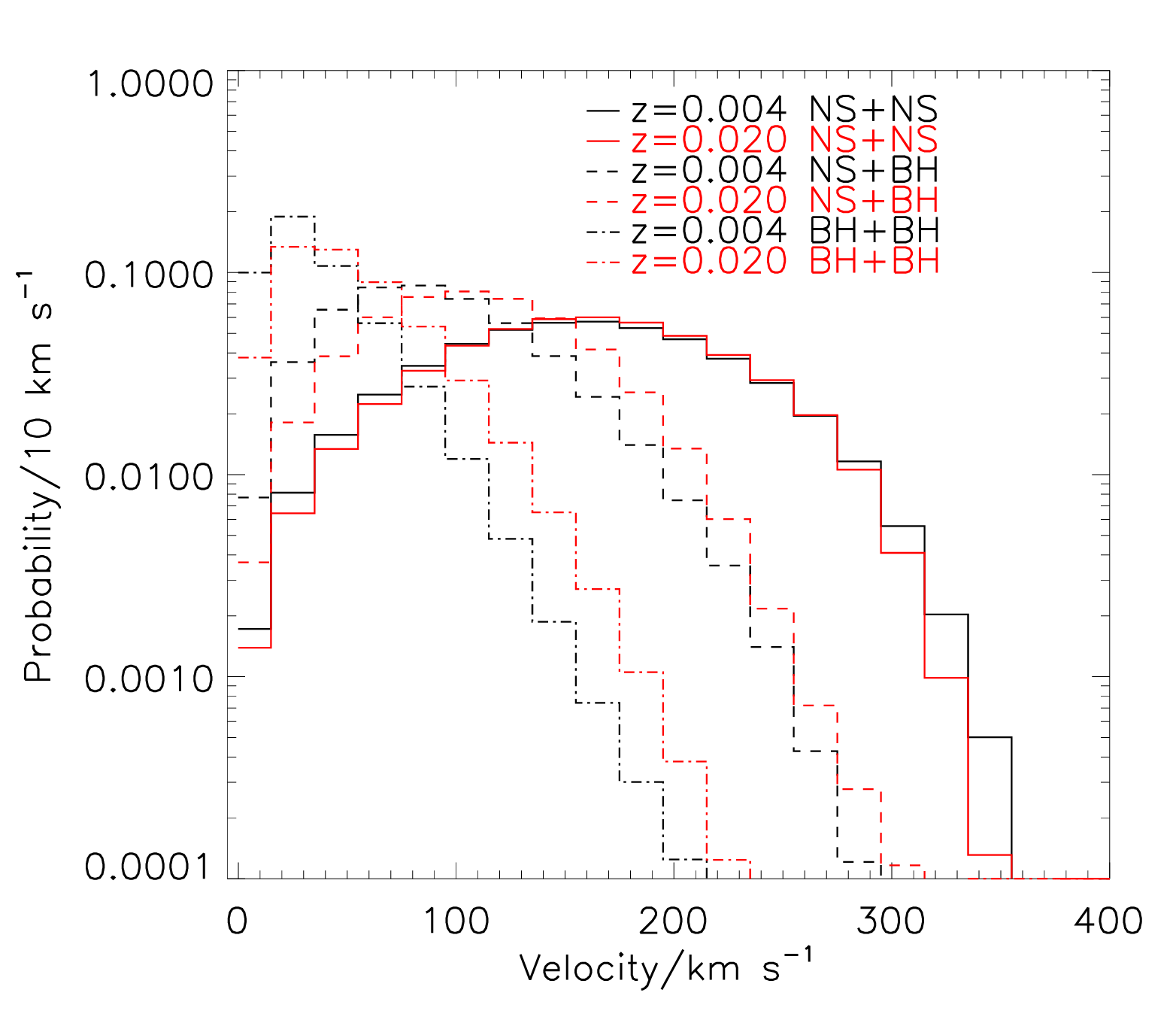}
\caption{Similar to Fig. \ref{NSvelocities} but for systems
  containing two compact objects.}
\label{doublevelocities}
\end{figure}

\begin{figure}
\includegraphics[angle=0,width=80mm]{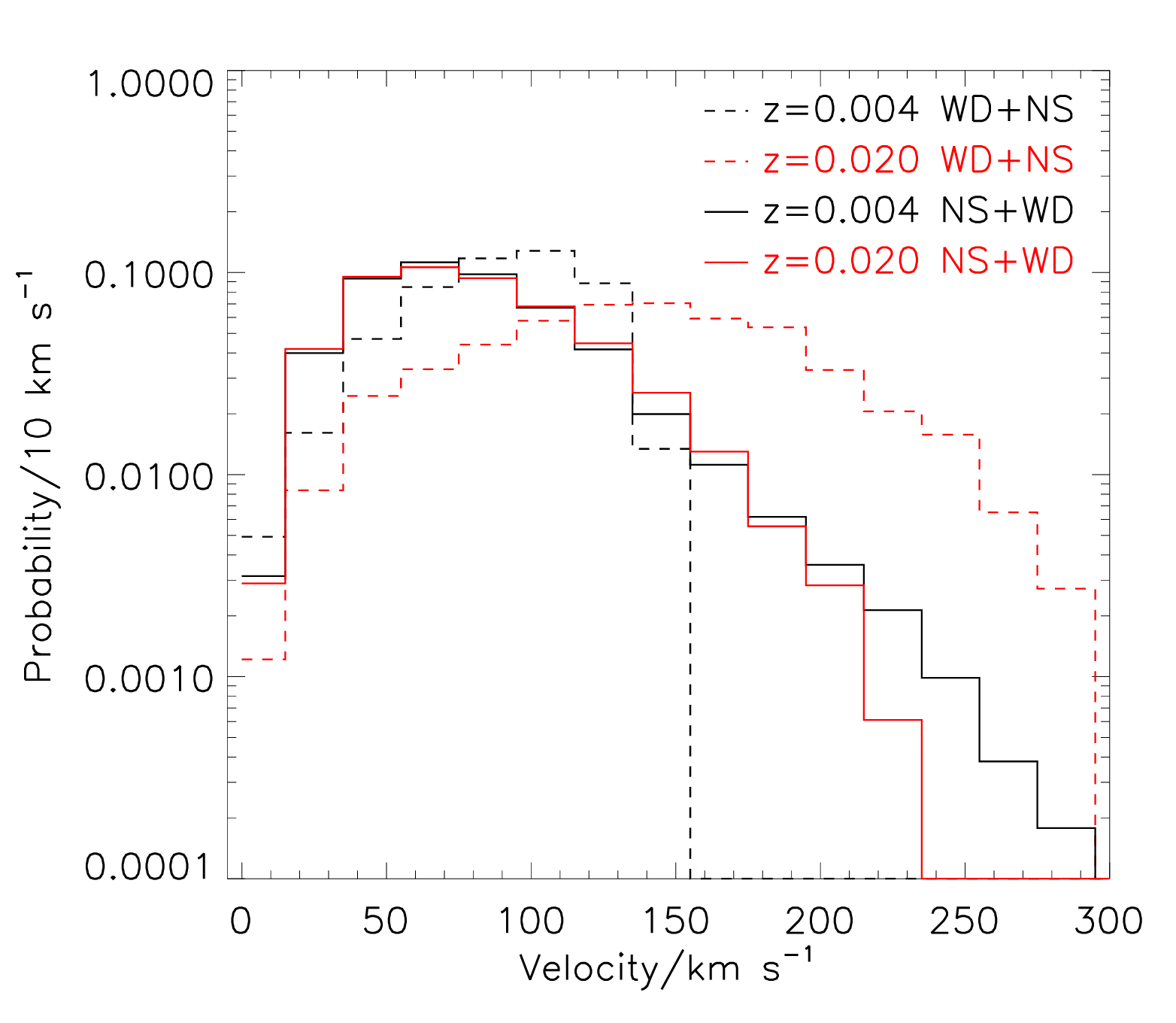}
\includegraphics[angle=0,width=80mm]{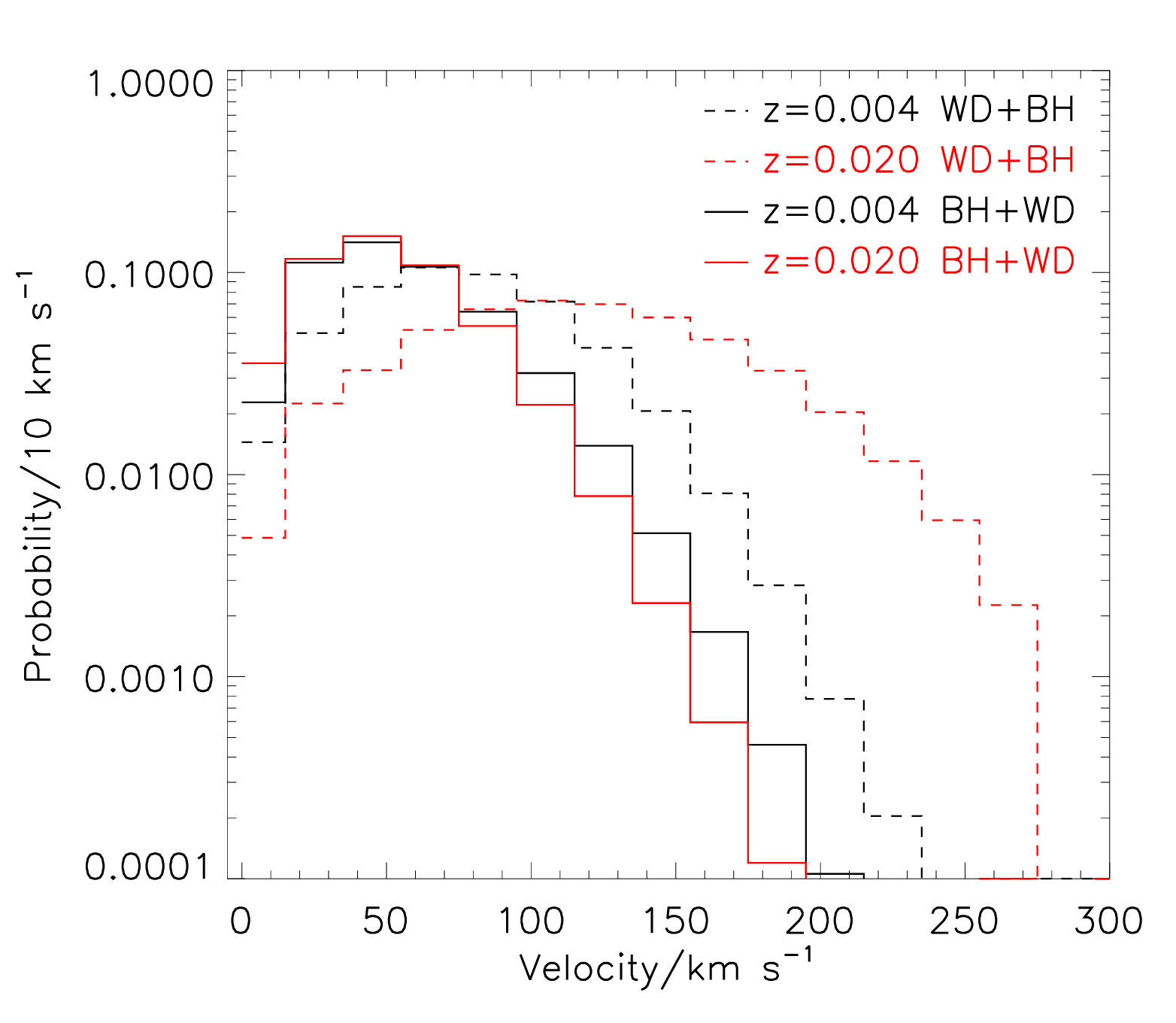}
\caption{Similar to Fig. \ref{NSvelocities} but for systems
  containing a compact object and a white dwarf.}
\label{dwarfvelocities}
\end{figure}

\subsubsection{Stars with a compact companion}

If a binary survives the first SN then its velocity relative to the
pre-SN centre of mass changes. The neutron star or black hole kick
determines this velocity. We see our results in
Fig. \ref{compactvelocities} and there are clear differences between
the two types of remnant. The larger neutron star kicks lead to
greater velocities so the peak velocity is slightly higher, $60\,{\rm
  km \, s^{-1}}$ rather than $20\,{\rm km \, s^{-1}}$ for black hole
binaries.  The maximum velocities possible by each binary reflect this
trend with most black hole binaries having velocities below $200\,{\rm
  km \, s^{-1}}$ while neutron star binaries reaching velocities up to
$300\,{\rm km \, s^{-1}}$.

For both compact remnants a reasonable number of these binaries have
velocities over $30\,{\rm km \, s^{-1}}$. Therefore a prediction of
our code is that such binaries are less likely to be observed within
their natal clusters and are more likely to be runaway stars
\citep[e.g.][]{mirabel,fragos}.

\subsubsection{Double compact systems}
\label{dcs}

Systems which experience two SNe are very unlikely to remain
bound. There is very little mass left in the system at the time of the
second SN, so the system is easily unbound. However the more massive
the remnant from the first SN the more likely the binary is to remain
bound but the slower the resultant binary will move.  The relative
birth rates of different single and binary compact remnants are listed
in Tables \ref{remz020} and \ref{remz004} and this trend is clear with
there being more black-hole primary compact binaries than neutron-star
primary binaries. Furthermore in Tables \ref{remvelz020} and
\ref{remvelz004} the mean velocities of neutron-star primary systems
is greater than that when the primary produces a black-hole.

Further insight into the nature of compact binaries can be seen in
Fig. \ref{doublevelocities}. There is a trend through the
distributions from only low space velocities for double black-hole
binaries to the highest velocities for double neutron-star
binaries. This demonstrates the most important fact for the remnant
velocities is the number of large neutron-star kicks a binary
receives.

As mentioned in Section \ref{wdnsbh}, these velocity distributions
imply that double neutron-star compact binaries may be able to escape
their host galaxy because of they experience at least one large kick
compared to double black hole binaries which should be retained by
most galaxies.

\subsubsection{White dwarf systems}

For completeness we have also included the systems that contain a
white dwarf and a neutron star or black hole. We find in most of these
systems the white dwarf forms second. The velocity distribution is
determined primarily by the kick given to the neutron star or black
hole at its formation.

Systems that form white dwarfs first and then neutron stars or black
holes due to mass transfer are rare. This is to be expected because
white dwarfs have low mass and, unless the kick is small or occurs in
a small restricted set of directions, the system becomes unbound. The
runaway velocity of such systems is greater than systems that
experience the SN before the WD is formed. This is because while the
systems are easier to unbind because of their low system mass they are
also easier to accelerate to high velocities. Therefore the peak
velocity of these systems is higher and provides an observational
signature of such systems. The fact that we model some such systems
indicates that we are producing SNe from secondary stars that accrete
material to end their life in SNe rather than as white dwarfs.

\subsection{Supernovae and gamma-ray bursts}

With the velocities of the runaway stars and their lifetimes as
runaways we calculate the distance that stars may travel before their
explosive deaths. This may help to qualitatively interpret recent
observations concerning the environments and locations within the
host-galaxy of core-collapse SNe and gamma-ray bursts. For example
\citet{fruchter,jamessnhalpha,kelly} and \citet{snhalpha} resolve the host galaxy at
resolutions of hundreds of parsec. Determining how different
supernovae types are distributed relative to the light of a
galaxy. \citet{hammer} studies the host galaxies with greater
resolution, calculating the distance between long-GRBs and the nearest
region containing WR stars. We calculate the distance a progenitor
travels by multiplying the runaway velocity of the progenitor by the
time spent as a runaway. To calculate the apparent distance observed
on the sky we also take account of the random orientation of the plane
of the binary and the phase of the two stars in the orbit. When
we calculate the apparent distance we also take into consideration the
three dimensional motion of the runaway relative to the original
binary orbit as given by the formulae of \citet{boundunbound}.

We analyze our results in two different forms. First we arrange the SN
progenitors by the type of SN we predict the star produces and
secondly we consider how distance varies with initial mass. Our
results are shown in Tables \ref{populationdetailsz020} and
\ref{populationdetailsz004} showing the predicted ratios, mean
\textit{effective} initial mass, runaway velocity, distance travelled
and ages before core-collapse for the different SN types and
progenitor sources. In calculating the distance travelled we neglect
the galaxy potential. Therefore the furthest distances that we predict
may be modified if this was considered \citep{voss}. We note the
distance distributions are non-Gaussian and thus the mean distances
give here should be considered with care. In these tables there are
also some surprises. For example in Table \ref{populationdetailsz004}
the mean mass of type Ic SNe decreases relative to the higher
metallicity mass in Table \ref{populationdetailsz020}. This is
primarily due to the inclusion of QHE changing the outcome of lower
mass stars and also that the highest mass stars retain enough helium
to become type Ib rather than type Ic (see Figure 1).

\begin{figure*}
\includegraphics[angle=0,width=80mm]{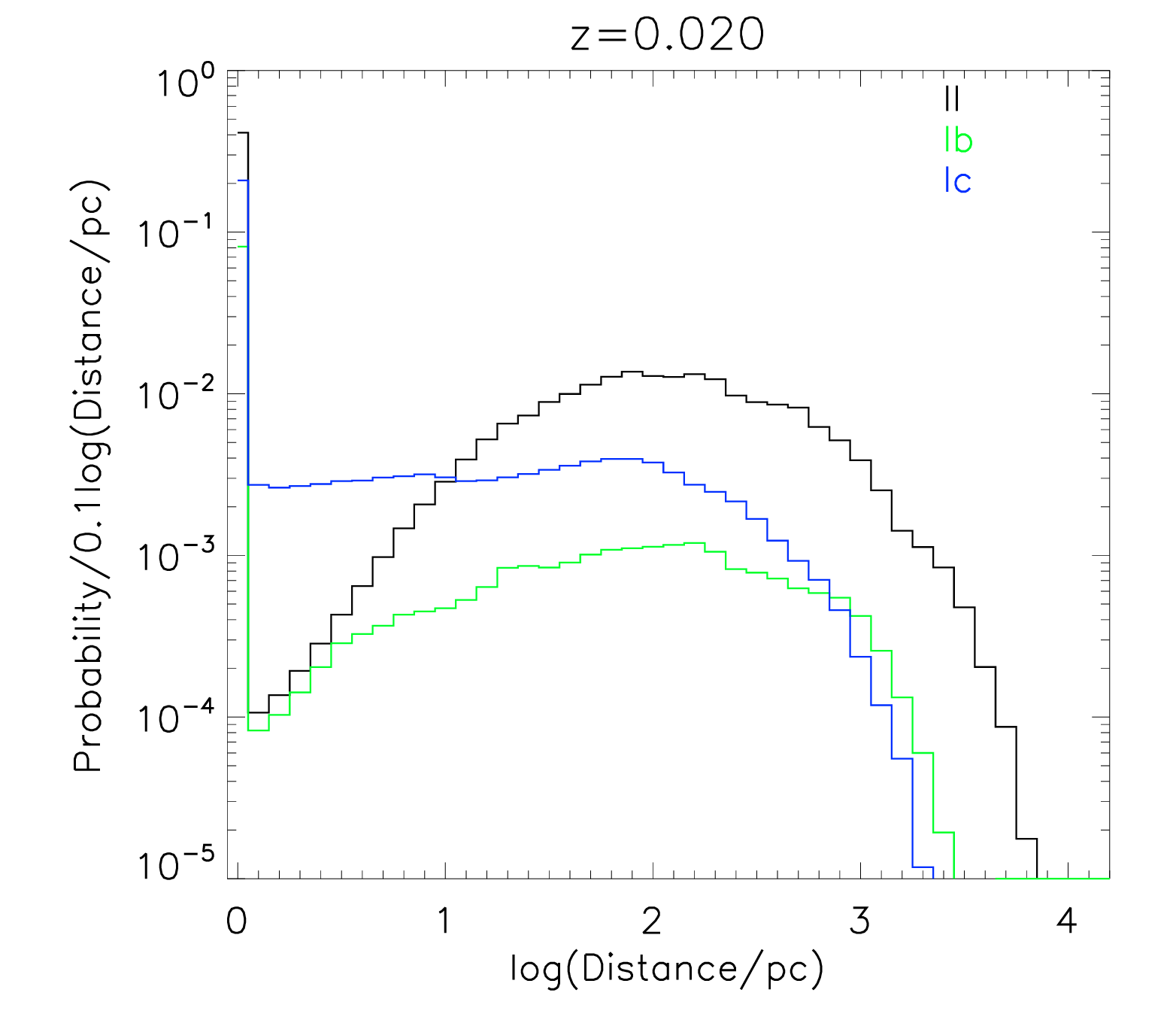}
\includegraphics[angle=0,width=80mm]{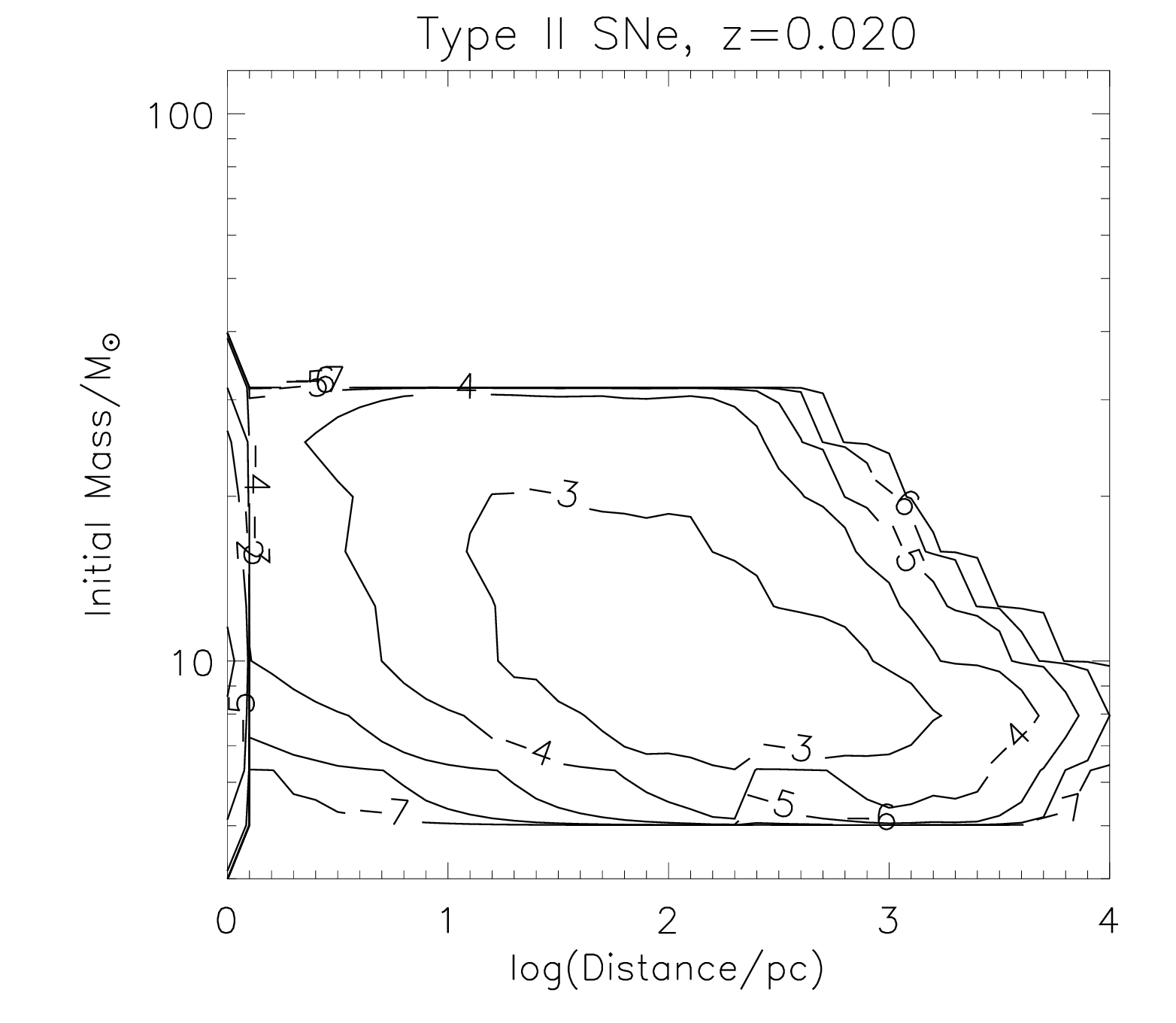}

\includegraphics[angle=0,width=80mm]{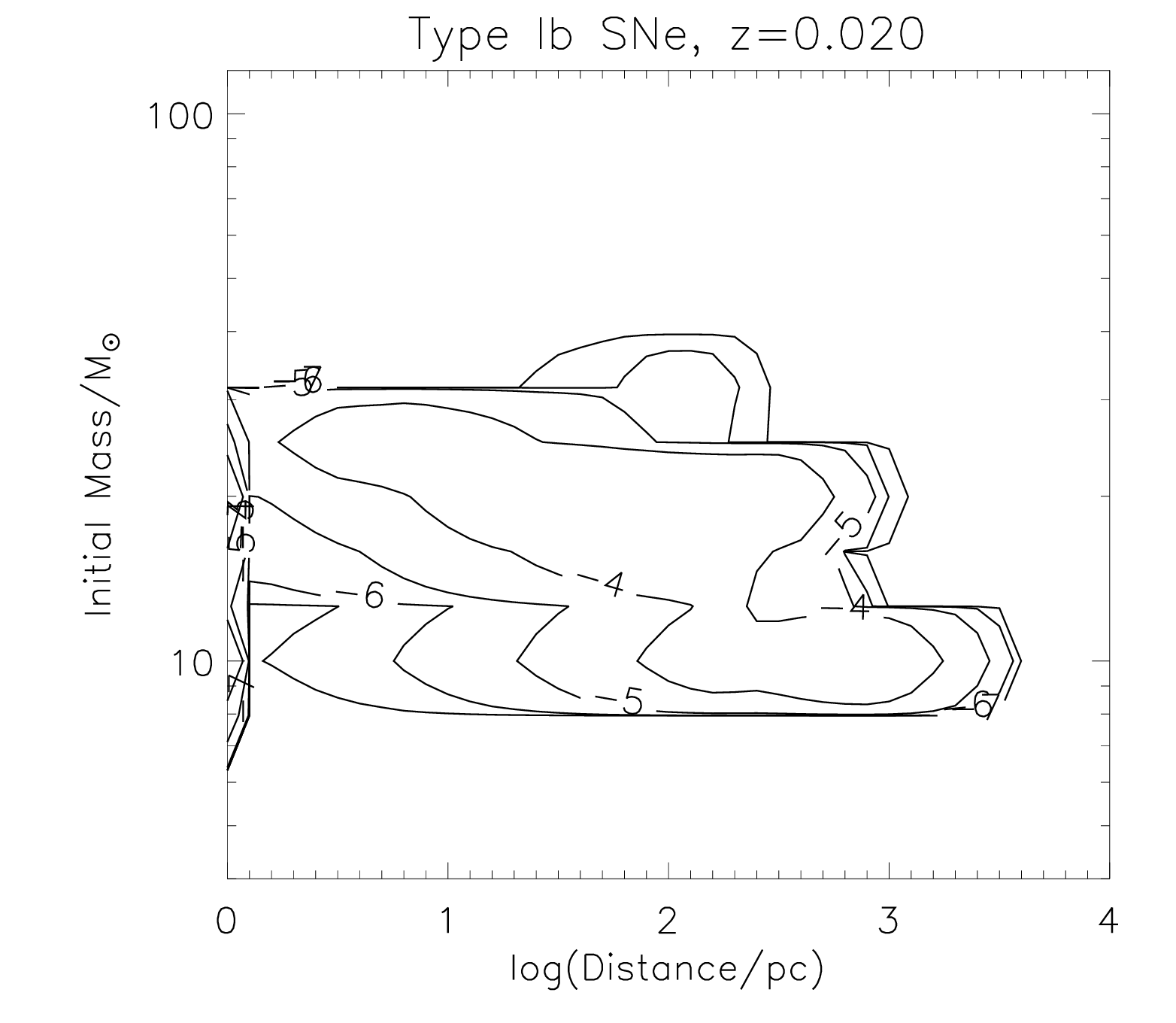}
\includegraphics[angle=0,width=80mm]{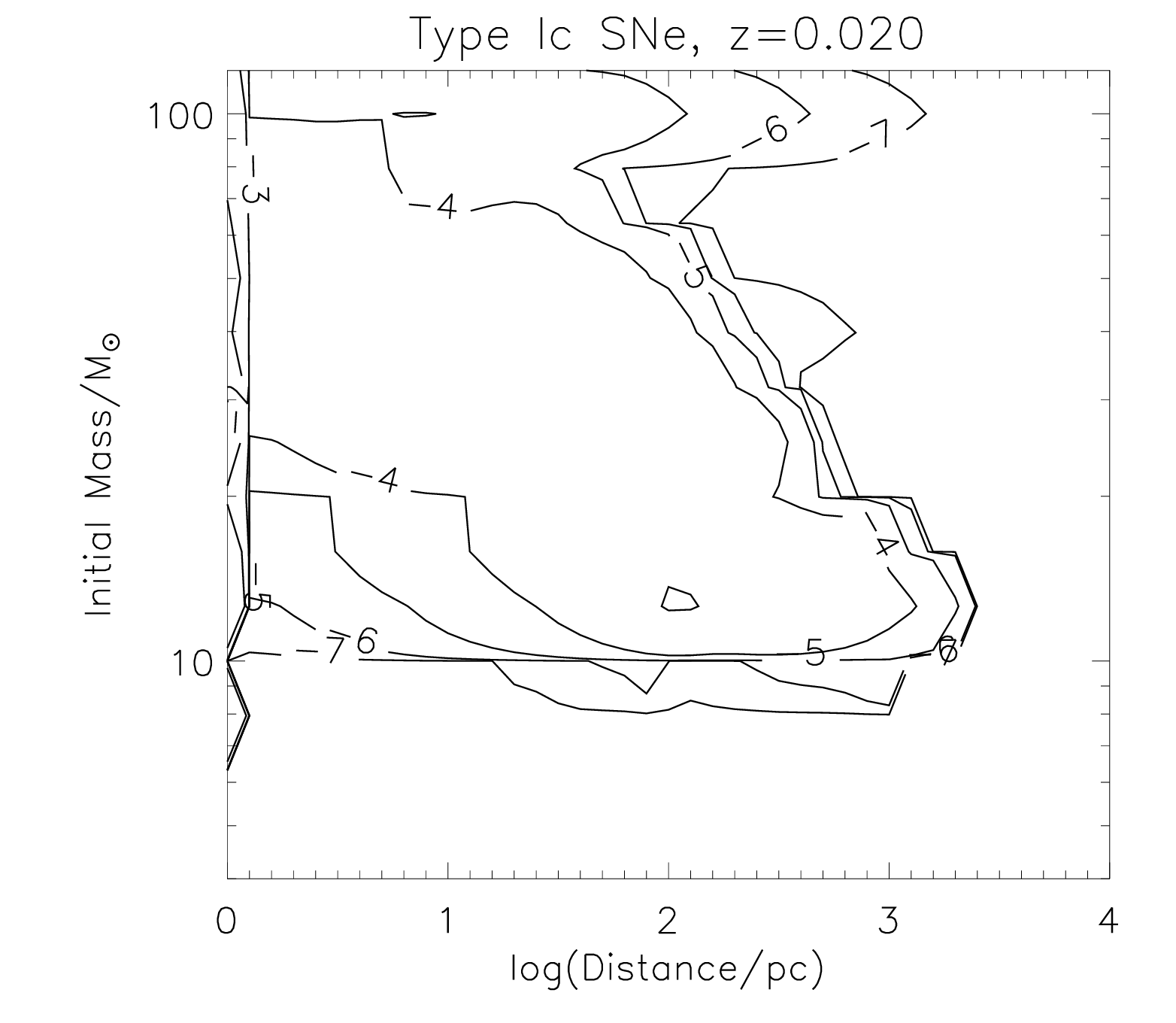}
\caption{The distribution of distances travelled by different SN
  progenitors arranged by SN type and initial mass from our binary
  models alone. The lines and contours are normalised so the total
  number of SNe is 1. The plots are for solar metallicity. Contours
  are $\log_{10}({\rm probability}/M_{\odot} {\rm \, pc})$}
\label{starvelocitiesTYPE}
\end{figure*}

Using the method outlined in Section 2.4 above we are able to create
Fig. \ref{starvelocitiesTYPE} and \ref{starvelocitiesTYPE2} showing
the distances travelled by progenitors of different SN types from our
binary models alone. There are also panels showing how the different
progenitor masses contribute to the combined distributions. We only
consider binaries because we have not included DES runaways in our
model. If the number of runaways is similar from DES as for BSS then
the distribution of SNe for the latter should be similar to that for
the combined runaway population.

On first inspection of Figures \ref{starvelocitiesTYPE} and
\ref{starvelocitiesTYPE2} we see that most SNe occur at their initial
location. This is no surprise because 60 per cent of SNe from binaries
are from the primary star, with only 40 per cent coming from secondary
stars which are the most likely to travel any distance from their
initial location. Primary stars dominate the statistics because this
also includes merged systems. These increase the number of primary
stars that have an effective initial mass great enough to produce a SN
and decrease the number of secondary stars available to produce a
supernova. The latter effect further boosts the apparent contribution
of single stars to the total number of SNe. 

The property that most affects the range of runaway distances for a
specific SN type is the mass range of the progenitors. Type IIP come
from stars with masses below $20M_{\odot}$, which travel great
distances due to their long lifetime.  The other SN types typically
have higher maximum initial masses and wider mass ranges (see Fig.
\ref{massranges}) which lead to the different distributions to those in
the Figs.  In the solar metallicity plot in
Fig.~\ref{starvelocitiesTYPE} we see that type Ib SNe have a similar
distribution to type IIP SNe because most progenitors come from the
same initial mass range. This similarity is weaker at the lower
metallicity because it becomes swamped by the stars with
QHE. This is also evident from the mean
initial mass of the different progenitors in Tables
\ref{populationdetailsz020} and \ref{populationdetailsz004}.

Type Ic SNe typically occur closer to their initial positions than
type IIP and Ib SNe. It is difficult to find observations with which
these predictions can be tested. Use of the observations of discussed
above would require some knowledge of the luminosity of the source
stellar population, star formation history and the galactic potentials
through which the runaway stars travel. Also, as yet, these studies do
not consider the selection effects of detecting supernovae in luminous
and dusty regions of galaxies. For example, type Ib/c supernovae are
generally more luminous than type IIP supernovae, although there is
significant diversity in the SN population \citep{snlum}. If this
difference was true then type Ib/c would be more easily discovered in
luminous regions of galaxies where type IIPs might be missed. Thus
some of the observed difference in the distributions of different
supernova types could be due to the intrinsic luminosities of these
objects and selection effects.

When we consider that the resolution which \citet{kelly} resolves
typical galaxies in their samples is only a few hundred
parsecs. Therefore it is the relative distributions beyond these
distances are important. Our results in Figures
\ref{starvelocitiesTYPE} and \ref{starvelocitiesTYPE2} do therefore
qualitatively agree with the findings of \citet{kelly} that, in
general terms, type IIP and Ib SNe should be less correlated with
star-formation in a galaxy than type Ic SNe, especially in
environments with solar metallicities.  This is because type II and Ib
SNe travel similar distances because the age and mass ranges at which
these SNe occur overlap. In comparison the ages for type Ic SNe are
considerably lower. The type Ic SNe do not travel so far from their
birthplaces because they arise from the most massive progenitors. This
is in addition to the lifetimes of the progenitors of typical IIP and
Ib SNe being similar and longer than those of type Ic SNe. 

\subsection{Long gamma-ray bursts}

In conjunction with studying SNe we consider the predicted
distribution for long-duration GRBs and we relate this to
quasi-chemically homogeneous evolution (Sect.~2.2). While
\citet{yoon2} relate long-duration GRBs to the quasi-chemically
homogeneous evolution of rapidly-rotating single stars, we consider
only the binary-induced quasi-chemically homogeneous evolution as
explored by \citet{cantiello}. This has the advantage that the
predictions do not depend on the initial distribution function of
stellar rotation rates.  This also means that we predict a wider range
of initial masses for long-GRB progenitors than \citet{yoon2}. For
their single star models stars would lose significant angular momentum
before core-collapse, even at lower metallicities than we study
here. In our model because the stars are spun up they can retain more
angular momentum until core-collapse. Because of this we may be
overestimating the number of very massive (greater than $60M_{\odot}$)
long-GRB progenitors. If we have fewer of these objects it would
increase the mean distance travelled by the progenitor stars given in
Table \ref{populationdetailsz004}.

\begin{figure*}
\includegraphics[angle=0,width=80mm]{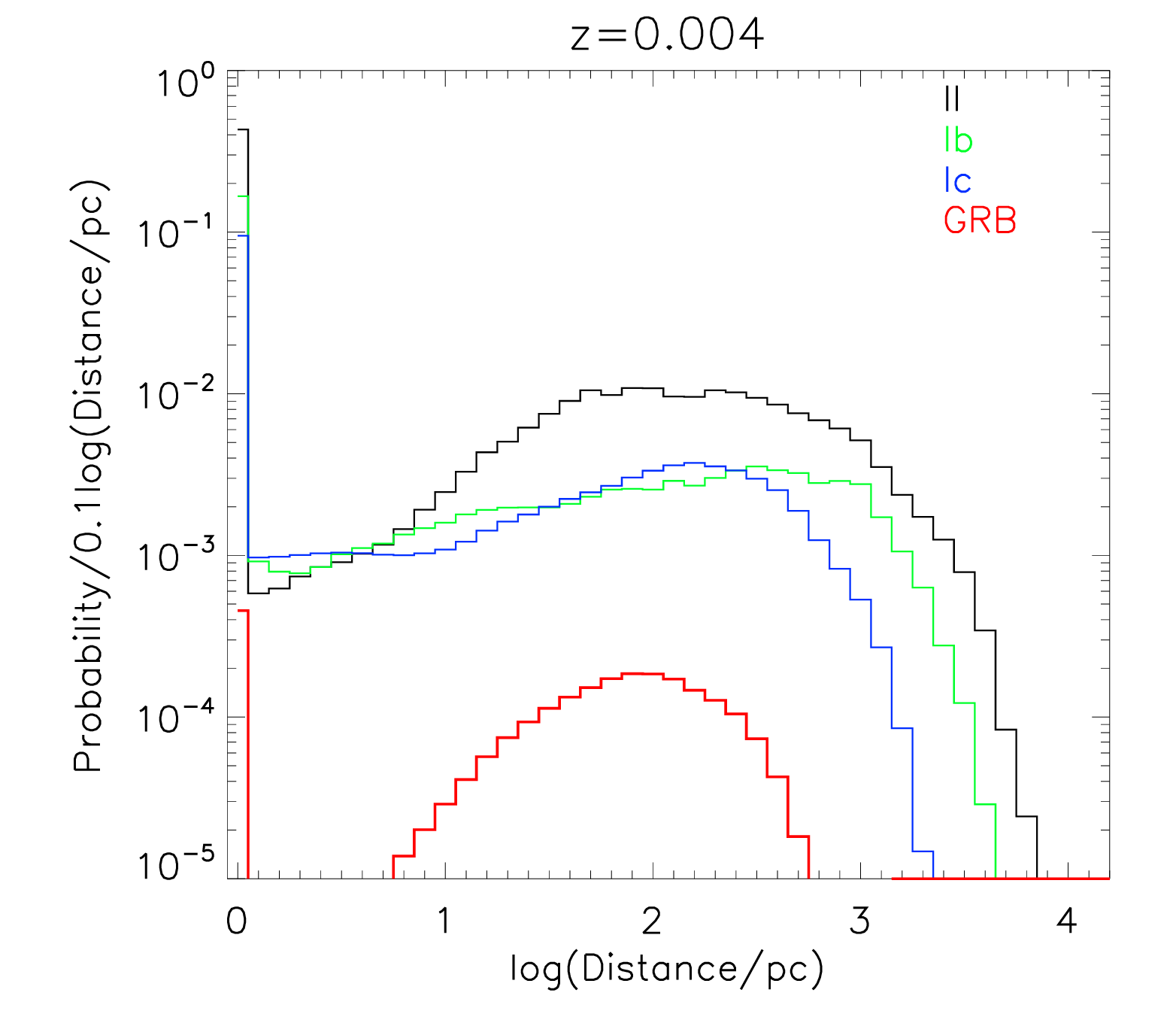}
\includegraphics[angle=0,width=80mm]{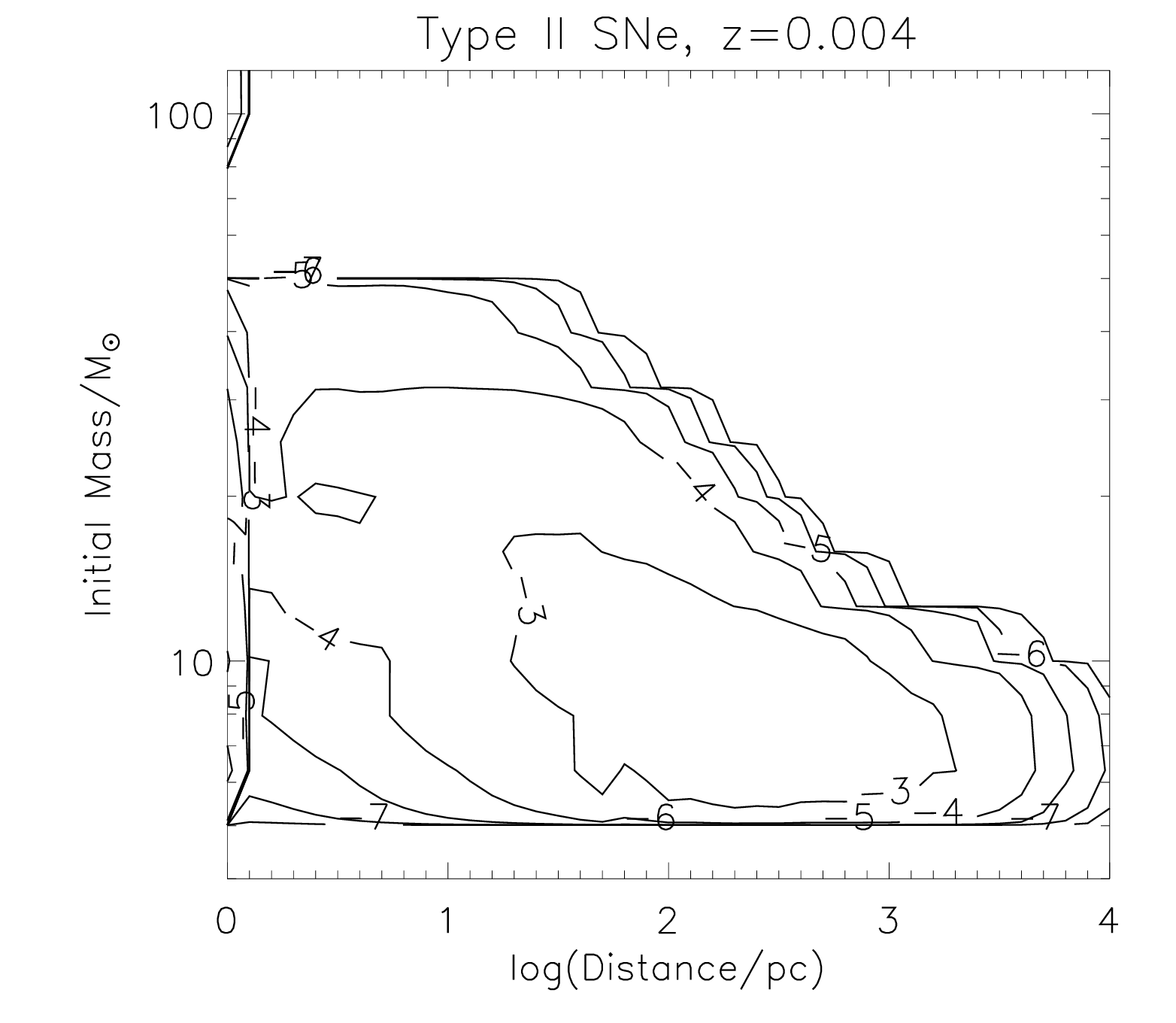}

\includegraphics[angle=0,width=80mm]{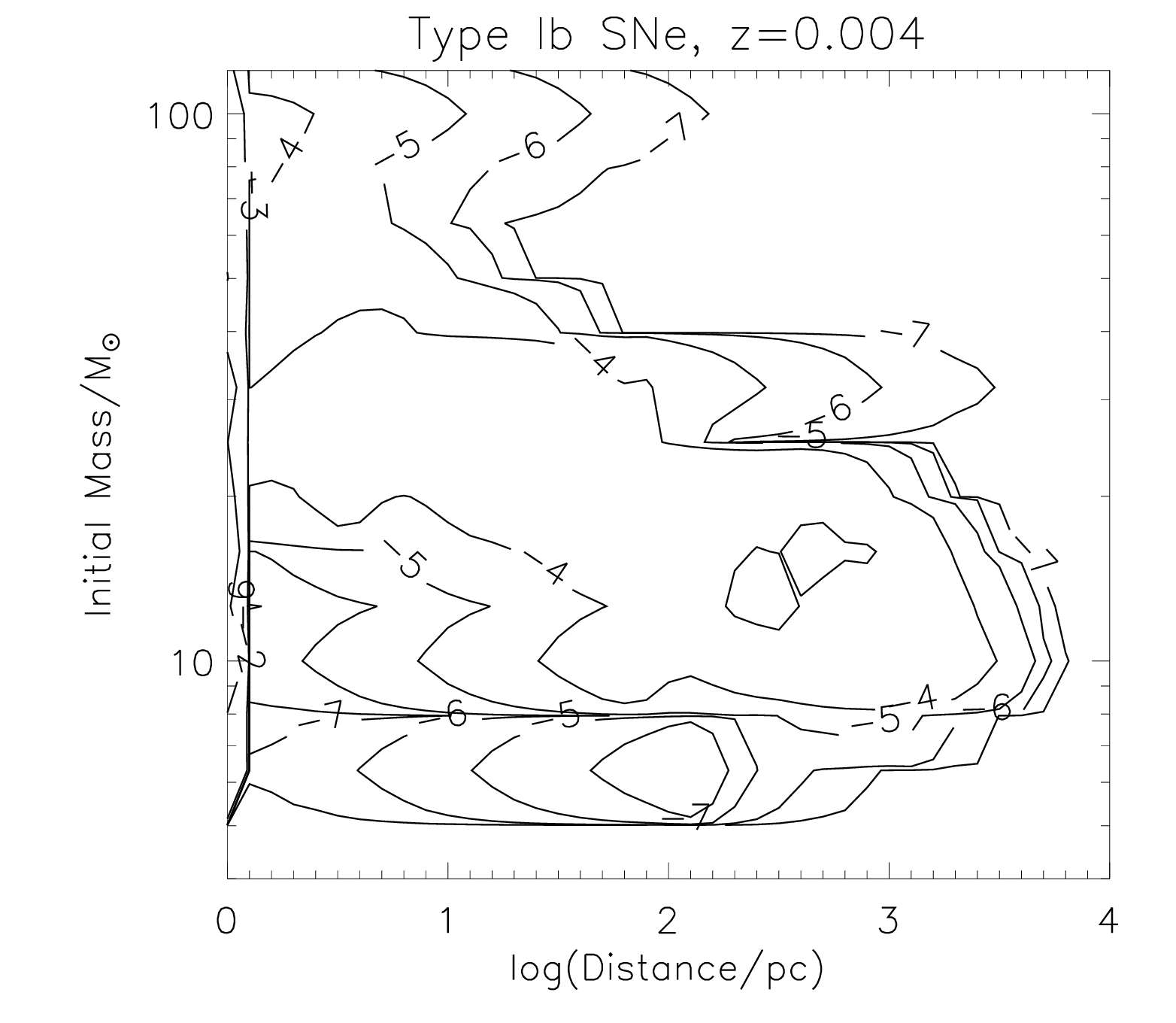}
\includegraphics[angle=0,width=80mm]{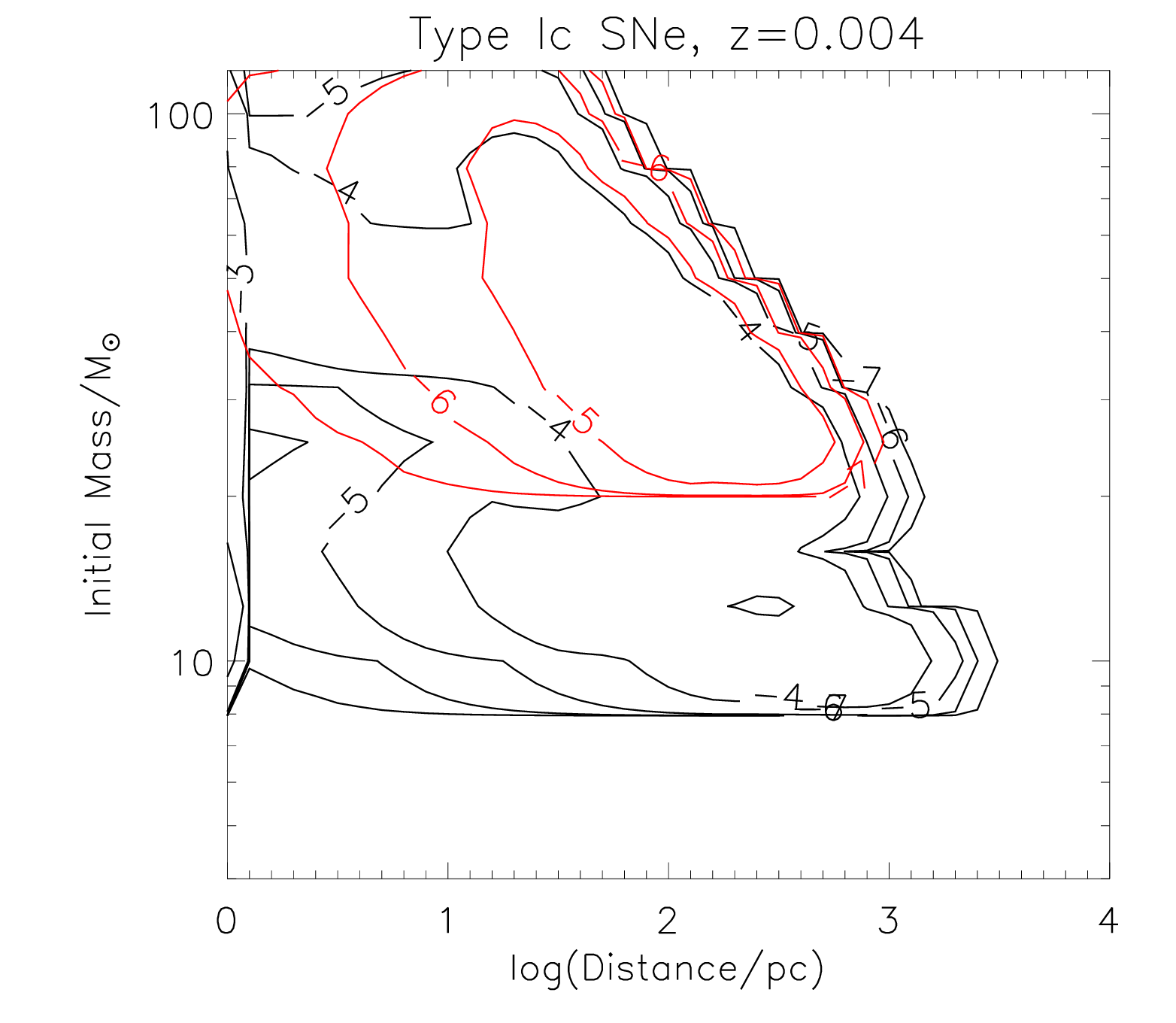}
\caption{Similar to Figure \ref{starvelocitiesTYPE} but now SMC-like
  metallicity. The red contours in the lower right panel indicate
  that distribution for long-GRB progenitors.}
\label{starvelocitiesTYPE2}
\end{figure*}

We find that long-duration GRBs have a bimodal distribution of
distances. First, 20 per cent explode without travelling any
distance. These are from binaries where the primary star does not
explode or the supernova order has been reversed. Then a second
distribution travels on average a few hundred parsecs before
exploding. This second group of long-GRBs are constrained to greater
apparent distances because of the requirement that the binary orbit
must be perpendicular to the line of sight for the GRB to be
observed. Their distribution is also similar to that for type Ic, as
suggested by \citet{fruchter} and \citet{kelly}. While at the same
time as shown in Figure \ref{starvelocitiesTYPE2} the maximum
distances possible are smaller than those of other SNe. In addition
many of our possible GRB progenitors travel a few hundred parsecs from
their initial location. This is in agreement with the observations of
\citet{hammer} who find such distances for some near by GRB
progenitors. Initially this result may appear at odds with the
observations of \citet{fruchter} but as mentioned the maximum distance
that can be travelled would be unresolved in their observations so
GRBs would be expected to occur associated with regions of the most
intense star formation.

Also we note from Fig. \ref{starvelocitiesTYPE2} that lower mass
long-GRB progenitors are more likely to travel further than high mass
progenitors. The nearby long-GRBs considered by \citet{hammer} are
relatively weak compared to normal cosmological GRBs
\citep{review}. It is possible that the nearby bursts come from lower
mass progenitors than the high redshift long-GRBs. However such a
deduction is uncertain and cannot be investigated with our simple
model. 

We have also varied our requirement that the final CO core mass must
be greater than $7M_{\odot}$ for a long-GRB to occur. For example if
we restrict long-GRB to have CO core masses at core-collapse of
$5M_{\odot}$ rather than $7M_{\odot}$ as in our fiducial model then
the mean distance travelled by long-GRBs increased to to
$240\pm380$pc. The mean effective initial mass and lifetimes also
change to $21\pm14M_{\odot}$ $43\pm25$Myrs. These differences are
because the long-GRB population is more dominated by the lower mass
QHE progenitors. We find qualitatively that if such a population
represented the observed population then long-GRBs would be the most
widely distributed explosive transient of all core-collapse events.
Therefore out predicted long-GRB progenitor population is strongly
dependent on the approximations included in our simulation.

Two studies which used the results of \citet{fruchter} to determine a
limit on the progenitors of long-GRBs were by \citet{larsson} and
\citet{raskin}. They found that they could explain the results of
\citet{fruchter} if the initial mass of GRB progenitors is greater
than $20M_{\odot}$ and $43M_{\odot}$ respectively (with lifetimes of
10Myrs and 5Myrs). They did not consider the effect of runaway stars
in this analysis. Our results on GRB progenitors in Table
\ref{populationdetailsz004} show that GRB progenitors may have
effective initial masses of $41\pm17M_{\odot}$ (or an initial mass of
$27\pm16M_{\odot}$ before mass accretion). Because the stars can
accrete mass from their primary companion the lifetime of these
possible progenitors is $15\pm5$\,Myrs, three times the age normally
expected for stars of this mass. This is greater than the maximum age
predicted by \citet{larsson} and \citet{raskin} for the progenitors of
long-GRBs. The greater age is due to the QHE increasing the
main-sequence lifetime and rejuvenation of the stars. Our result
therefore agrees with the mass of \citet{raskin} but disagrees with
their age as it is too low. While the mass of \citet{larsson} agrees
with the initial mass of our progenitors before mass accretion is
considered and is closer to our age estimate.

Qualitatively, because of the uncertainties of what is required to
cause a long-GRB and the effect of DES runaways on our results, our
model is consistent with the observations of the locations of SNe and
long-GRBs. Quantitative analysis however require involving our runaway
models with a more detailed galactic model that also considers the
stellar populations (that consider binaries) that give rise to these
events. 

\subsection{Merger events of compact-object binaries}

\begin{figure*}
\includegraphics[angle=0,width=80mm]{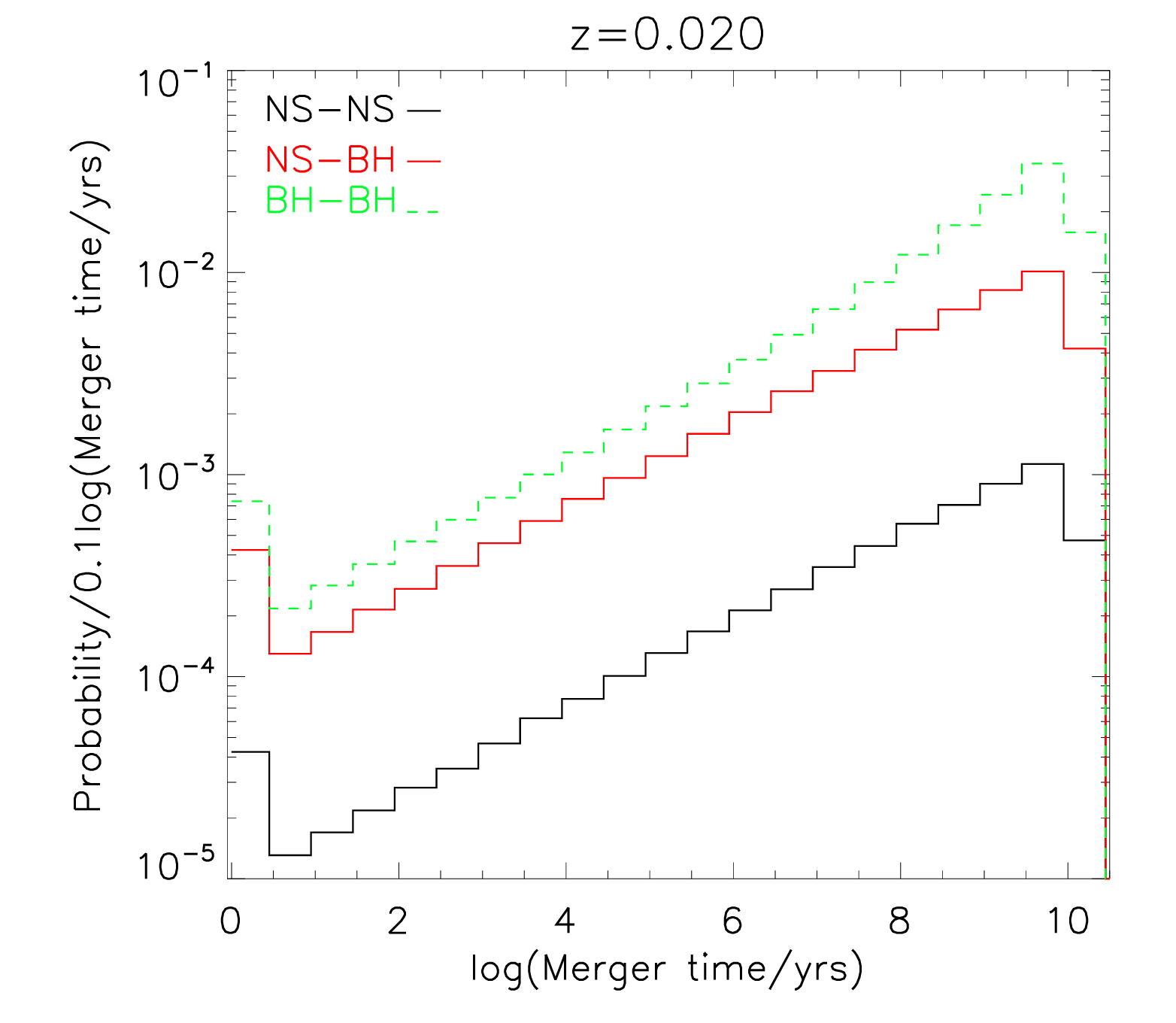}
\includegraphics[angle=0,width=80mm]{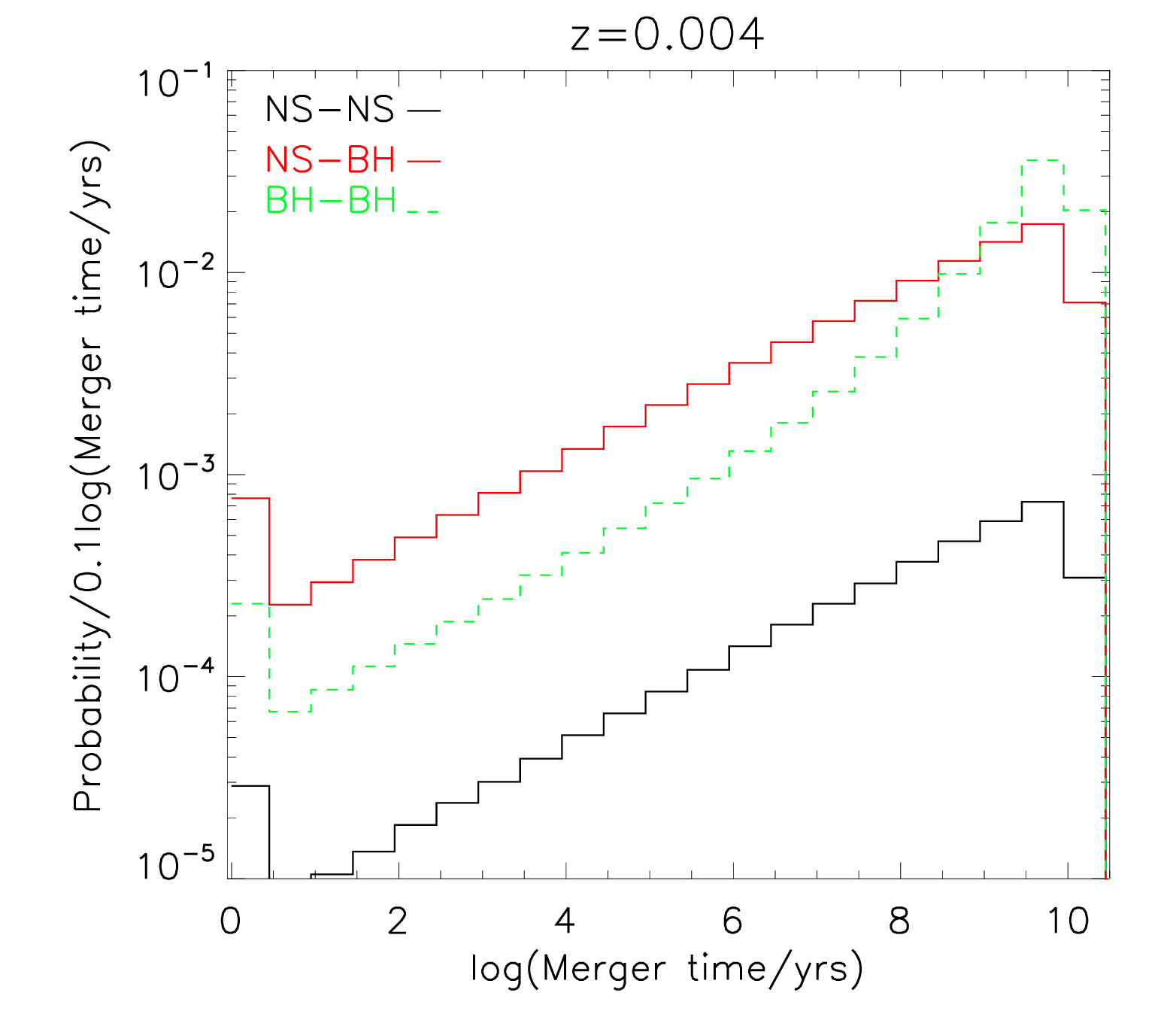}
\caption{The distribution of time required for compact binaries to
  merge. The lines are normalised so the total number of mergers is
  one.}
\label{shortgrbs2}
\end{figure*}

We have also calculated similar plots as in the previous subsection
for mergers of the compact remnant binaries we discuss in Section
\ref{dcs}. We use the general relativistic weak-field approximation
equation employed by \citet{HPT02} to calculate the time for such
binaries to merge via gravitational radiation. We then only count it
as a merger if the time we estimate is less than 13.5 Gyrs.

Our calculations are based on the present day population. I.e., they
do not include a realistic simulation of the changing populations over
cosmological times. Our results should thus hold as long as most
systems merge within the age of the Universe. As shown below, this is
mostly the case.

We show the distribution of merger times in Fig. \ref{shortgrbs2}. We
see that it is more probable that a system takes a long time to
merge. However there are also systems that have extremely short merger
times and so could merge within star formation regions. These results
mirror those of \citet{dewipols}, \citet{voss}, \citet{kris} and
\citet{shaughnessy}.

Over a similar range of merger times \citet{kris} have two sharply
defined peaks in the merger rate at around $10^5$ and $10^9$
years. They found that the two peaks correspond to two different
compact binary formation scenarios. In the first each star experiences
one mass transfer episode while in the second the secondary
experiences an extra mass transfer episode when it becomes a helium
giant. This second scenario leads to the tightest binaries and thus
the shortest merger times. We do not find such a sharply defined peaks
however our range of possible times is comparable. We may be
incorrectly estimating some of the merger times due to our basic
treatment of the gravitational radiation and assumption of circular
orbits in the stellar models.

\begin{table}
\caption{Predicted rates of binary compact object mergers and births
  relative to the rate of core collapse supernovae, at the two
  considered metallicities.}
\label{merger-rates}
\begin{tabular}{cccccc}
\hline
\hline
Z    &   WD-NS & WD-BH & NS-NS & NS-BH &    BH-BH \\
\hline
\multicolumn{5}{l}{Compact-system merger rate} \\
0.004 &  -- &  --          &  0.00007 &  0.0017  &    0.0019\\
0.020 &  -- &  --          &  0.00009 &  0.0008  &    0.0021\\
\hline
\multicolumn{5}{l}{Compact-system birth rate} \\
0.004 &  0.0045 &  0.0078 &  0.0003 &  0.0052  &    0.0292\\
0.020 &  0.0054 &  0.0074 &  0.0004 &  0.0029  &    0.0181\\
\hline
\hline
\end{tabular}
\end{table}

The merger rates, shown in Table \ref{merger-rates} and
Fig. \ref{shortgrbs2}, indicates that merging black holes and neutron
stars are the most common visible mergers. However we are not able to
estimate the merger rates for systems containing white dwarfs due to
limitations of our grid of models for the secondaries evolution. We
are able to calculate the birthrate and give this in Table
\ref{merger-rates}. If a similar number of these systems merge within
the age of the universe as for the other compact binaries then the
rate of white dwarf--neutron star or black hole mergers may be
greatest. The rate of such mergers must be less than 1/80th of the
core collapse supernova rate. \citet{thompson} suggest a rate between
1/20th and 1/40th of the SN rate from the observation of one such
system. Because the observable outcome of such events is not yet well
studied \citep{wdmergers}, it seems unclear whether they have
observable counterparts. We suggest that it is possible for these
systems the merger is most likely to be an extended period of mass
transfer. When the white dwarf is less massive than the neutron star
any mass transfer is sufficiently stable \citep{toutrlof} so these
binaries might be observed as X-ray binaries with mass-transfer driven
by gravitational radiation.
 
When we consider where compact binary merger events might be observed
we should bear in mind the results displayed in
Fig. \ref{doublevelocities}. Here we find that many compact binaries
have high space velocities and can escape their host galaxies,
especially double neutron-star systems. Therefore we should expect
some compact-object binary merger events to occur some distance away
from any nearby galaxy. There is growing evidence that some such
events do occur \citep{levangrb,grboffset}.  Merging double-neutron
star systems as well as neutron star-black hole mergers are possibly
related to short gamma-ray bursts. In this case these rates should be
considered upper limits on the observable rate because the progenitor
rotation axis must be within a small angle to the line of sight to be
observed, just as for the long-GRBs.

\section{Discussion \& Conclusions}

Our population synthesis of runaway stars is able to predict a
velocity distribution similar to that observed for OB stars. We
predict that in the Galaxy the number of all O stars that are runaways
from BSS is between 0.5 to 2.2 per cent. Using the catalogue of
\citet{ocat} we estimate the observed number is $4\pm1$ per cent,
however both numbers are uncertain and very sensitive to the total
number of O stars in the sample and how runaways are identified.

While we do not consider dynamical ejection from clusters here, we
suggest that this distribution may be similar to that predicted for
the binary supernova ejection scenario. This is because stars ejected
by dynamical interactions in a cluster are normally the result of
binary interactions so that the determining factor for both scenarios
is the initial separation distribution of binary stars. The initial
binary period/separation distribution has a direct effect on the
velocities of runaway stars. A wide range of binary separations is
required to explain the observed velocity distribution. A distribution
that is flat in the logarithm of the separation reproduces the
observed range of runaway velocities. However this separation
distribution may underestimate the number of close binaries.

To match the fact that there are a number of observed WR runaways with
velocities greater than $30\, {\rm km \, s^{-1}}$ we include black
hole kicks when black holes are formed by core-collapse. This has
implications for the velocities of binaries that include a black
hole. These should be more often observed as runaways. For example the
X-ray binary Nova Sco which has a space velocity of $150 \pm 19 \,{\rm
  km \, s^{-1}}$ provides evidence that at least some black holes may
have kicks at their formation. \citet{nelemans} suggest that such
large velocities for black-hole binaries do not require a large
kick. From the black-hole binaries they list we find the mean velocity
is $34 \pm 32 {\rm km \, s^{-1}}$, this is similar to the distribution
of synthetic black hole binaries shown in Figure
\ref{compactvelocities}.

We consider the final outcome of our stars and the velocity
distribution of compact stellar remnants. Our models predict that
there should be a small surplus of neutron stars with space velocities
less than $200\, {\rm km \, s^{-1}}$. These require a system not to be
unbound by the first SN but by the second. If these were not taken
into account when the neutron star kick is determined from
observations it would lead to an underestimate of the kick
velocity. Most white dwarfs and black hole runaways have velocities
typically below those expected for neutron stars. The final outcome of
evolution of all our binaries is that most compact stellar remnants
are single. Only 20 per cent of all binaries that experience at least
one SN remain bound. The rarest are white dwarf and neutron star
binaries where the white dwarf formed first. Normally such systems can
be identified by their highly eccentric orbits, for example
PSR~B2303+46 and PSR~J1141-6545 \citep{nswd1,nswd2}. There is a
predicted difference in the space velocity between such binaries and
those in which the neutron star forms first.

Whether a SN progenitor is a runaway star is an important factor in
where a SN might occur in a galaxy. We have shown that less massive
stars travel further from their initial positions than more massive
stars. However the majority of stars explode without travelling a
large distance.  When we consider the different SN types we find that
because type II and type Ib SNe come from progenitors less massive
than $20M_{\odot}$ they can travel furthest from their initial
positions, $48\pm210$pc and $34\pm150$pc respectively on
average. While type Ic progenitors come from more massive stars and
therefore have shorter lifetimes and travel shorter distances,
$17\pm74$pc. This picture changes at low metallicity with the
inclusion of quasi-homogeneous evolution, which prolongs the life and
the distance to which type Ic and type Ib progenitors can travel.

Nearly one third of long-GRB progenitors may travel a few hundred pc
from their birthplaces before experiencing core-collapse. But, they
can travel shorter distances and/or explode before type IIP and Ib SNe
and so should be the most tightly correlated with star-formation in
the host galaxies. These distances however are dependent on the
parameters taken for a model to give rise to a long-GRB.

Type IIP supernova progenitors that are secondaries in a binary are
distributed throughout a galaxy due to BSS more than progenitors of
long-GRBs. These remain closer to their birth places and have shorter
lifetimes so explode when the star-formation episode they belong to is
still visible \citet{larsson,raskin}. Therefore these effects in
combination can explain the observed distribution of SN within
galaxies. To infer the nature of long-GRB progenitors from such
observations runaway stars must be considered.  We find that of the
supernovae in Tables \ref{populationdetailsz020} and
\ref{populationdetailsz004}, 20 per cent of the SN come from BSS
runaway stars. If there are at least as many DES runaway progenitors
then between 20 to 40 per cent of SNe has a progenitor that was
runaway star. A similar fraction was suggested from observations by
\citet{jamessnhalpha}. Without considering the effect of runaway
stars, the mass of long-GRB progenitors is likely to be overestimated.

Neutron star--black hole binaries and more so double neutron star
systems, which are candidate short-GRB progenitors, are found to reach
large runaway velocities. We estimate their merger timescales and find
that they should often merge outside of their host galaxy. Compact
mergers of a white dwarf with either a neutron star or black hole are
estimated to be at least as common as other compact binary
mergers. Due to their low runaway velocities and merger timescales,
they should mostly merge within their host galaxy.  The observational
signature of such events appears to be as yet unclear.

\begin{table*}

\caption[]{The relative birth rates of binary-produced unbound compact
  objects (white dwarfs, neutron stars and black holes), and compact
  objects which remain part of a binary system, compared to the birth
  rate of single compact objects (last line). For compact objects
  remaining in binaries, it is distinguished whether they form from
  the primary or the secondary star. These results refer to our
  population synthesis model at a metallicity of Z=0.020. The columns
  indicate the remnant from the primary star of the binary with the
  rows referring to the remnant from the secondary star.}
\label{remz020}
\begin{tabular}{lllll}
\hline
\hline
             &  Unbound       &  Primary WD    &  Primary NS    &  Primary BH \\
\hline
Unbound      &           & 0.0176           & 0.1397             & 0.0747 \\
Secondary WD &   0.0280  & 0.1634           & 0.0032             & 0.0045 \\
Secondary NS &   0.0916  & 0.0002           & 0.0003             & 0.0013 \\
Secondary BH &   0.0354  & 0.0001           & 0.0005             & 0.0112  \\ 
\hline
Single &                 & 0.1831           & 0.1481            &  0.0972 \\
\hline
\hline
\end{tabular}

\caption[]{The mean space velocities of different single and binary compact remnants at a metallicity of Z=0.020.}
\label{remvelz020}
\begin{tabular}{lcccc}
\hline
\hline
             &  Unbound       &  Primary WD    &  Primary NS   &   Primary BH \\
\hline
Unbound      &                  & $ 11\pm3  $  & $ 420\pm180$ & $100\pm 70$ \\
Secondary WD &  $ 24\pm18 $     &           0  & $ 81\pm38 $  & $ 51\pm27$ \\
Secondary NS &  $ 390\pm180$    & $140\pm 55$ &  $160\pm62 $   & $110\pm45$ \\
Secondary BH &  $100\pm67$      &  $120\pm53$  & $110\pm52 $ & $ 53\pm32$ \\
\hline
Single &                        &  0           & $ 430\pm190$ & $96 \pm61$ \\
\hline
\hline
\end{tabular}

\caption[]{The relative birth rates of different single and binary compact remnants at a metallicity of Z=0.004.}
\label{remz004}
\begin{tabular}{lcccc}
\hline
\hline
             &  Unbound       &  Primary WD    &  Primary NS    &  Primary BH \\
\hline
Unbound      &        & 0.0230     & 0.1681 & 0.0763 \\
Secondary WD & 0.0220 & 0.1175     & 0.0032 & 0.0048 \\
Secondary NS & 0.0897 & 0.00005    & 0.0002 & 0.0024 \\
Secondary BH & 0.0461 & 0.0010     & 0.0014 & 0.0214 \\
\hline
Single &              & 0.1279     & 0.1887 & 0.1063 \\
\hline
\hline
\end{tabular}

\caption[]{The mean space velocities of different single and binary compact remnants at a metallicity of Z=0.004.}
\label{remvelz004}
\begin{tabular}{lcccc}
\hline
\hline
             &  Unbound       &  Primary WD    &  Primary NS    &  Primary BH \\
\hline
Unbound      &                & $ 10 \pm 2 $   &$  420\pm180$ & $ 95\pm 72$ \\
Secondary WD &  $  26\pm20$   &            0   &  $ 81\pm39$  &$  56\pm30$ \\
Secondary NS &  $ 380\pm180$  & $ 89 \pm29$    & $160 \pm65 $ & $ 97\pm46$ \\
Secondary BH &  $  93\pm55 $  &  $ 76\pm36$  &  $ 87\pm55$  & $ 37\pm28$ \\
\hline
Single &                      &              0 & $ 430\pm180$ & $ 94\pm71$ \\
\hline
\hline
\end{tabular}

\end{table*}

\begin{table*}

\caption[]{The relative rates of the different SN types and their
  progenitors mean effective initial mass, mean space velocity and mean
  distance travelled from point of origin at a metallicity of
  Z=0.020. The relative rates and means are also split up into the
  values for single stars, primaries and secondaries.}
\label{populationdetailsz020}
\begin{tabular}{clccccc}
\hline
\hline
 &   &  IIP   &     non-IIP   &        Ib      &    Ic   &        L-GRB\\
\hline
Fraction & All        & 0.582 & 0.119 & 0.068 & 0.231 & 0.0\\
         & Single     & 0.287 & 0.046 & 0.007 & 0.058 & 0.0\\
         & Primary    & 0.150 & 0.055 & 0.045 & 0.123 & 0.0\\
         & Secondary  & 0.144 & 0.018 & 0.016 & 0.050 & 0.0\\
\hline
$<M_{\rm i}>$              & All       &  $10.7\pm3.0$ & $ 17.7\pm6.3 $&  $ 16.8\pm5.8$ & $ 37.1\pm22.8 $    & ~ \\
$\left[ M_{\odot}\right]$  & Single    &  $11.1\pm2.9$ & $ 21.9\pm2.3 $&   $27.5\pm0.5$ & $ 50.7\pm21.8 $   & \\
                          & Primary   &  $10.1\pm2.9$ & $ 13.2\pm5.4 $&   $15.6\pm4.5$ & $ 32.7\pm22.5 $    & \\
                          & Secondary &  $10.5\pm3.2$ & $ 20.3\pm6.6$ &  $ 15.6\pm5.8$ & $ 32.3\pm18.0 $    & \\
\hline
$<v_{\rm run}>$                   & All      & $ 4.4\pm12.0$ &  $2.0\pm6.1 $& $ 5.0\pm13.4$ & $ 4.0\pm10.8 $    &  \\
$\left[ \rm km \, s^{-1}\right]$ & Single   &     0.0       &   0.0        &   0.0          &  0.0            &\\
                                & Primary  & $  0.2\pm1.2$ & $ 1.0\pm3.5$ &$  0.3\pm1.8$   & $0.4\pm1.9  $   &  \\
                                & Secondary& $ 17.5\pm18.7$& $10.0\pm 11.6$&$ 20.3\pm21.0$ & $17.4\pm17.4  $  & \\
\hline
$<d_{\rm run}>$         & All      &  $  48\pm210$& $ 10\pm 39$& $35\pm150$ & $ 17\pm 74 $   &  \\
$\left[ \rm pc \right]$& Single   &      0.0     &     0.0    &     0.0     &     0.0      &  \\
                       & Primary  & $    2\pm16 $& $ 8\pm36$& $   2\pm15 $ & $  1\pm 10  $  &  \\
                       & Secondary& $  190\pm380$& $  38\pm71 $& $150\pm280$ & $ 74\pm150 $  &  \\
\hline
$<t_{\rm run}>$     & All        & $ 36\pm31$ & $ 22\pm38$ &$  21\pm21$   &$ 9\pm7$  &  \\
$\left[ \rm Myr \right]$& Single& $ 26\pm10$ &  $ 9\pm1  $ &$ 7.0\pm0.1 $ &$ 5\pm1$   &  \\
                   & Primary   &   $33\pm15$ &  $ 26\pm20$ & $ 17\pm9$ &   $10\pm5$    &  \\
                   & Secondary &  $ 57\pm52$ & $  46\pm85$ & $ 37\pm34$&   $13\pm11$   &  \\
\hline
\hline
\end{tabular}

\caption[]{The relative rates of the different SN types and their
  progenitors mean effective initial mass, mean space velocity and
  mean distance travelled from point of origin at a metallicity of
  Z=0.004. The relative rates and means are also split up into the
  values for single stars, primaries and secondaries. The mean mass in
  brackets for the long-GRBs is the actual initial mass before the
  stars have accreted material from the primary.}
\label{populationdetailsz004}
\begin{tabular}{clccccc}
\hline
\hline
 &    &  IIP   &   non-IIP   &        Ib      &    Ic   &        L-GRB\\
\hline
Fraction & All        &   0.605 &  0.141  &  0.164 &  0.091  &  0.001 \\
         & Single     &   0.324 &  0.055  &  0.025 &  0.000  &  0.0 \\
         & Primary    &   0.163 &  0.079  &  0.079 &  0.055  &  0.0 \\
         & Secondary  &   0.118 &  0.008  &  0.060 &  0.036  &  0.001 \\
\hline
$<M_{\rm i}>$            & All    & $  9.9 \pm3.4 $ & $ 21.4\pm10.0$ & $ 30.0\pm25.3$ &  $25.6\pm18.5$ &  $40.7\pm16.8 (15.2\pm12.1)$\\
$\left[ M_{\odot}\right]$& Single & $  10.6\pm3.5$ &  $ 28.4\pm 5.9$ & $ 66.2\pm20.4$ &     --        &      \\
                      & Primary  & $  9.2 \pm3.1 $ & $ 16.3\pm 9.5$ & $ 28.5\pm24.0$ & $ 22.8\pm17.9$ &      \\
                      & Secondary& $  9.0 \pm2.8$  &  $24.4\pm 5.4$ & $ 17.3\pm11.1$ &  $30.0\pm18.4$ & $ 40.7\pm16.8 ( 15.2\pm12.1)$\\
\hline
$<v_{\rm run}>$                   &All  & $    4.2\pm12.1 $ & $ 1.0\pm3.6$ &  $ 6.0\pm14.0 $&$  8.6\pm16.3$ &$  12.0\pm10.9$\\
$\left[ \rm km \, s^{-1}\right]$ & Single &     0.0   &            0.0 &           0.0  &      --        &      \\
                                & Primary & $ 0.1\pm0.9$  &  $ 0.6\pm2.1 $& $  0.3\pm1.6$  &$0.2\pm1.2 $  &      \\
                                & Secondary&$21.2\pm19.9$ &  $12.7\pm6.5 $&  $15.8\pm19.4$ & $21.3\pm19.9$&$   12.0\pm10.9$ \\
\hline
$<d_{\rm run}>$       & All      & $  54 \pm240   $ & $ 3 \pm 19  $ &$  75  \pm 250 $ & $60\pm150 $ & $88\pm100$\\
$\left[ \rm pc \right]$& Single&         0.0  &         0.0              &  0.0            &  0.0      &      \\
                & Primary         & $0.5\pm  4  $  & $3  \pm 21  $ &$  3   \pm  21 $ &   $1\pm6$ &      \\
                & Secondary&        $280\pm 480$ & $ 27  \pm 35 $  & $200  \pm 390 $ & $150\pm210 $&$  88\pm100$\\
\hline
$<t_{\rm run}>$           & All  &  $ 39\pm25$ &$  15\pm12$ & $ 27\pm30$&$ 16\pm7 $ & $15\pm5$ \\
$\left[ \rm Myr \right]$& Single& $  31\pm14$ & $  8\pm1  $  & $4\pm1 $&    --        &  \\
                      & Primary &  $ 41\pm20$ & $ 21\pm14$ & $ 17\pm15$& $15\pm7 $  &  \\
                     & Secondary& $  58\pm40$ & $  9\pm2$  & $ 49\pm36$& $17\pm7$  & $15\pm5$ \\
\hline
\hline
\end{tabular}
\end{table*}

\section{Acknowledgements}
We would like to thank the constructive input from the annoynmous
referee. JJE is supported by the Institute of Astronomy's STFC Theory
Rolling Grant. CAT thanks Churchill College for his Fellowship. Thanks
to V. Gvaramadze and Thomas Tauris for discussions.

\section{Appendix}

\begin{figure*}
\includegraphics[angle=0,width=80mm]{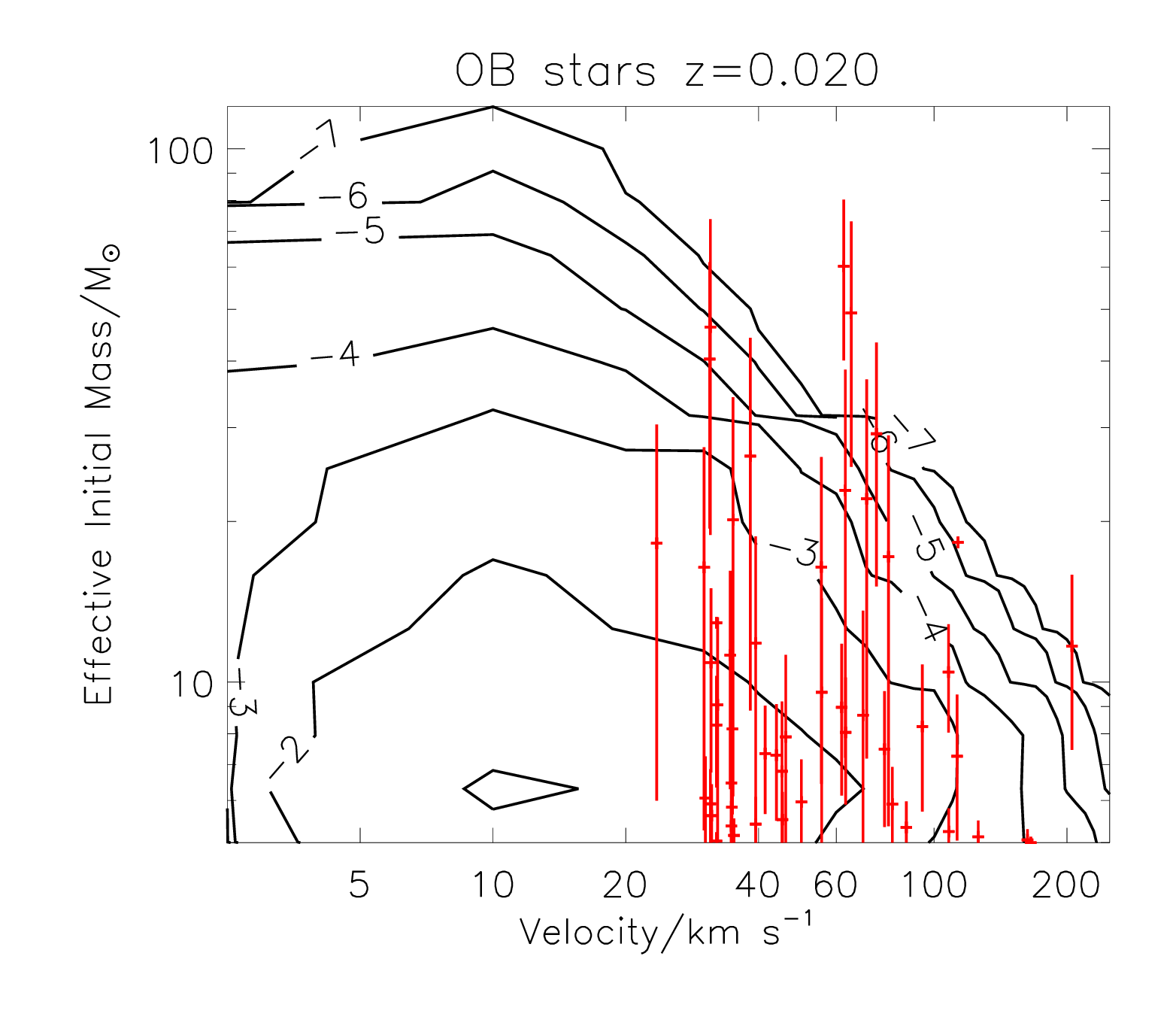}
\includegraphics[angle=0,width=80mm]{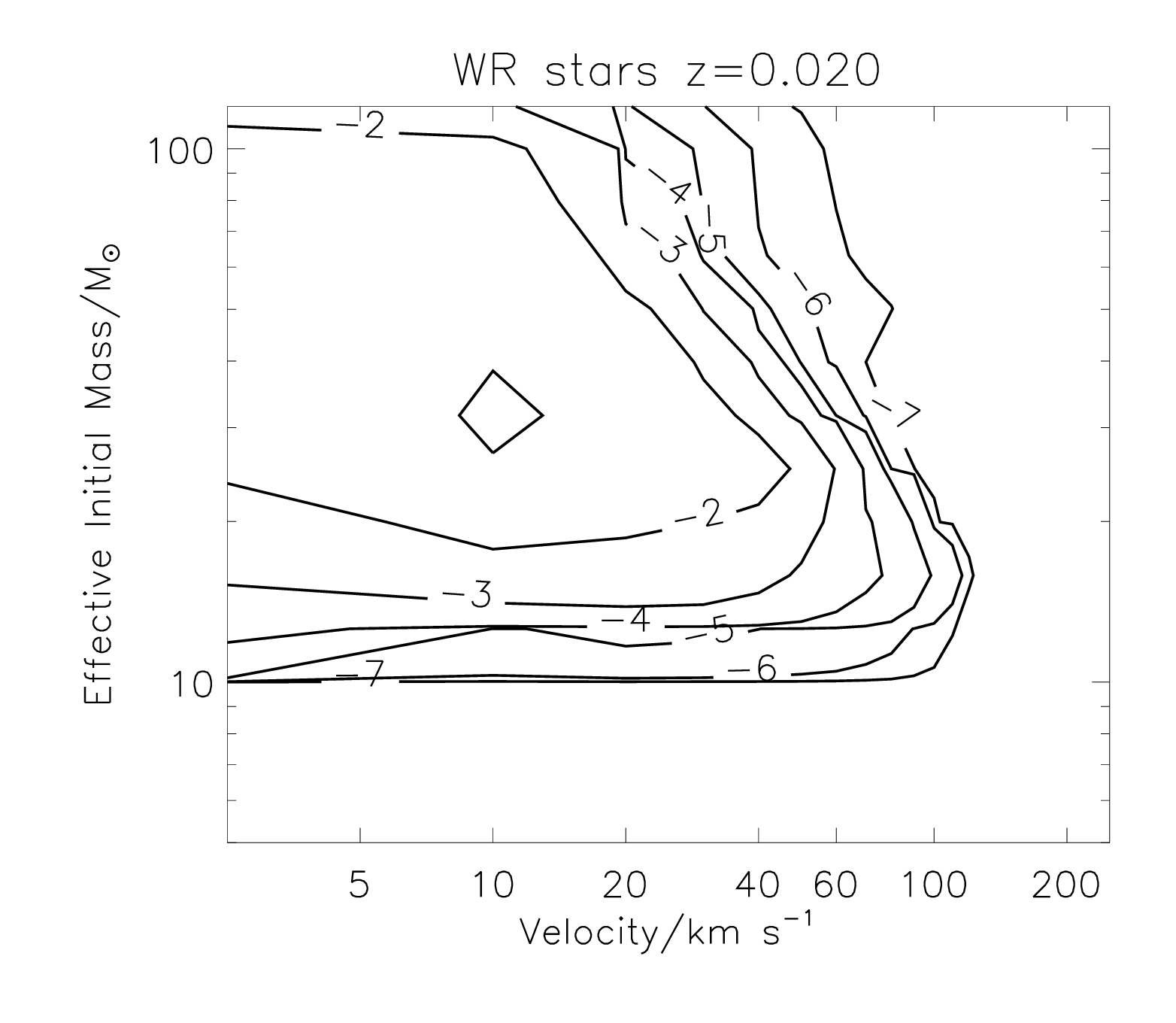}

\includegraphics[angle=0,width=80mm]{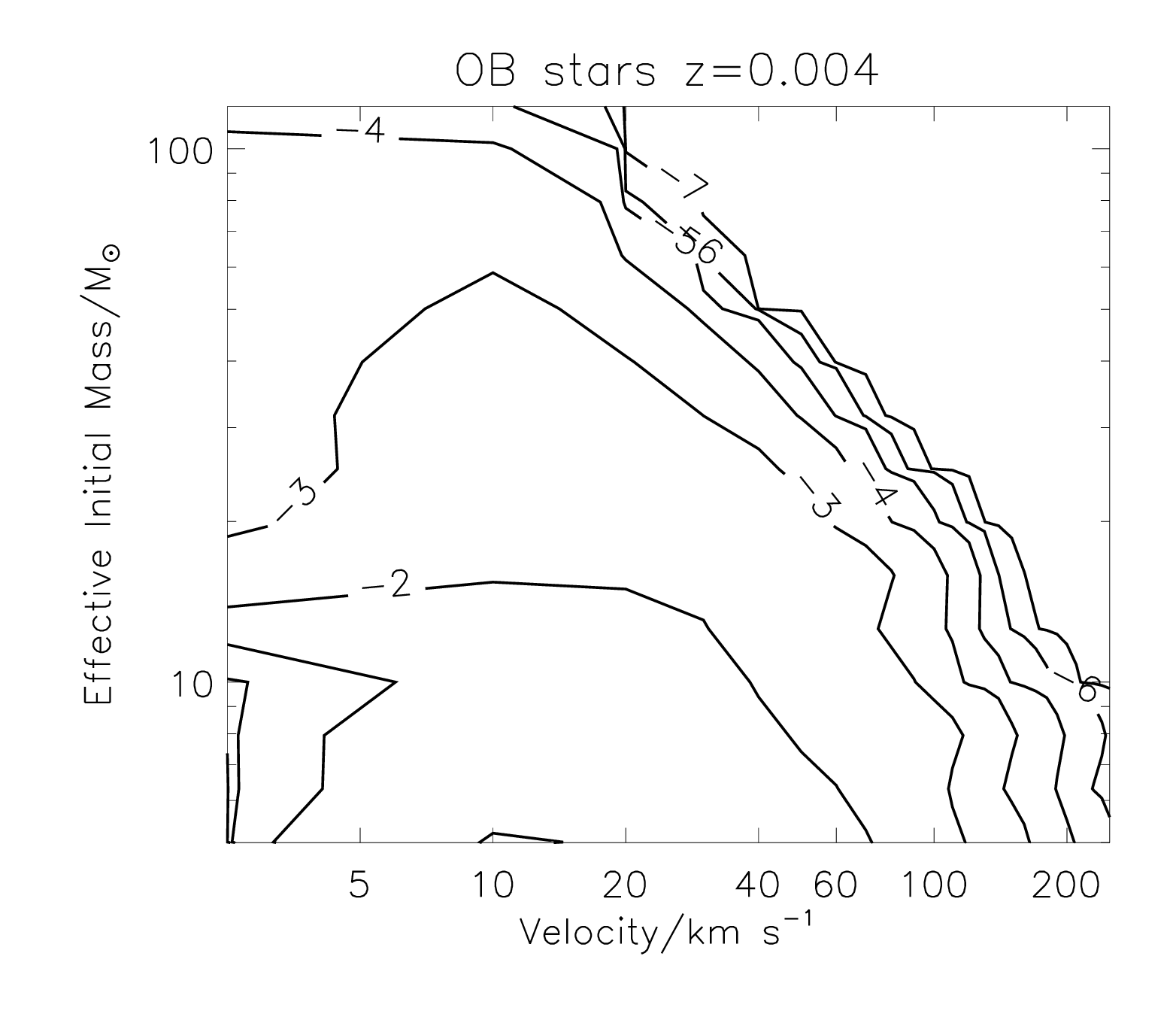}
\includegraphics[angle=0,width=80mm]{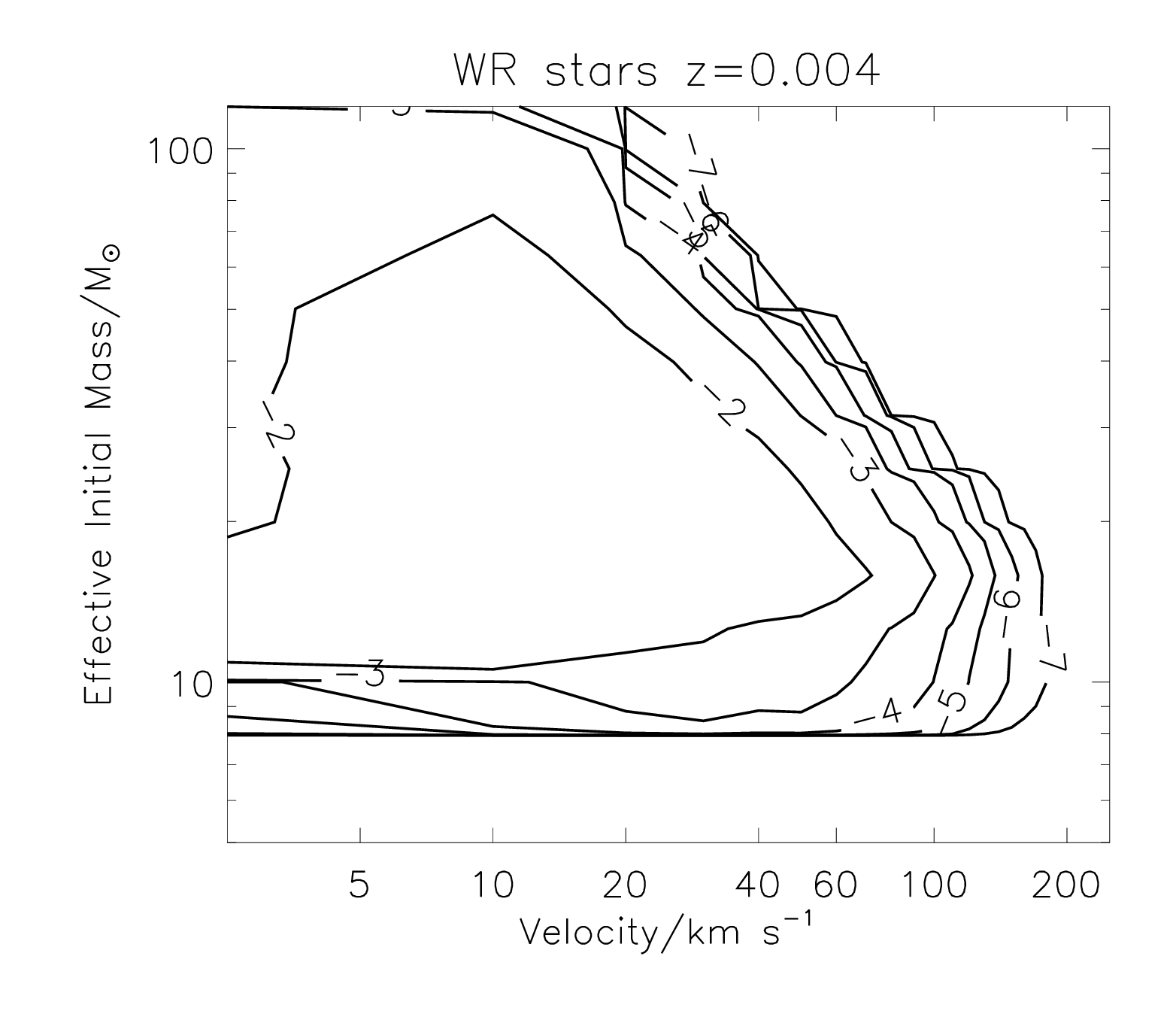}

\caption{Similar to Figure \ref{mainrunaways2} but with the velocity
  range below the normal velocity limit required for a star to be
  identified as a runaway.}
\label{mainrunaways2appendix}
\end{figure*}

\begin{figure*}
\includegraphics[angle=0,width=80mm]{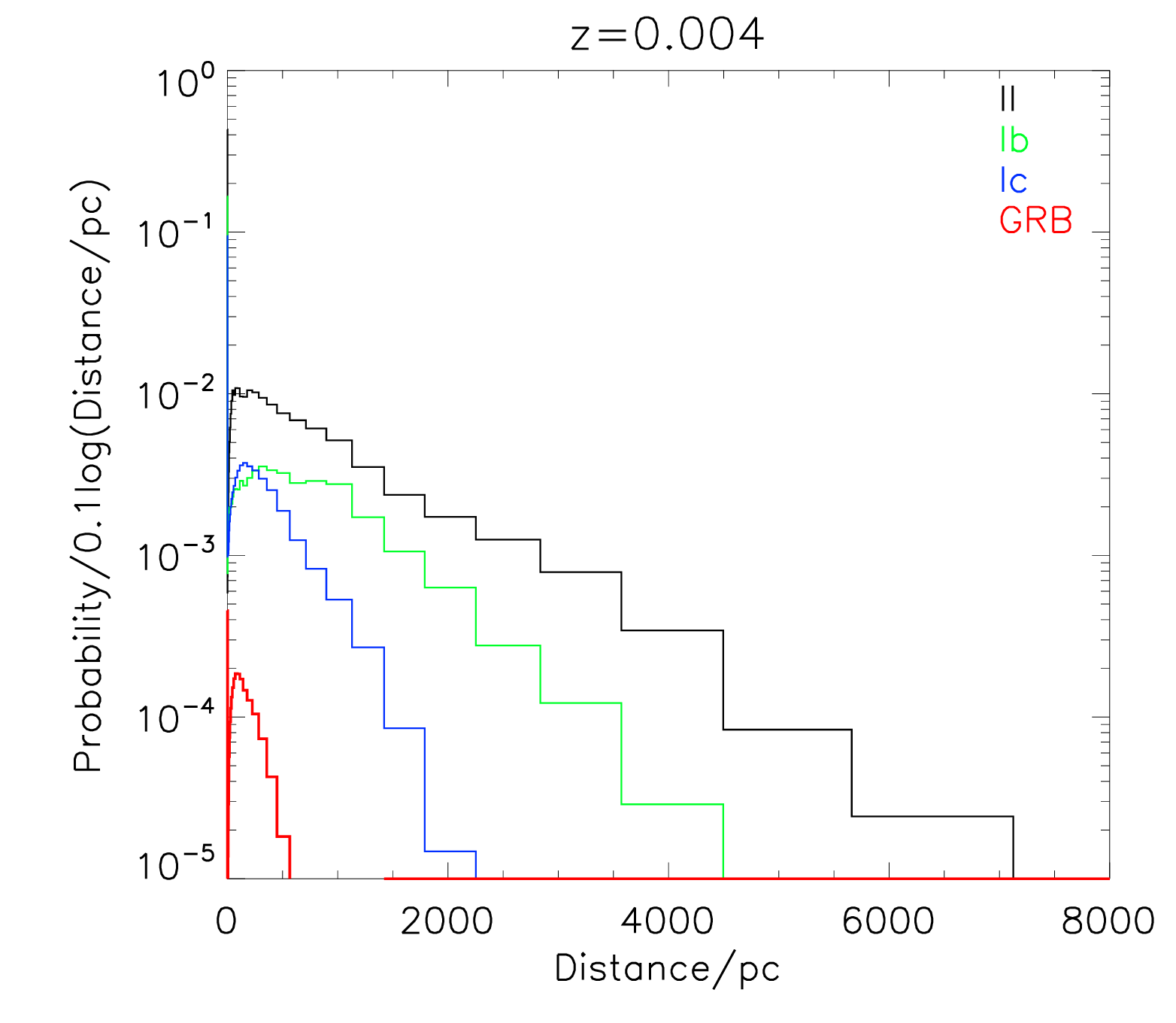}
\includegraphics[angle=0,width=80mm]{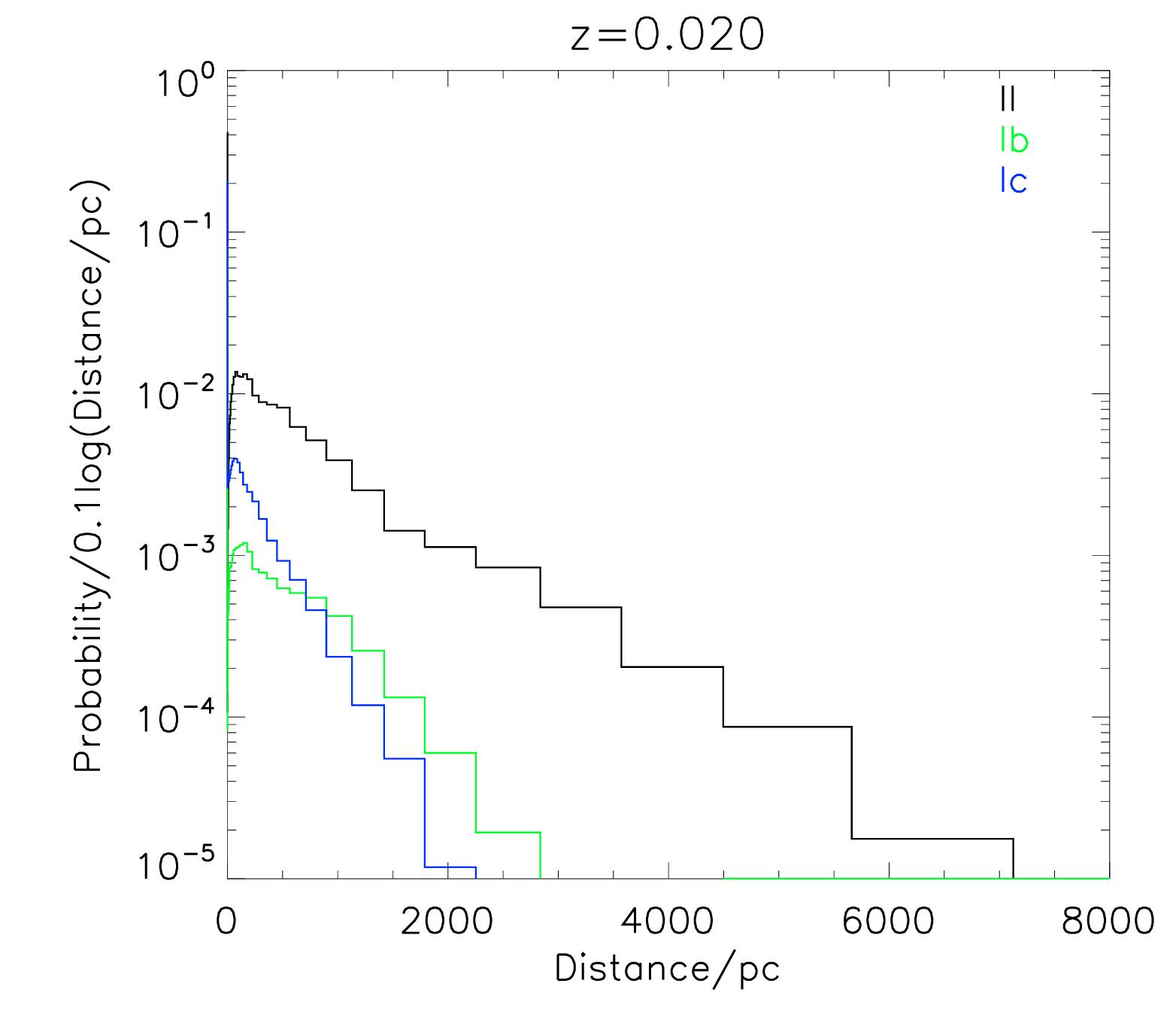}
\caption{Similar to the upper left panels in Figures
  \ref{starvelocitiesTYPE} and \ref{starvelocitiesTYPE2} but with a
  linear scale on the x-axis.}
\label{s2}
\end{figure*}

\begin{figure*}
\includegraphics[angle=0,width=80mm]{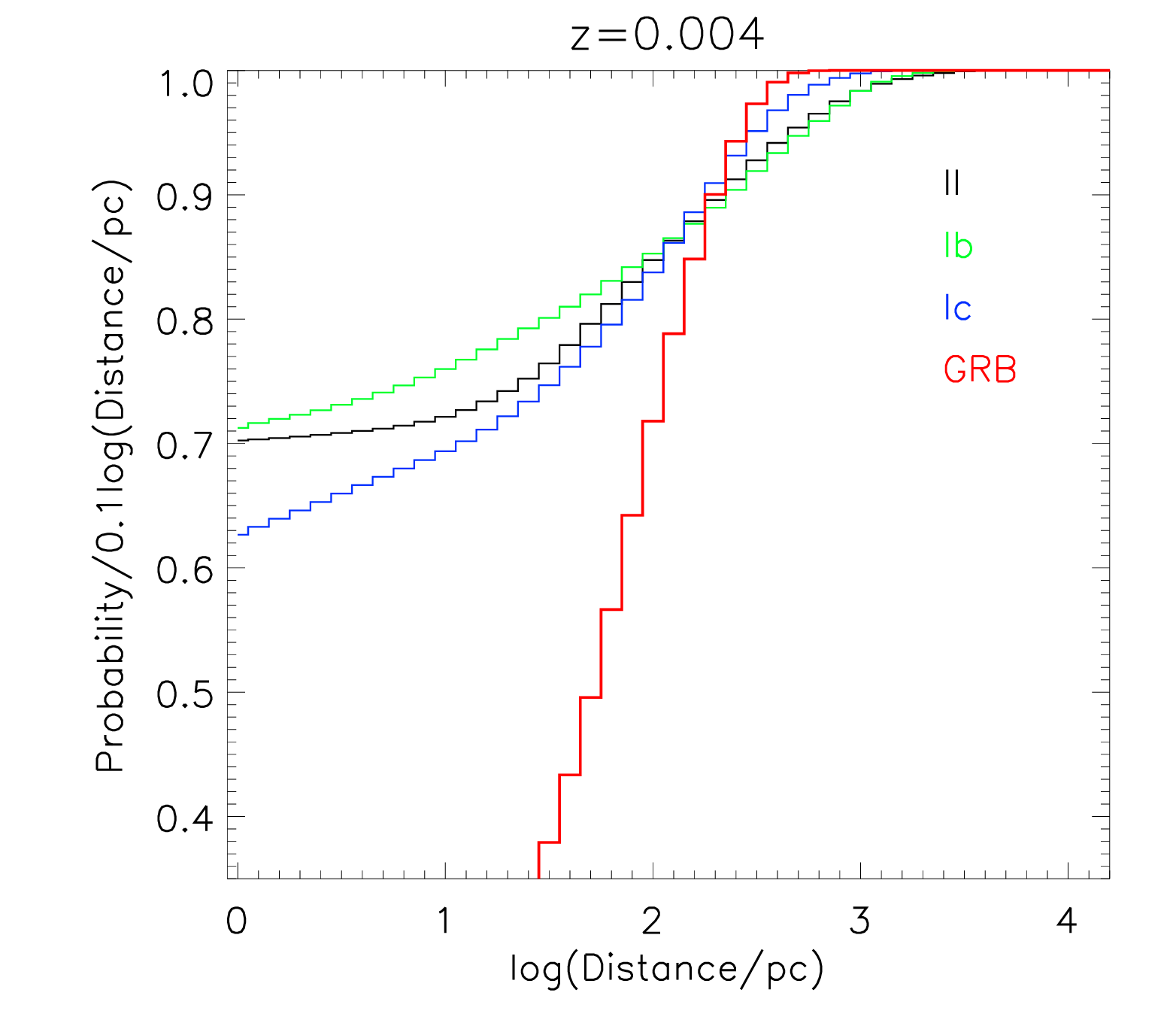}
\includegraphics[angle=0,width=80mm]{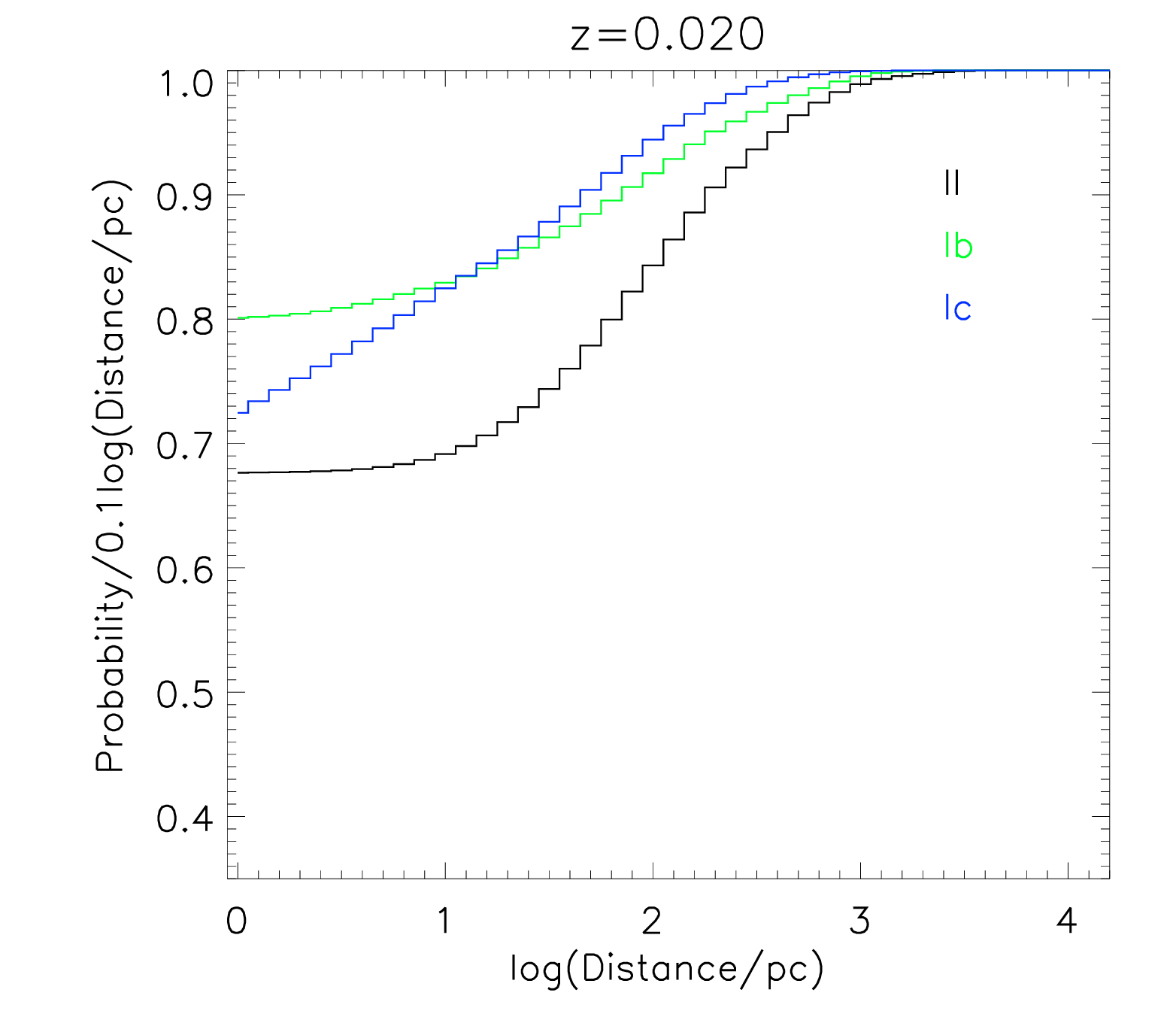}
\caption{Similar to the upper left panels in Figures
  \ref{starvelocitiesTYPE} and \ref{starvelocitiesTYPE2} but with the
  cummulative probability with increasing progenitor distance on the
  y-axis with a linear scale.}
\label{s5}
\end{figure*}

\begin{figure*}
\includegraphics[angle=0,width=80mm]{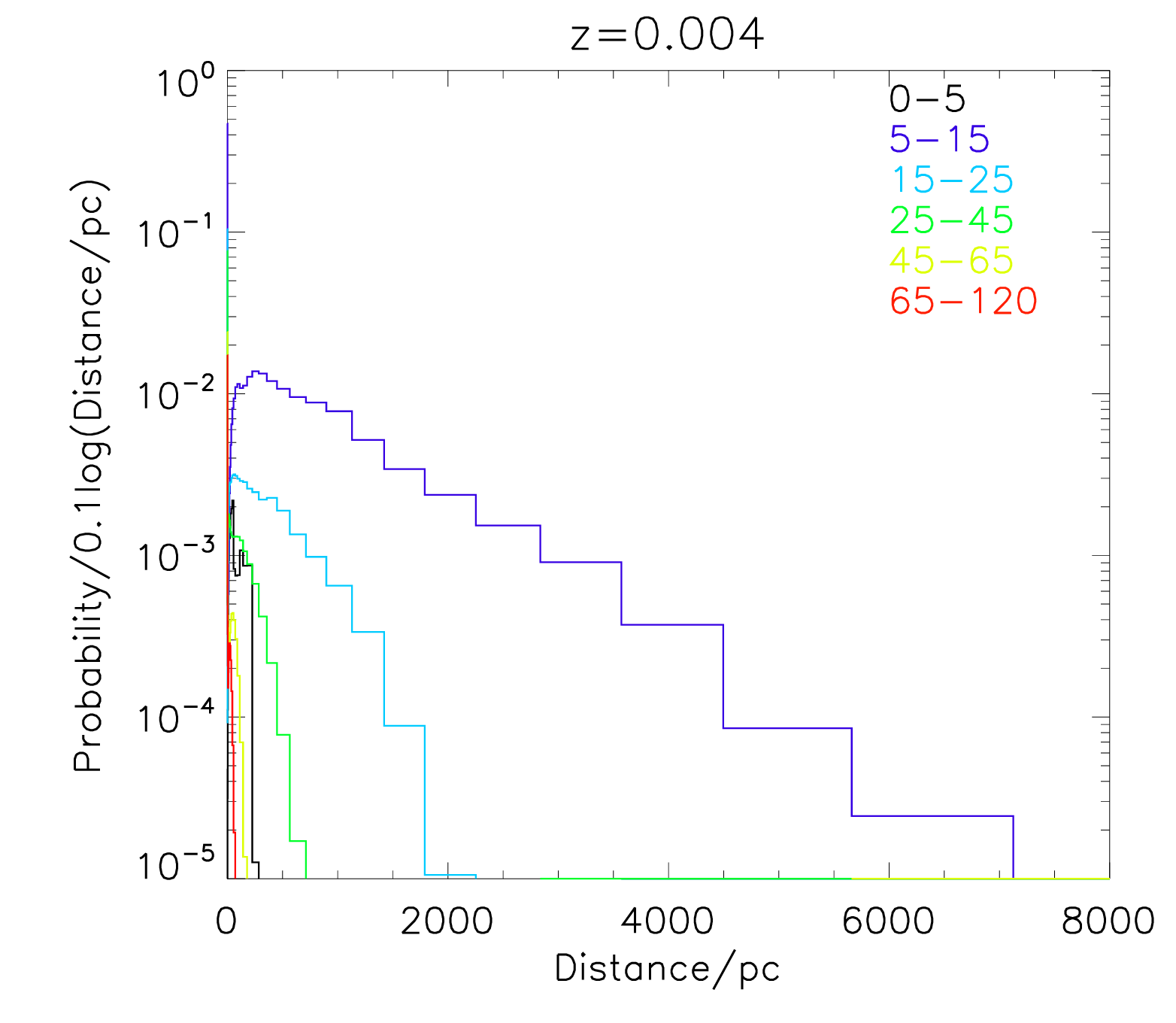}
\includegraphics[angle=0,width=80mm]{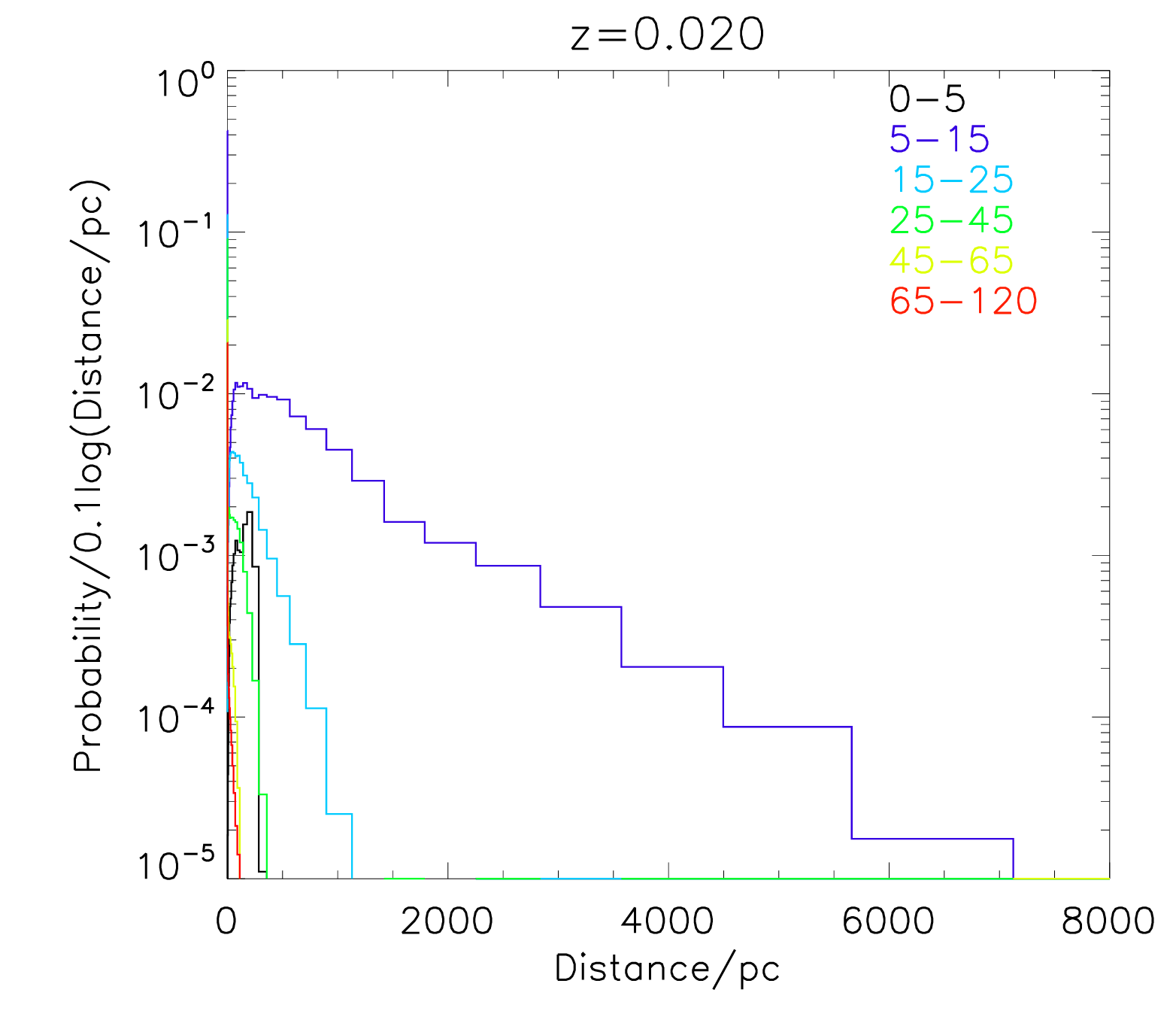}
\caption{Similar to the upper left panels in Figures
  \ref{starvelocitiesTYPE} and \ref{starvelocitiesTYPE2} but with a
  linear scale on the x-axis and the lines now represent different
  initial mass ranges for the progenitor stars.}
\label{s3}
\end{figure*}

\begin{figure*}
\includegraphics[angle=0,width=80mm]{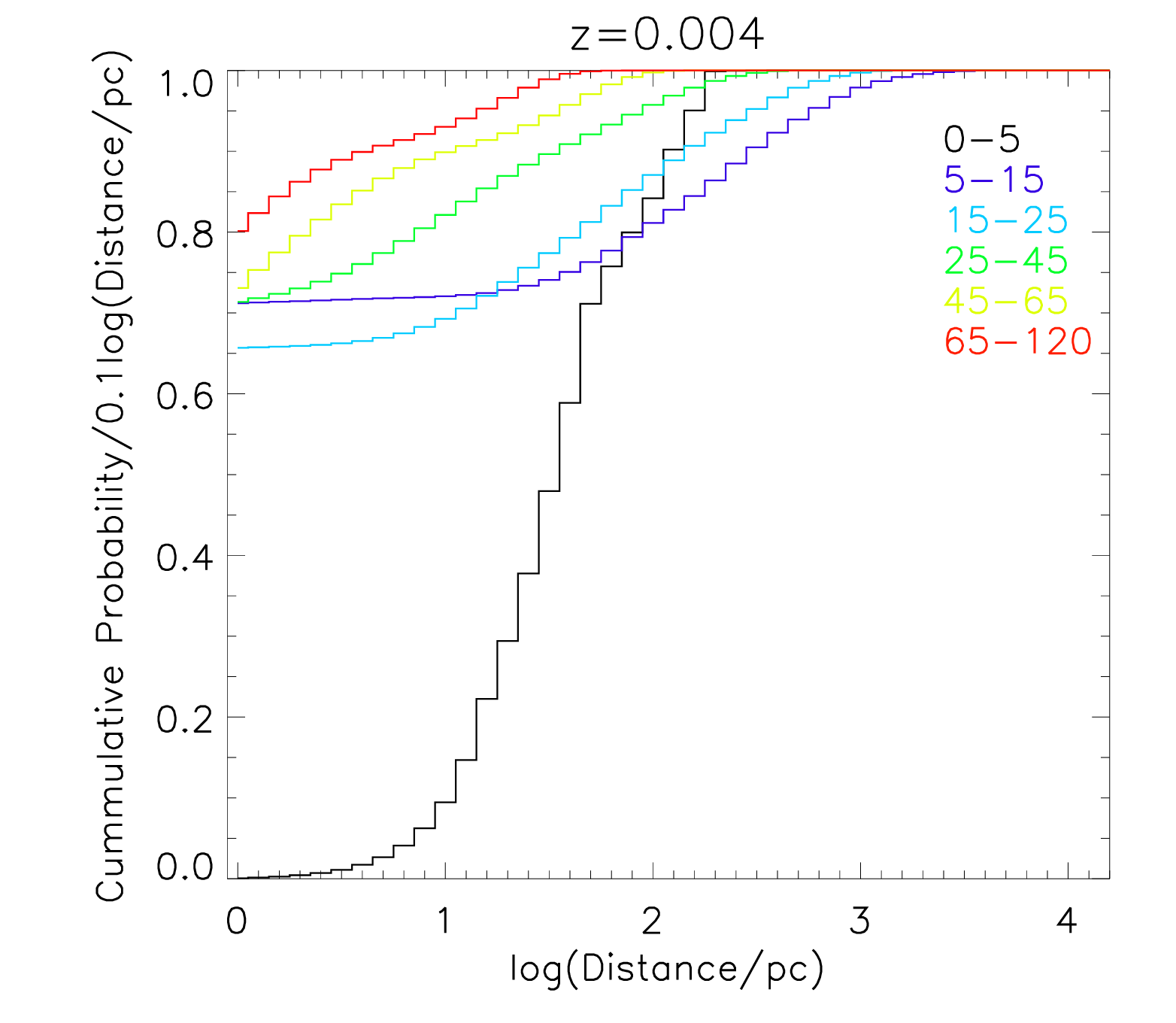}
\includegraphics[angle=0,width=80mm]{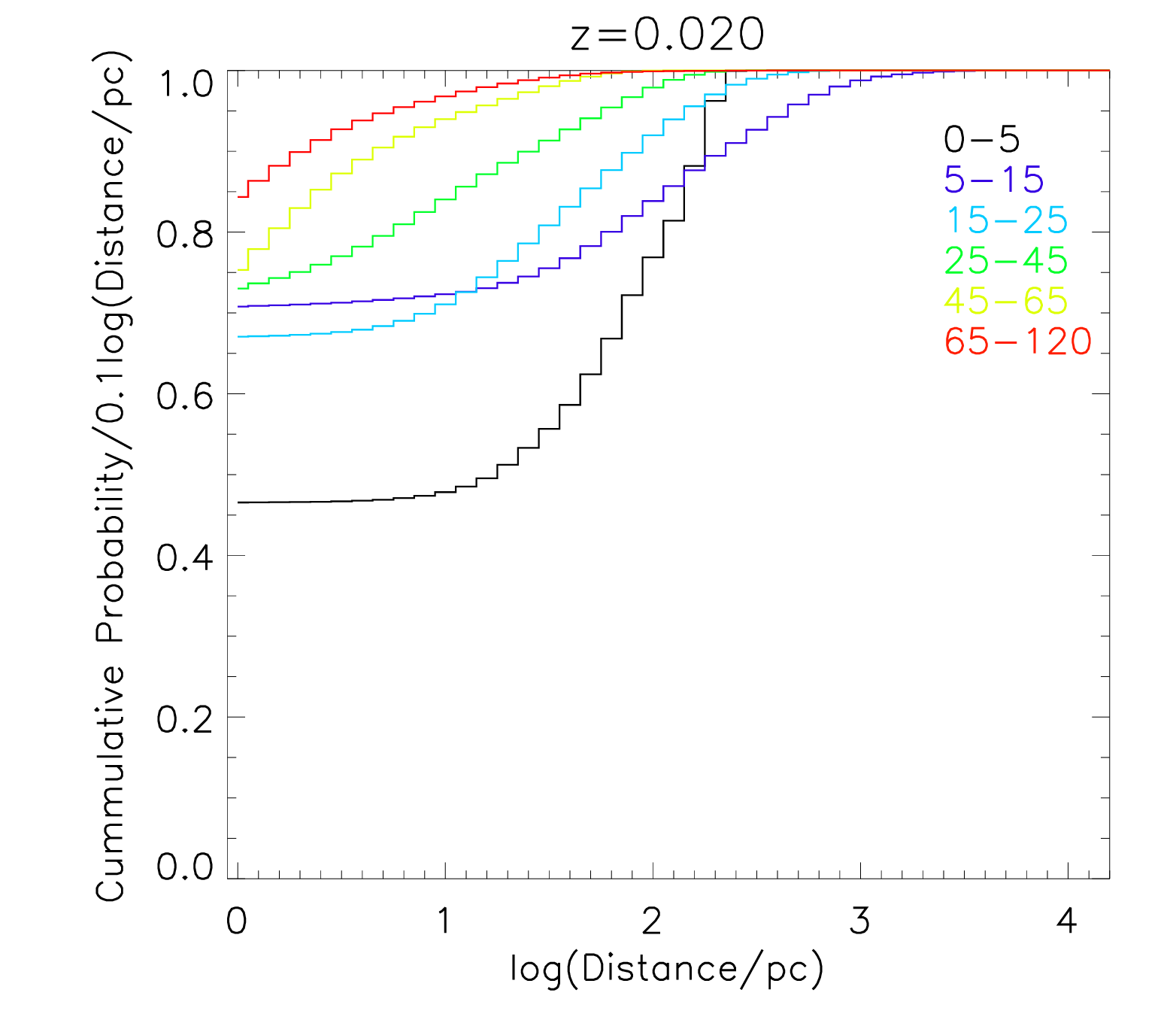}
\caption{Similar to the upper left panels in Figures
  \ref{starvelocitiesTYPE} and \ref{starvelocitiesTYPE2} with the
  cummulative probability with increasing progenitor distance on the
  y-axis with a linear scale with a linear scale. }
\label{s6}
\end{figure*}

\begin{figure*}
\includegraphics[angle=0,width=80mm]{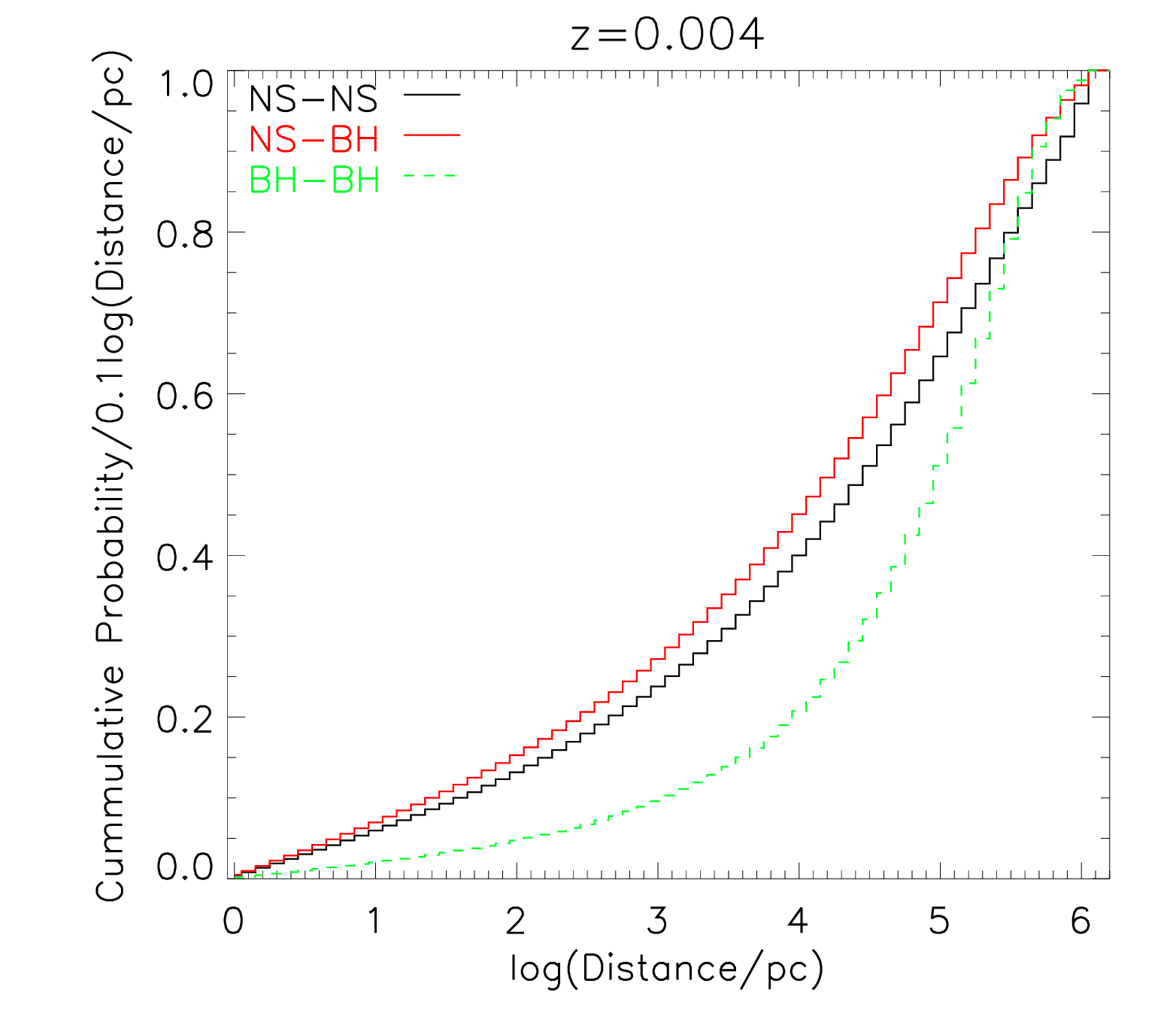}
\includegraphics[angle=0,width=80mm]{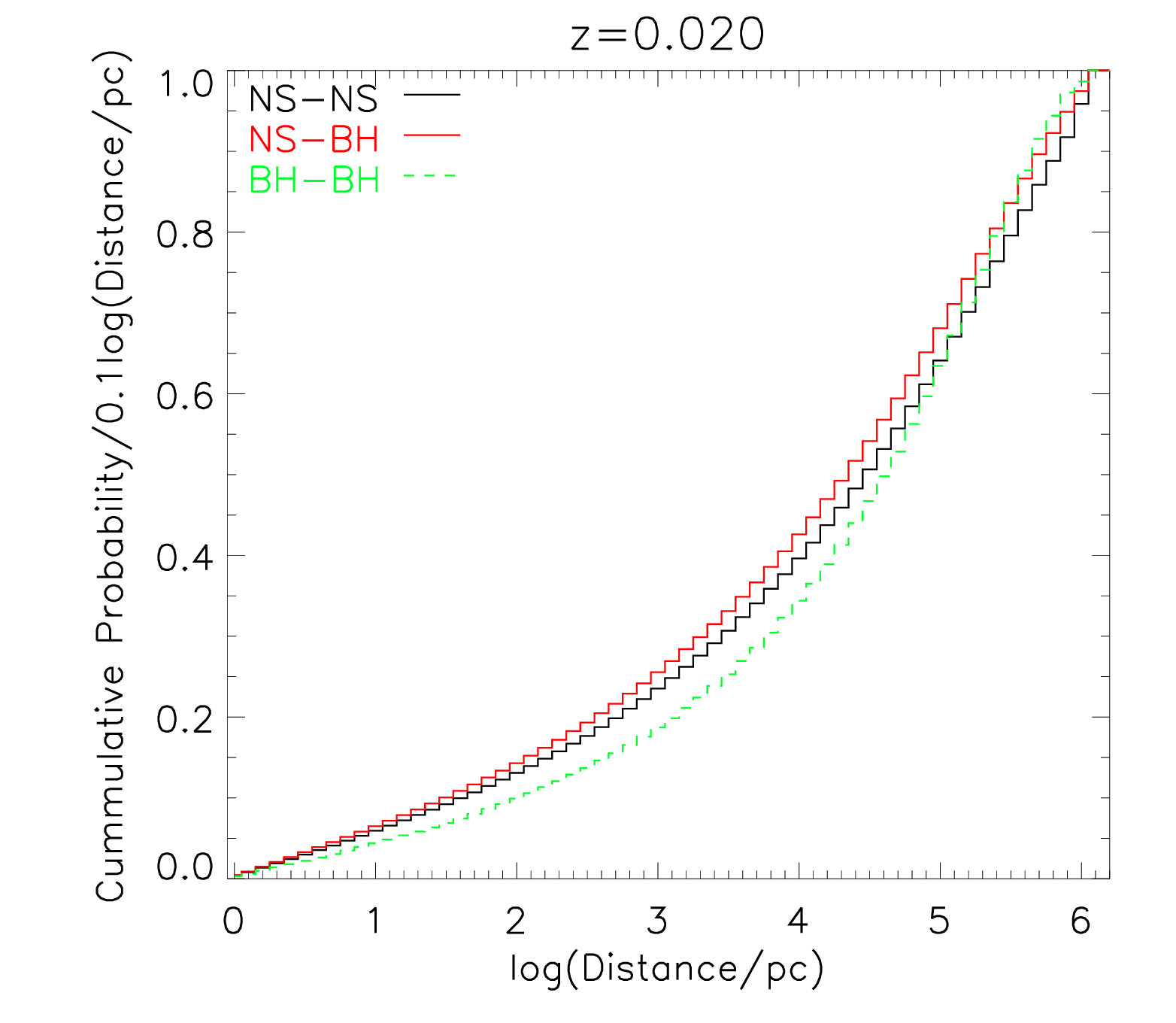}
\caption{Similar to the upper left panels in Figures
  \ref{starvelocitiesTYPE} but now for the mergers of compact objects
  shown in Figure \ref{shortgrbs2} with the cummulative probability of
  a merger versus increasing progenitor distance on the y-axis with a
  linear scale with a linear scale.}
\label{s4}
\end{figure*}

In this appendix, we present additional results from our simulations.
These are placed here not to divert the main results we wished to
cover. These Figures contain similar information to that presented in
the main paper. However these provide greater clarity on some specific
results of the simulation.  Fig. \ref{mainrunaways2appendix} is
similar to Fig. \ref{mainrunaways2} but with the range of velocities
included widened to cover velocities less than those typically
expected for runaway stars. Fig. \ref{s2} is similar to
Fig. \ref{starvelocitiesTYPE} but is over a linear distance
scale. Fig. \ref{s5} is also similar to
Fig. \ref{starvelocitiesTYPE} but is for cummulative probability. Then
Figs. \ref{s3} and \ref{s6} are also similar to these plots
but split the SN events up into different initial-mass ranges rather
than SN types. Finally Figs. \ref{s4} are similar to the
other plots but show results for the compact-object mergers rather
than SNe.

\label{lastpage}
\bsp

\end{document}